\documentclass[prd,reprint,superscriptaddress,altaffillsymbol,amsmath,nofootinbib,longbibliography]{revtex4-1}
\pdfoutput=1

\usepackage{ifthen}
\usepackage{epsfig}
\usepackage{booktabs} 
\usepackage{natbib}
\usepackage{wasysym}

\usepackage{slashed} 
\usepackage{bbm}
\usepackage{mathrsfs}
\usepackage{amssymb}
\usepackage{dsfont}
\usepackage[export]{adjustbox}

\renewcommand{\vec}[1]{\mathbf{#1}}
\newcommand{\unitvector}[1]{\hat{\mathbf{#1}}}
\newcommand{\norm}[1]{\left|#1\right|}
\newcommand{\vesc}[1]{v_\mathrm{esc}\left(#1\right)}

\usepackage{color}

\definecolor{ColorSun1}{rgb}{1.0,0.75,0.11}
\definecolor{ColorSun2}{rgb}{0.72,0.09,0.01}

\usepackage[
colorlinks=true, 
linkcolor=ColorSun2,
citecolor=ColorSun2,
urlcolor=ColorSun2,
]{hyperref}

\allowdisplaybreaks

\newcommand{\erf}{\mathop{\mathrm{erf}}}
\newcommand{\dd}{\mathrm{d}}

\begin{document}

\title{Solar reflection of light dark matter with heavy mediators}

\author{Timon Emken}
\email{timon.emken@fysik.su.se}
\affiliation{Chalmers University of Technology, Department of Physics, SE-412 96 G\"oteborg, Sweden}
\affiliation{The Oskar Klein Centre, Department of Physics, Stockholm University, AlbaNova, SE-10691 Stockholm, Sweden}

\begin{abstract}
The direct detection of sub-GeV dark matter particles is hampered by their low energy deposits.
If the maximum deposit allowed by kinematics falls below the energy threshold of a direct detection experiment, it is unable to detect these light particles.
Mechanisms that boost particles from the galactic halo can therefore extend the sensitivity of terrestrial direct dark matter searches to lower masses.
Sub-GeV and sub-MeV dark matter particles can be efficiently accelerated by colliding with thermal nuclei and electrons of the solar plasma respectively.
This process is called `solar reflection'.
In this paper, we present a comprehensive study of solar reflection via electron and/or nuclear scatterings using Monte Carlo simulations of dark matter trajectories through the Sun.
We study the properties of the boosted dark matter particles, obtain exclusion limits based on various experiments probing both electron and nuclear recoils, and derive projections for future detectors.
In addition, we find and quantify a novel, distinct annual modulation signature of a potential solar reflection signal which critically depends on the anisotropies of the boosted dark matter flux ejected from the Sun.
Along with this paper, we also publish the corresponding research software.
\end{abstract}

\maketitle

\section{Introduction}
\label{sec:introduction}

Around the globe, a great variety of direct detection experiments are searching for the dark matter~(DM) of our galaxy.
These experiments attempt to verify the hypothesis that the majority of matter in the Universe consists of new particles which occasionally interact and scatter with ordinary matter through a non-gravitational portal interaction~\cite{Goodman:1984dc,Wasserman:1986hh,Drukier:1986tm}.
They are motivated by a collection of astronomical observations of gravitational anomalies on astrophysical and cosmological scales, which point to the presence of vast amounts of invisible matter which governs the dynamics of galaxies, galaxy clusters, and the cosmos as a whole~\cite{Bertone:2004pz,Bertone:2016nfn}.

Originally, direct detection experiments were guided by the so-called `WIMP-miracle', looking for nuclear recoils caused by Weakly Interacting Massive Particles (WIMPs) with electroweak scale masses and interactions.
Large-scale experiments such as XENON1T have been very successful in probing and excluding large portions of the parameter space, setting tight constraints on the WIMP paradigm~\cite{Aprile:2018dbl}.
Over the last decade, the search strategy has broadened more and more to include detection for new particles with masses below a GeV.
With the exception of a few low-threshold direct detection experiments such as CRESST~\cite{Angloher:2015ewa,Angloher:2017sxg,Abdelhameed:2019hmk}, nuclear recoils caused by sub-GeV~DM particles are typically too soft to be observed.
One way to probe nuclear interactions of low-mass DM is to exploit the Migdal effect or Bremsstrahlung, i.e. the emission of an observable electron or photon respectively after an otherwise unobservable nuclear recoil~\cite{Kouvaris:2016afs,Ibe:2017yqa}.
This strategy has been applied for different experiments to extend the experimental reach such as CDEX-1B~\cite{Liu:2019kzq}, EDELWEISS~\cite{Armengaud:2019kfj}, LUX~\cite{Akerib:2018hck}, liquid argon detectors~\cite{GrillidiCortona:2020owp}, or XENON1T~\cite{Aprile:2019jmx}.

However, the main shift of strategy that enabled direct searches for sub-GeV~DM was to look for DM-electron interactions instead of nuclear recoils~\cite{Kopp:2009et,Essig:2011nj}.
At this point, the null results of a number of electron scattering experiments constrain the sub-GeV~DM paradigm.
Some of these experiments probe ionization in liquid noble targets, namely XENON10~\cite{Angle:2011th,Essig:2012yx,Essig:2017kqs}, XENON100~\cite{Aprile:2016wwo,Essig:2017kqs}, XENON1T~\cite{Aprile:2019xxb}\footnote{Most recently, XENON1T has reported an excess of electron recoil events, whose unknown origin is currently being studied~\cite{Aprile:2020tmw}. Solar reflection has been among the many proposed explanations~\cite{Chen:2020gcl}.}, DarkSide-50~\cite{Agnes:2018oej}, and PandaX~\cite{Cheng:2021fqb}.
Other experiments look for electron excitations in (semiconductor) crystal targets setting constraints on DM~masses as low as 500~keV~\cite{Graham:2012su,Essig:2015cda,Lee:2015qva}, most notably SENSEI~\cite{Tiffenberg:2017aac,Crisler:2018gci,Abramoff:2019dfb,Barak:2020fql}, DAMIC~\cite{Aguilar-Arevalo:2019wdi}, EDELWEISS~\cite{Arnaud:2020svb}, and SuperCDMS~\cite{Agnese:2018col,Amaral:2020ryn}.
For even lower masses, these experiments are not able to observe DM-electron interactions, since the kinetic energy of halo DM~particles would not suffice to excite or ionize electrons.
One approach to achieve sensitivity of terrestrial searches to sub-MeV DM~mass is the development of new detection strategies and detector technologies, and the application of more exotic condensed matter systems with low, sub-eV energy gaps as target materials, see e.g.~\cite{Hochberg:2017wce,Griffin:2019mvc,Geilhufe:2019ndy}.

An alternative approach, which requires neither new detection technologies nor additional theoretical assumptions, is the universal idea to identify processes in the Milky Way which \emph{accelerate} DM~particles, and predict their phenomenology in direct detection experiments.
Such particles could have gained enough energy to trigger a detector, which can thereby probe masses far below what was originally expected based on standard halo DM alone.
In addition, the high-energy DM~population often features a distinct phenomenology and new signatures in direct detection experiments, which could help in distinguishing them from both ordinary halo~DM and backgrounds.
One way for a strongly interacting DM~particle to gain kinetic energy is to `up-scatter' by a cosmic ray particle~\cite{Yin:2018yjn,Bringmann:2018cvk,Ema:2018bih,Cappiello:2019qsw,Bondarenko:2019vrb,Wang:2019jtk}.
Existing experiments can search for the resulting (semi-)relativistic particle population.
Similarly, the Sun can act as a DM~accelerator boosting halo particle into the solar system~\cite{Kouvaris:2015nsa,An:2017ojc,Emken:2017hnp,Zhang:2020nis}.

The idea of solar reflection, i.e. the acceleration of DM~particles of the galactic halo by scatterings with thermal solar nuclei or electrons was first proposed in~\cite{An:2017ojc, Emken:2017hnp}.
At any given time, a large number of DM~particles are falling into the gravitational well of the Sun and passing through the solar plasma.
Along their underground trajectories there is a certain probability of scattering on an electron or nucleus.
This way, a light DM~particle can gain kinetic energy through one or multiple scatterings before getting ejected from the Sun.
As a consequence, we expect a flux of solar reflected DM (SRDM)~particles streaming radially from the Sun.
Terrestrial DM~detectors can look for these particles the same way as for halo~DM with the difference that SRDM particles can have more kinetic energy than even the fastest of the halo particles.
Solar reflection does not require additional assumptions, as the interaction probed in the detector is typically the same as the one in the Sun boosting the DM~particle.
Therefore, an additional, often highly energetic population of solar reflected DM~particles in the solar system is an intrinsic feature of any DM~particle scenario with sizeable DM-matter interaction rates.

\begin{figure}[t!]
    \centering
    \includegraphics[width=0.35\textwidth]{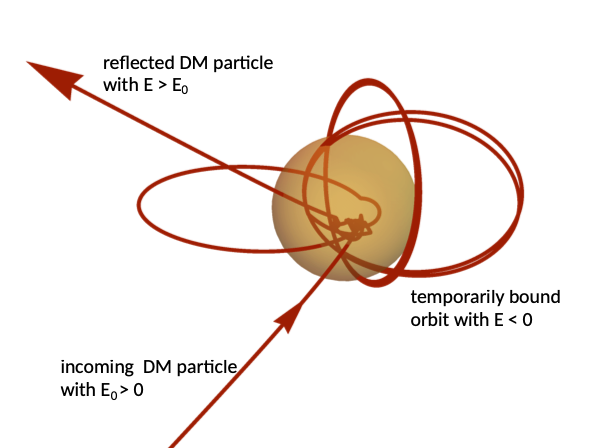}
    \caption{An illustrative example of a 100~MeV mass DM particle's trajectory through the Sun obtained with MC~simulations. After having scattered around 60 times on solar nuclei, getting decelerated, accelerated, and temporarily captured gravitationally, it finally gets reflected with around 10 times its initial kinetic energy.} 
    \label{fig: trajectory}
\end{figure}

In a previous work~\cite{Emken:2017hnp}, we established and studied the SRDM~flux caused by a single nuclear scattering inside the Sun with analytic methods.
We explored the prospects of detecting SRDM by extending the theory of gravitational capture of WIMPs developed by Press, Spergel, and Gould~\cite{Press:1985ug,Gould:1987ir,Gould:1987ju,Gould:1991hx}.
These analytic results were understood to be conservative, as contributions to the SRDM flux caused by multiple scatterings are not accounted for by the analytic formalism.
Their contributions are best described using Monte Carlo~(MC) simulations of DM~trajectories through the Sun.
Figure~\ref{fig: trajectory} depicts an example trajectory of a DM~particle traversing the solar plasma, where we assumed a DM~mass of 100~MeV and spin-independent nuclear interactions.
It can be seen that the particle scatters many times losing and gaining energy in the process.
While it gets gravitationally bound temporarily, in the end the DM~particle escapes the Sun with increased speed.

The first simulations of DM~traversing the Sun were performed in the context of energy transport and WIMP evaporation~\cite{Nauenberg:1986em}.
Here, the MC~approach was used to evaluate the distribution of~DM inside the Sun in comparison to the analytic work by Press and Spergel~\cite{Spergel:1984re,Press:1985ug}.
Similar DM~simulations were also used to study the impact of underground scatterings inside the Earth and to quantify daily signal modulations~\cite{Collar:1993ss,Collar:1992qc,Hasenbalg:1997hs,Emken:2017qmp,Kavanagh:2020cvn} and the loss of underground detectors' sensitivity to strongly-interacting~DM~\cite{Emken:2017erx,Mahdawi:2017cxz,Emken:2018run,Mahdawi:2018euy,Emken:2019tni}.
In the context of solar reflection, MC~simulations were used for the scenario where~DM interacts with electrons only~\cite{An:2017ojc}.

In this paper, we extend this work and present the most comprehensive MC~study to date of solar reflection of sub-GeV and sub-MeV DM~particles and their detection in terrestrial laboratories.
In addition to leptophilic~DM, solar reflection via nuclear scatterings is studied using MC~simulations for the first time.
We also consider a DM~model where nuclear and electron interactions are present simultaneously, which has not been considered before.
With respect to the MC~simulations, this paper includes an exhaustive documentation of the involved physical processes with a high level of detail.
Furthermore, our simulations improve upon previous studies and represent the most general and thorough description of solar reflection so far by using advanced numerical methods and following high software engineering standards.

Using these simulations, we generate precise MC~estimates of the DM~flux emitted from the Sun and passing through the Earth, and we investigate various aspects and properties of the SRDM~particles.
The inclusion of these particles in direct detection analyses can extend the sensitivity of the respective detectors to lower masses.
Both for nuclear and electron recoil searches, we derive SRDM~exclusion limits of existing experiments and compare those to the standard halo limits.
Furthermore we obtain projections for next-generation experiments and study the prospects of future searches for SRDM.
Beyond constraints, we also focus on the phenomenology of a potential DM~signal from the Sun.
In particular, we predict a novel, annual signal modulation resulting from a non-trivial combination of the anisotropy of the solar reflection flux and the eccentricity of the Earth's orbit.

Lastly, a central result of this work is the simulation code itself.
The MC simulation tool \textit{Dark Matter Simulation Code for Underground Scattering - Sun Edition} (\texttt{DaMaSCUS-SUN}) which was used to obtain our results is the first publicly available and ready to use code describing solar reflection~\cite{Emken2021}.

Concerning the paper's structure, we start with a general discussion and description of the idea and phenomenology of solar reflection in Sec.~\ref{sec:solar reflection}.
Section~\ref{sec:monte carlo} describes the MC~simulations implemented in the~\texttt{DaMaSCUS-SUN} code.
This is followed by a review of the DM~models and interactions considered in this work in Sec.~\ref{sec:DM models}.
The last two chapters,~\ref{sec:results} and~\ref{sec:discussion}, discuss and summarize our findings.
Lastly, two appendices contain more details on the trajectory simulation.
Appendix~\ref{app: simulation details} reviews the equations of motion and their analytic and numeric solutions.
This appendix also covers the generation of initial conditions and sampling of target velocities in greater detail than the main body of the paper.
The solar model used for this work is described and reviewed in App.~\ref{app:solar model}.
Throughout this paper, we use natural units with~$c=\hbar=k_B=1$.

\section{Solar reflection of light DM}
\label{sec:solar reflection}

In this section, we present the idea of solar reflection as a process of accelerating DM~particles and how these boosted particles could be detected.
We start by reviewing the Standard Halo Model~(SHM) and how the hard speed cutoff of the SHM translates into a lower bound on observable DM~masses, a limit that can be circumvented by taking solar reflection into account.

\subsection{Dark matter in the galactic halo}
\label{ss: dark matter in halo}

A central source of uncertainty in making predictions for direct detection experiments is the halo model, i.e. the assumptions about the local properties of the DM~particles of the galactic halo~\cite{Nesti:2013uwa}.
The conventional choice is the SHM, which models the local~DM as a population of particles with constant mass density~$\rho_\chi\approx 0.4\text{ GeV cm}^{-3}$~\cite{Catena:2009mf,Read:2014qva} following a Maxwell-Boltzmann velocity distribution truncated at the galactic escape velocity~$v_\mathrm{gal}\approx 544\text{ km sec}^{-1}$~\cite{Smith:2006ym},\footnote{The results of this paper do not depend critically on these choices, as we will demonstrate in Sec.~\ref{ss: results spectrum}.}
\begin{subequations}
\label{eq: SHM velocity distribution}
\begin{align}
    f_{\rm halo}(\mathbf{v}) &= \frac{1}{N_\mathrm{esc}\pi^{3/2}v_0^3} \exp\left(- \frac{\mathbf{v}^2}{v_0^2} \right)\Theta(v_\mathrm{gal}-|\mathbf{v}|)\, ,
    \intertext{where the normalization constant reads}
    N_\mathrm{esc}&\equiv \erf\left({v_\mathrm{gal} \over v_0}\right)  - {2 \over \sqrt{\pi}} \frac{v_\mathrm{gal}}{v_0}\exp\left(-\frac{v_\mathrm{gal}^2}{v_0^2}\right)\, .
\end{align}
\end{subequations}
Here, the velocity dispersion~$v_0\approx 220\text{ km sec}^{-1}$ is set to the Sun's circular velocity~\cite{Kerr:1986hz}, and~$\Theta(x)$ is the unit step function.

To describe the DM~distribution in the rest frame of a direct detection experiment moving with velocity~$\mathbf{v}_\mathrm{obs}$, this distribution needs to be transformed via a Galilean boost,
\begin{align}
    f_\mathrm{obs}(\mathbf{v}) &= f_\mathrm{halo}(\mathbf{v} + \mathbf{v}_\mathrm{obs})\, . \label{eq: SHM boost}
\end{align}
As such, the maximum speed a DM~particles can pass through the detector with is~$v_\mathrm{max} = v_\mathrm{gal} +|\vec{v}_\mathrm{obs}|$, which directly corresponds to the maximum nuclear recoil energy they can possibly induce in a collision with a nucleus of mass~$m_N$, given by~$E_R^\mathrm{max}=\frac{2\mu^2_{\chi N}}{m_N}v_\mathrm{max}^2$.
Assuming a DM~mass of~$m_\chi$, $\mu_{\chi N}$ denotes the reduced mass of the DM~particle and the nucleus.
If $E_R^\mathrm{max}$ falls below the experimental nuclear recoil threshold~$E_\mathrm{thr}$, the DM~particles do not have enough kinetic energy to trigger the detector.
Hence, DM can only be detected by a nuclear recoil experiment if
\begin{align}
    m_\chi \gtrsim \frac{m_N}{\sqrt{2 m_N \over E_\mathrm{thr}}v_\mathrm{max}-1}\, . \label{eq: minimum mass nucleus}
\end{align}
Experimental measures to probe lower DM~masses are therefore the construction of detectors with lower thresholds and lighter nuclear target masses, a strategy followed e.g. by the CRESST experiments~\cite{Angloher:2015ewa,Angloher:2017sxg,Abdelhameed:2019hmk}.
A similar argument applies to DM-electron scattering experiments, where a DM~particle can essentially transfer all its kinetic energy to the electron provided that it exceeds the energy gap~$E_{\rm gap}$ for excitations or ionizations,
\begin{align}
    m_\chi \gtrsim {2 E_\mathrm{gap}\over v_\mathrm{max}^2}\, . \label{eq: minimum mass electron}
\end{align}

In this paper, instead of considering the possibility to detect standard halo DM, we explore the detection of a different population of particles with higher kinetic energies, generated and accelerated by~\emph{solar reflection}.
Considering a DM~population with a higher maximum speed~$v_\mathrm{max}$ than present in the SHM extends the discovery reach of direct detection experiments to lower masses following Eqs.~\eqref{eq: minimum mass nucleus} and~\eqref{eq: minimum mass electron}.
Assuming that DM interacts with ordinary matter through non-gravitational interactions (i.e. the general assumption of direct detection experiments), the solar reflected dark matter (SRDM) component is an intrinsic part of the DM population in the solar system, and does not rely on further assumptions.

It is crucial for its description to understand the orbits of DM~particles as they pass through the Sun.

\subsection{Dark matter in the Sun}
\label{ss: dark matter in sun}

\noindent\textbf{Falling into the Sun: }
Incoming DM~particles approach the Sun following a hyperbolic Kepler orbit (see App.~\ref{app: kepler}), getting focused and accelerated by the gravitational pull.
For a given DM~speed~$u$ asymptotically far away from the Sun, energy conservation determines the particle speed~$w(u,r)$ at a finite distance~$r$,
\begin{align}
    w(u,r) &= \sqrt{u^2 + \vesc{r}^2}\, , \label{eq: blue shift}
\end{align}
where~$\vesc{r}$ is the local escape velocity from the Sun's gravitational well (not to be confused with the galactic escape velocity~$v_\mathrm{gal}$).
Outside the Sun, it is given by~$\vesc{r}=\sqrt{\frac{2G_N M_\odot}{r}}$, where~$G_N$ is Newton's constant and~$M_\odot$ is the total solar mass.
Inside the Sun, the local escape velocity is given by Eq.~\eqref{eq: sun escape velocity} of App.~\ref{app:solar model}, which summarizes the chosen solar model~\cite{Serenelli:2009yc}.

\begin{figure}[h!]
    \centering
    \includegraphics[width=0.45\textwidth]{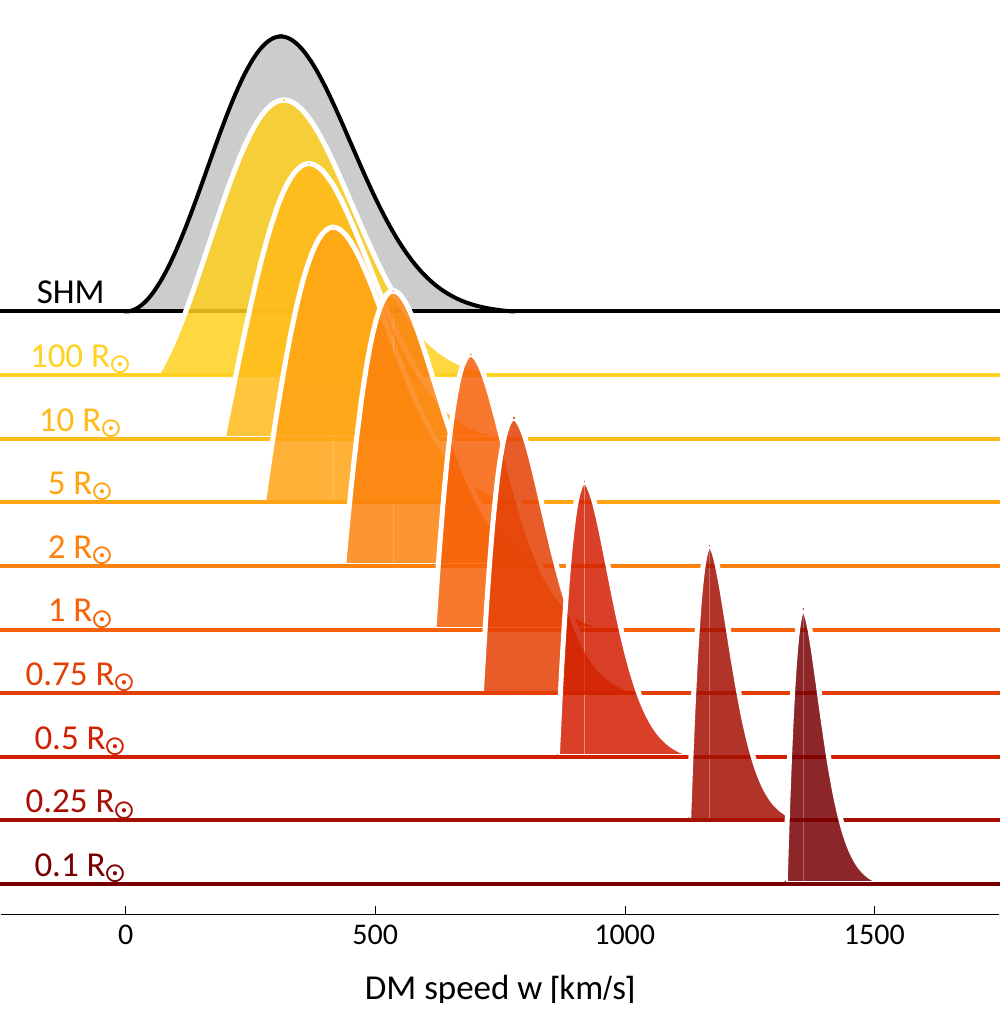}
    \caption{Speed distributions~$f_\mathrm{obs}(w,r)$ of accelerating halo DM falling into the Sun's gravitational well as described by Eq.~\eqref{eq: liouville}.}
    \label{fig: SHM blueshift}
\end{figure}

While the determination of the particle's full velocity vector~$\mathbf{w}$ close to the Sun requires a few additional steps (see App.~\ref{app: kepler}), the infalling particles' local speed distribution in the solar neighbourhood can be found using Liouville's theorem~\cite{Gould:1991hx},
\begin{align}
    f_\odot(w,r) = f_\odot(u(w,r))\, , \label{eq: liouville}
\end{align}
where~$u(w,r)$ is the inversion of Eq.~\eqref{eq: blue shift}, and~$f_\odot(v)$ is the speed distribution in the Sun's rest frame given by Eq.~\eqref{eq: SHM boost} with~$\mathbf{v}_\mathrm{obs}=\mathbf{v}_\odot$.
Here, $\mathbf{v}_\odot$ is the Sun's velocity in the galactic rest frame given in Eq.~\eqref{eq: Sun velocity}.
The resulting distributions are shown in Fig.~\ref{fig: SHM blueshift} for selected distances from the solar center.

If the periapsis of a DM~particle's orbit is smaller than the Sun's radius~$R_\odot$, it will pass the solar surface and propagate through the hot plasma of the Sun.
We can compute the total rate~$\Gamma$ of DM~particle passing into the Sun to be~\cite{Gould:1987ir,Emken:2017hnp} 
\begin{subequations}
\label{eq: total entering rate}
\begin{align}
    \Gamma(m_\chi) &= n_\chi \pi R_\odot^2 \int \dd u\; f_\odot(u)\left(u + {\vesc{R_\odot}^2 \over u} \right) \\
    &= \frac{\rho_\chi}{m_\chi}\pi R_\odot^2 \left( \langle u  \rangle +  \vesc{R_\odot}^2 \langle u^{-1}  \rangle\right)\\
    &\approx 1.1 \cdot 10^{33} \left(\frac{m_\chi}{\text{MeV}}\right)^{-1} \text{ s}^{-1}\, ,
\end{align}
\end{subequations}
where $\vesc{R_\odot} \approx 618 \text{ km sec}^{-1}$~\cite{Serenelli:2009yc}.

\vspace{10pt}\noindent\textbf{Scattering rate: }
While the particle moves along its no-longer Keplerian orbit through the Sun's bulk mass, there is a probability of scattering on constituents of the solar plasma, either thermal nuclei or electrons, depending on the DM-matter interaction cross sections and the particle's location and speed.
The total scattering rate as a function of the particles radial distance~$r$ to the solar center, and its velocity~$\vec{v}_\chi$ is given by~\cite{Press:1985ug,Gould:1987ir,Gould:1987ju},
\begin{align}
	\Omega(r,\vec{v}_\chi) = \sum_i n_i(r) \langle \sigma_i|\vec{v}_\chi-\mathbf{v}_{T,i}|\rangle\, . \label{eq: total scattering rate}
\end{align}
In this expression, the index~$i$ runs over all solar targets, i.e. electrons and the various nuclear isotopes that make up the star.
We obtain their number densities~$n_i(r)$ as part of the solar model, in our case the Standard Solar Model (AGSS09)~\cite{Serenelli:2009yc}, which is reviewed in App.~\ref{app:solar model}.
Furthermore, $\sigma_i$ denotes the total DM scattering cross section with the $i^\mathrm{th}$ target and depends on the assumed DM~particle model (see Sec.~\ref{sec:DM models}), and~$\mathbf{v}_T$ is the target's velocity.
The brackets~$\langle \cdot\rangle$ denote a thermal average.
In cases where the total cross section is independent of the relative speed~$|\vec{v}_\chi-\mathbf{v}_T|$, as is the case e.g. for spin-independent contact interactions, we can exploit that~$\langle \sigma|\vec{v}_\chi-\mathbf{v}_T|\rangle = \sigma\langle |\vec{v}_\chi-\mathbf{v}_T|\rangle$ applies.
Assuming the solar targets' speed follows a Maxwell-Boltzmann distribution, we can evaluate the thermal average of the relative speed analytically,
\begin{align}
    \langle |\vec{v}_\chi-\mathbf{v}_{T,i}|\rangle &=\int\dd^3\mathbf{v}_T\;|\vec{v}_\chi-\mathbf{v}_T| f_i(\mathbf{v}_T,T)\nonumber\\
    &=\frac{1+2\kappa_i^2v_\chi^2}{2\kappa_i^2v_\chi}\erf{(\kappa_i v_\chi)} + \frac{1}{\sqrt{\pi}\kappa_i}e^{-\kappa_i^2v_\chi^2}\, ,\label{eq: thermal average of relative speed}
\end{align}
where~$v_\chi\equiv|\vec{v}_\chi|$.
We used the probability density function~(PDF) of the Maxwell-Boltzmann distribution for a target particle of mass~$m_i$ and temperature~$T$,
\begin{align}
    f_i(\mathbf{v}_T,T)&= \frac{\kappa^3}{\pi^{3/2}}e^{-\kappa_i^2 v_T^2}\, , \label{eq: maxwell boltzmann}
\end{align}
with $\kappa_i\equiv \sqrt{m_i \over 2 T}$.

\begin{figure*}
    \centering
    \includegraphics[width=0.4\textwidth]{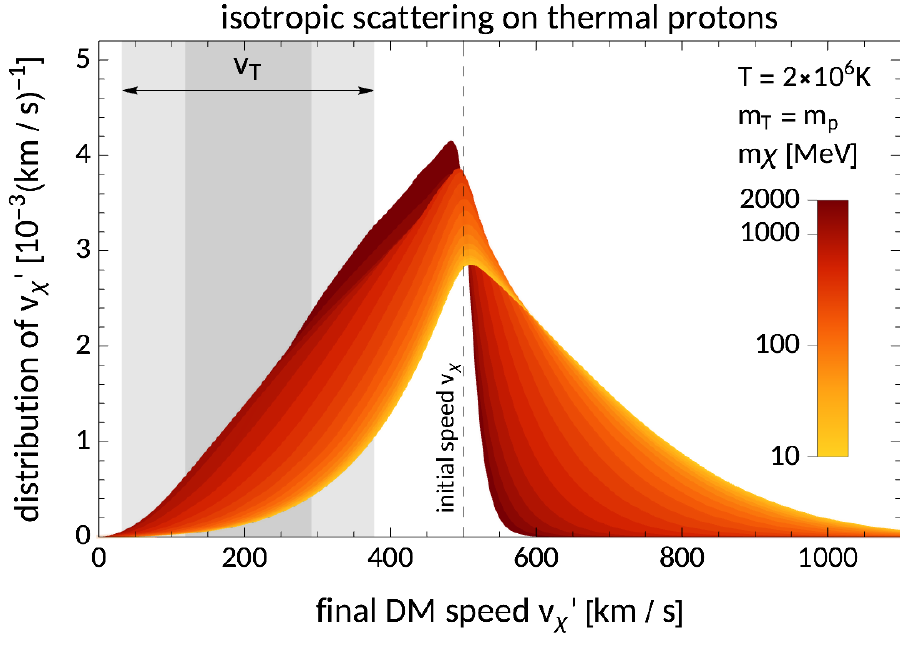}
    \includegraphics[width=0.41\textwidth]{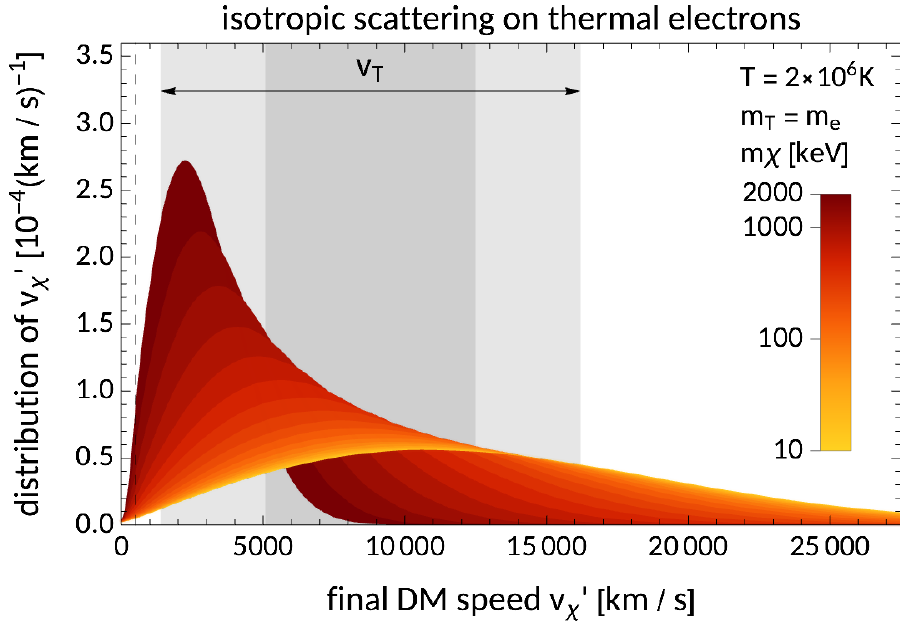}
    \caption{Distribution of the DM speed~$v_\chi^\prime$ after scattering isotropically on a thermal proton (left) and electron (right). The vertical dashed line shows the assumed initial speed~$v_\chi=500\text{ km/s}$. The colors denote various DM~masses. The intervals shaded in (dark) gray indicate the $1\sigma$($2\sigma$) variation of the target speeds around their mean and thereby illustrate the typical speeds of the thermal targets.}
    \label{fig: kinematics}
\end{figure*}

\vspace{10pt}\noindent\textbf{Kinematics: }
By scattering on a thermal target, a DM~particle may lose or gain kinetic energy, depending on the relation of its kinetic energy to the thermal energy of the plasma.
Assuming its mass and velocity to be~$m_\chi$ and ~$\mathbf{v}_\chi$ respectively, the DM~particle's new velocity after scattering on a target of mass~$m_T$ and velocity~$\mathbf{v}_T$ is given by
\begin{align}
    \mathbf{v}^\prime_\chi &={m_T |\mathbf{v}_\chi-\mathbf{v}_T| \over m_T+m_\chi}\mathbf{n} + {m_\chi \mathbf{v}_\chi + m_T \mathbf{v}_T \over m_T+m_\chi}\, . \label{eq: kinematics}
\end{align}
Here, we introduced the unit vector~$\vec{n}$, which points toward the new DM~velocity in the center-of-mass frame of the scattering process.
The angle~$\alpha$ between~$\vec{n}$ and~$\vec{v}_\chi$ is called the scattering angle.
Hence, the new velocity, and also the question if the DM~particle got accelerated or decelerated, is determined by the scattering angle~$\alpha$ and the target's velocity~$\vec{v}_T$.
In the case of spin-independent contact interactions, the scattering is approximately isotropic, i.e.~$\cos\alpha$ follows the uniform distribution~$\mathcal{U}_{[-1,1]}$, whereas we sample~$\mathbf{v}_T$ from a thermal Maxwell-Boltzmann distribution given by Eq.~\eqref{eq: maxwell boltzmann} weighted by the velocity's contribution to the overall scattering probability (for more details see App.~\ref{app: target velocity sampling}).

The distribution of the final DM~speed~$v_\chi^\prime$ after scattering on thermal protons (electrons) with temperature~$T=2\times 10^6\text{ K}$ is shown in the left (right) panel of Fig.~\ref{fig: kinematics} depending on the DM~mass.
\footnote{These distributions are obtained by MC sampling the isotropic scattering angle~$\alpha$ and target velocity~$\vec{v}_T$, generating a large sample of final speeds~$v_\chi^\prime$. We obtain a smooth estimate of the PDFs~$f(v_\chi^\prime)$ via kernel density estimation. However, it should be noted that the distribution can also be expressed analytically for isotropic scatterings~\cite{Emken:2017hnp}.}
In the case of sub-GeV DM~masses and nuclear interactions with protons, we find that deceleration becomes more likely for heavier masses.
Above the proton mass, most DM~particles lose kinetic energy by scattering on a thermal proton.
In contrast, a lighter DM~particle e.g. with~$m_\chi=10 \text{ MeV}$ has a~$\sim50\%$ chance of getting accelerated.

A more efficient process to speed up low-mass DM~particles is a collision with a thermal electron.
Due to the lower target mass, thermal electrons are faster than the protons of the plasma.
Consequently, a scattering between a sub-MeV mass DM~particle and a thermal electron almost always results in an acceleration as seen in the right panel of Fig.~\ref{fig: kinematics}.
Just as for proton targets, lighter particles are much more likely to get accelerated to higher speeds.

\begin{figure*}
    \centering
    \includegraphics[width=0.4\textwidth]{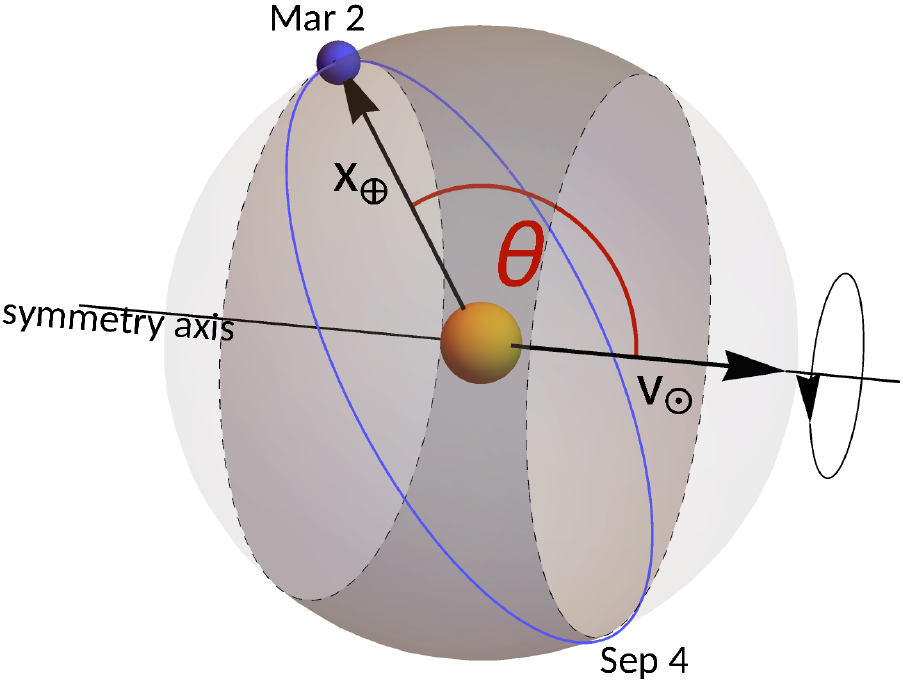}\hspace{2cm}
    \includegraphics[width=0.45\textwidth]{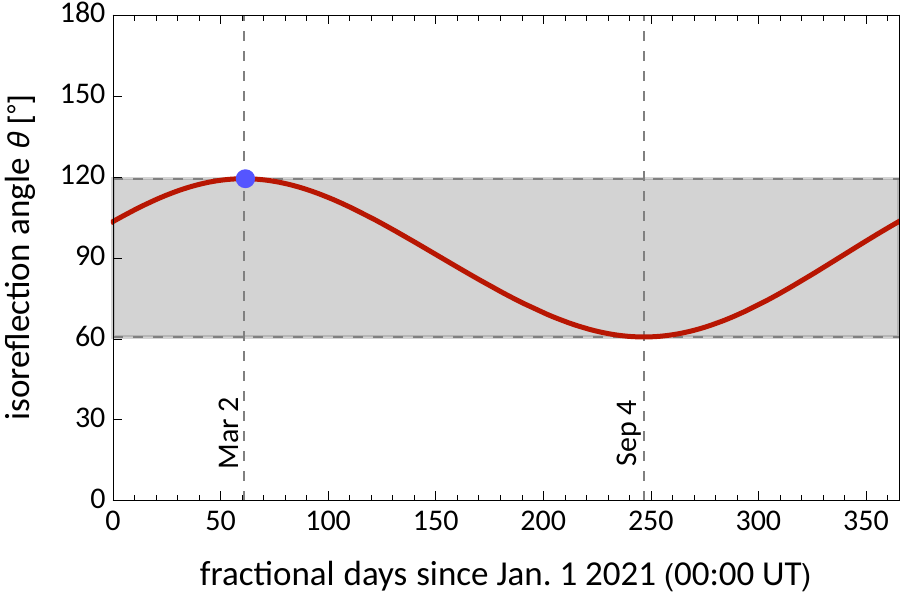}
    \caption{The left panel depicts the symmetry axis along the Sun's velocity and the isoreflection angle~$\theta$ of the Earth, i.e. the polar angle of the symmetry axis defined in Eq.~\eqref{eq: isoreflection angle}. Its time-evolution~$\theta(t)$ over the course of one year is shown in the right panel. Along its Kepler orbit, the Earth approximately covers the interval~$\theta\in[60^\circ,120^\circ]$, reaching the maximum and minimum on Mar. 2 and Sep. 4 respectively.}
    \label{fig: isoreflection angle}
\end{figure*}

\vspace{10pt}\noindent\textbf{Reflection flux estimate: }
In order to get a first idea about the resulting particle flux we can expect in our terrestrial laboratories, we can, for the moment, absorb our ignorance about the details of the reflection process into an average probability~$p_\mathrm{refl}$ of a DM~particle entering the Sun to get reflected, i.e. to scatter once or multiple times on solar nuclei and electrons and escaping the Sun afterward.
Using the total rate~$\Gamma(m_\chi)$ of halo DM~particles falling into the Sun given by Eq.~\eqref{eq: total entering rate}, we can estimate the total reflection rate~$\mathcal{R}_\odot$ and SRDM flux~$\Phi_\odot$ on Earth,
\begin{align}
\mathcal{R}_\odot(m_\chi)&=p_\mathrm{refl}\Gamma(m_\chi)\, , \label{eq: srdm rate estimate} \\
    \Phi_\odot(m_\chi)& = \frac{\mathcal{R}_\odot}{4\pi\ell^2}\nonumber\\
    &\approx  3.8\cdot 10^5\;p_{\rm refl}\left(\frac{m_\chi}{\text{MeV}}\right)^{-1}\text{ s}^{-1}\text{cm}^{-2} \label{eq: srdm flux estimate}
\end{align}
where we substituted~$\ell=1\text{ AU}$.
It goes without saying that this is a highly-suppressed particle flux, compared to the standard halo flux,
\begin{align}
    \Phi_\mathrm{halo}(m_\chi)\approx 1.3 \cdot 10^{10} \left(\frac{m_\chi}{\text{MeV}}\right)^{-1} \text{ s}^{-1}\text{cm}^{-2}\, .
\end{align}
However, the reflected particles might have higher kinetic energy than any particle of the galactic halo.
Our MC simulations will reveal not just the value of~$p_\mathrm{refl}$ and thereby the total flux~$\Phi_\odot$, but also its speed distribution encoded in the differential SRDM~flux~${\dd \Phi_\odot \over \dd v_\chi}$.

\subsection{Anisotropy of solar reflection flux}
\label{ss: anisotropy}

One central question with important implications is if the SRDM flux gets ejected by the Sun isotropically.
While the velocity distribution of the DM~particles of the galactic halo is indeed assumed to be isotropic (in the Standard Halo Model), the boost into the Sun's rest frame moving with velocity~$\vec{v}_\odot$ introduces a dipole commonly called the ``DM~wind''.
This boost breaks the isotropy and reduces the spherical symmetry of the distribution to an axisymmetric distribution with symmetry axis along~$\mathbf{v}_\odot$, as illustrated in the left panel of Fig.~\ref{fig: isoreflection angle}.
In earlier works, it was assumed that the SRDM particle flux was ejected isotropically by the Sun~\cite{An:2017ojc,Emken:2017hnp}.
However, this was not verified, and we want to study if traces of the initial anisotropy survive the reflection process.
Such anisotropies would have important implications for the time-dependence of potential SRDM signals in terrestrial detectors, as we will discuss in Sec.~\ref{sss: expected signal modulation}.

Figure~\ref{fig: isoreflection angle} also introduces the \textit{isoreflection angle}~$\theta$ as the angle between the Sun's velocity~$\vec{v}_\odot$ and the Earth's location~$\vec{x}_\oplus$ in heliocentric coordinates,
\begin{align}
    \theta = \sphericalangle(\vec{x}_\oplus,\vec{v}_\odot)\, , \label{eq: isoreflection angle}
\end{align}
which will be a useful quantity in studying anisotropies.
The isoreflection angle~$\theta$ is the polar angle of our symmetry axis, and as such quantities like the SRDM flux or detection signal rates are constant along constant values of~$\theta$ justifying the name.\footnote{The angle~$\theta$ is similar to the \textit{isodetection angle} defined and used in the context of daily signal modulations~\cite{Collar:1992qc,Collar:1993ss,Hasenbalg:1997hs,Kavanagh:2016pyr,Emken:2017qmp,Kavanagh:2020cvn}.}

As the Earth orbits the Sun, its local isoreflection angle oscillates between~$\sim60^\circ$ (around September 4) and~$\sim120^\circ$ (around March 2) as shown in the right panel of Fig.~\ref{fig: isoreflection angle}.
\footnote{The time-dependent position vector of planet Earth in heliocentric coordinates is given e.g. in~\cite{McCabe:2013kea}.}
If the SRDM flux shows anisotropies, the Earth might at certain periods of the year pass through regions of the solar system with increased or decreased DM~flux from the Sun.
This would lead to a new type of annual modulation caused by the anisotropy of the SRDM particle flux.
We discuss signal modulations further in Sec.~\ref{sss: expected signal modulation} and present corresponding results in Sec.~\ref{ss: results direct detection}.

\subsection{Direct detection of solar reflected DM}
Once we know the total flux and spectrum of SRDM particles passing through Earth, we can start making predictions for the signal events they can trigger in terrestrial direct detection experiments and their phenomenological signatures.
As a first step, we express the well-known nuclear and electron recoil spectra in terms of an unspecified differential flux of DM~particles~${\dd \Phi_\chi \over \dd v_\chi}$, where~$v_\chi$ is the DM speed,.

\vspace{10pt}\noindent\textbf{Signal event spectra: }
Given a generic differential DM~particle flux~${\dd \Phi_\chi \over \dd v_\chi}$ through a detector target consisting of~$N_T$ nuclei, the resulting nuclear recoil spectrum is given by~\cite{Lewin:1995rx}
\begin{align}
    {\dd R \over \dd E_R} &= N_T \int_{v_\chi>v_\mathrm{min}} \dd v_\chi {\dd \Phi_\chi \over \dd v_\chi} {\dd \sigma_{\chi N}\over\dd E_R} \, . \label{eq: nuclear recoil spectrum}
\end{align}
Here, we introduced the differential cross section~${\dd \sigma_{\chi N}\over\dd E_R}$ for elastic DM-nucleus scatterings with recoil energy~$E_R$.
This cross section is determined by the underlying model assumptions, and we will go into further details on the DM~models studied in this paper in Sec.~\ref{sec:DM models}.
Furthermore, the minimum speed~$v_\mathrm{min}$ of a DM~particle required by the scattering kinematics to make a nucleus of mass~$m_N$ recoil with energy~$E_R$ is given by
\begin{align}
    v_\mathrm{min} &= \sqrt{m_N E_R \over 2 \mu_{\chi N}^2}\, , \label{eq: vMin nucleus}
\end{align}
where~$\mu_{ij}$ again denotes the reduced mass.

When searching for nuclear recoils caused by halo~DM alone, we would have to substitute the corresponding differential halo particle flux,
\begin{align}
 {\dd \Phi_\chi \over \dd v_\chi}\longrightarrow   {\dd \Phi_\text{halo} \over \dd v_\chi} &= \frac{\rho_\chi}{m_\chi} v_\chi f_\mathrm{obs}(v_\chi)\, , \label{eq: diff halo flux}
\end{align}
where $f_\mathrm{obs}(v_\chi)$ is the local DM~speed distribution corresponding to Eq.~\eqref{eq: SHM boost}.
\footnote{In this work, we are not interested in directional detection and always integrate over the directions of the DM velocity~$\vec{v}_\chi$ when computing event rates. The DM~speed distribution~$f_\mathrm{obs}(v_\chi)$ of the galactic halo is obtained as the marginal distribution of the full velocity distribution of Eq.~\eqref{eq: SHM boost}, i.e. $f_\mathrm{obs}(v_\chi) = \int \dd\Omega_{\mathbf{v}_\chi} v_\chi^2 f_\mathrm{obs}(\vec{v}_\chi)$.}
By substituting Eq.~\eqref{eq: diff halo flux} into~\eqref{eq: nuclear recoil spectrum}, we re-obtain the nuclear recoil spectrum in its usual form.

Because we are interested in the recoil spectrum caused by DM~particles reflected and accelerated by the Sun, the differential DM flux~$ {\dd \Phi_\chi \over \dd v_\chi}$ in Eq.~\eqref{eq: nuclear recoil spectrum} is the sum of the halo and the solar reflection flux.
Hence, we obtain the nuclear recoil spectrum caused by both DM~populations of the solar system as
\begin{align}
    {\dd R \over \dd E_R} &= N_T \int \dd v_\chi \underbrace{\left({\dd \Phi_\text{halo} \over \dd v_\chi}+{\dd \Phi_\odot \over \dd v_\chi}\right)}_{\equiv {\dd \Phi_\chi \over \dd v_\chi}} {\dd \sigma_{\chi N}\over\dd E_R} \, . \label{eq: nuclear recoil spectrum halo + SRDM}
\end{align}
However, considering both fluxes simultaneously is rarely necessary.
We discussed earlier how the flux of halo~DM completely dominates the DM~flux reflected from the Sun.
Hence, if the halo particles are detectable, the contributions of the first term of Eq.~\eqref{eq: nuclear recoil spectrum halo + SRDM} will always dominate the total spectrum, and the contribution of SRDM can safely be neglected.
In contrast, if the DM~mass is too low, even the most energetic halo particle is unable to leave a trace in the detector, and the first term vanishes.
In that case, the only observable DM particles are the ones boosted by the Sun, and only the second term of Eq.~\eqref{eq: nuclear recoil spectrum halo + SRDM} contributes. 

Besides nuclear recoils, we can also compute the energy spectra of DM-electron scatterings in atoms and semiconductors.
Starting with atoms, the differential event rate in a target of~$N_T$ atoms of ionizing the atomic shells with quantum numbers~$(n,\ell)$, is given by \cite{Kopp:2009et,Essig:2011nj,Essig:2015cda}
\begin{align}
    {\dd R_\mathrm{ion}^{n\ell}\over\dd E_e} &= N_T\int {\dd q^2\over 4E_e}\int_{v>v_\mathrm{min}}\dd v_\chi\; {\dd \Phi_\chi\over\dd v_\chi} \frac{\dd \sigma_{\chi e}}{\dd q^2}\left|f_\mathrm{ion}^{n\ell}(E_e ,q)\right|^2\, , \label{eq: energy spectrum ionization}
\end{align}
where~$E_e$ is the final energy of the released electron,~$q$ is the momentum transfer, and~$f_\mathrm{ion}^{n\ell}(E_e ,q)$ is the ionization form factor of the atomic~$(n,\ell)$-orbital.
Regarding the ionization form factor, we use the tabulated values obtained in~\cite{Catena:2019gfa}, which also describes the details of the initial and final electron states.
For low momentum transfers, we use the dipole approximation,
\begin{align}
    |f_\mathrm{ion}^{n\ell}(E_e ,q)|^2 &= \left(\frac{q}{q_0}\right)^2 |f_\mathrm{ion}^{n\ell}(E_e,q_0)|^2\, ,
\end{align}
which is valid for $q, q_0 \lesssim 1\text{ keV}$ \cite{Essig:2019xkx}.

As opposed to nuclear recoils, DM-electron scatterings are inelastic.
The kinematic threshold on the DM~speed to transfer an energy~$\Delta E$ to a bound electron, similar to Eq.~\eqref{eq: vMin nucleus} for nuclear recoils, is given by
\begin{align}
    v_\mathrm{min} = {\Delta E \over q} + {q \over 2m_\chi}\, .\label{eq: vMin electron}
\end{align}
In the case of ionization, the energy deposit~$\Delta E$ consists of both the binding energy and the final electron's kinetic energy, i.e.~$\Delta E=|E_B^{n\ell}|+E_e$.
We will specify the differential cross section~$\frac{\dd \sigma_{\chi e}}{\dd q^2}$ of elastic DM-electron scatterings in Sec.~\ref{sec:DM models}.

For direct detection experiments with crystal semiconductor targets, we can express the event spectrum in terms of the DM~particle flux in a similar way.
The differential rate of DM-induced electron transitions in a crystal of~$N_\mathrm{cell}$ unit cells with a total energy deposit of~$\Delta E=E_e$ can be derived to be~\cite{Essig:2015cda}
\begin{align}
    &{\dd R_\mathrm{crys}\over\dd E_e} =\nonumber\\
    &4\alpha m_e^2\, N_\mathrm{cell} \int \frac{\dd q}{q^2}\int_{v>v_\mathrm{min}} \dd v_\chi\;{\dd \Phi_\chi\over\dd v_\chi} \frac{\dd \sigma_{\chi e}}{\dd q^2}\left|f_\mathrm{crys}(E_e ,q)\right|^2\, . \label{eq: energy spectrum semiconductor}
\end{align}
Here,~$\alpha$ is the fine-structure constant, and $f_\mathrm{crys}(E_e ,q)$ is the crystal form factor.
For its evaluation use the tabulated values obtained with the \texttt{QEdark} module of \texttt{QuantumESPRESSO}~\cite{Giannozzi_2009,Essig:2015cda}.

\subsection{Signatures of a solar reflection signal}
\label{sss: expected signal modulation}
In the event of a successful DM~detection with solar reflection as origin, it is an important question how this signal can be distinguished from both background or (heavier) halo~DM.
The most obvious way would require a directional detection experiment, which could identify the Sun as the origin of the observed DM~flux.

However, most experiments are insensitive to directional information of the incoming DM~particle.
In this case, a key signature of a SRDM~signal in a terrestrial detector is its time modulation.
This is not different from conventional direct DM~searches, where we generally expect an annual modulation due to the Earth's motion relative to the Sun~\cite{Drukier:1986tm,Freese:1987wu} and also daily modulations for stronger interactions~\cite{Collar:1993ss,Hasenbalg:1997hs,Kavanagh:2016pyr,Emken:2017qmp}.
But for reflected~DM particles the modulations' origin and phenomenology differs substantially ~\cite{Emken:2017hnp}, and we can identify three sources of modulations:
\begin{itemize}
    \item \textbf{Orbital modulation: } The Earth-Sun distance~$\ell$ varies slightly over the year due to the eccentricity of the Earth's elliptical orbit. Since the solar reflection flux is proportional to~$\ell^{-2}$, the resulting signal rate will feature an annual modulation with the maximum  (minimum) at the perihelion (aphelion) on $\sim$Jan. 3 ($\sim$Jul. 4). The orbital modulation is fundamentally different from the standard annual modulation, which peaks in June and is not a feature of an SRDM signal. 
    \item \textbf{Anisotropy modulation:} As we will show, the Sun does not emit the reflected particles isotropically into the solar system. Even in the case of isotropic scatterings in the Sun, the initial anisotropy of the incoming DM~wind manifests itself as anisotropies in the resulting solar reflection flux. As the Earth moves around the Sun, it passes regions of increased and decreased particle fluxes, which is the source of a second annual modulation.
    \item \textbf{Daily modulation:} For interaction cross sections relevant for solar reflection, we typically also expect frequent underground scatterings inside the Earth, which causes a daily modulation due to the Earth's rotation around its axis, see e.g.~\cite{Emken:2017qmp}. While this modulation fundamentally works the same way for halo and reflected~DM, the fact that the reflected particles arrive exclusively from the Sun means that we can expect a higher rate during the day, and lower counts in our detectors during the night, where the detector is (partially) shielded off the Sun by the Earth's bulk mass.
\end{itemize}
In this paper, we focus on the superposition of the two annual modulations, which we will study in Sec.~\ref{ss: results direct detection}.
Nonetheless, the day-night modulation is a second powerful signature of a SRDM signal.

\section{Describing solar reflection with Monte Carlo simulations}
\label{sec:monte carlo}

By simulating DM~particle trajectories passing through and scattering inside the Sun, we can obtain solid estimates of the SRDM~flux reflected by the Sun through a single or multiple collisions with solar targets.
Similar simulations have been applied already in the '80s in the context of WIMP evaporation and energy transport~\cite{Nauenberg:1986em} testing the analytic groundwork of Press and Spergel~\cite{Spergel:1984re}, as well as more recently for the study of solar reflection via electron scatterings~\cite{An:2017ojc}.
In this section, we lay out the principles of the MC~simulations we performed using the public \texttt{DaMaSCUS-SUN} tool, and how we use them to estimate the SRDM~flux with good precision.
These partially build on foundations laid out in~\cite{Emken:2019hgy}.
For the interested reader, App.~\ref{app: simulation details} contains reviews and summaries of even more simulation details.

\subsection{Simulation of a trajectory}
\label{ss: trajectory simulation}
    \noindent\textbf{Initial conditions: }
The first step of simulating a particle trajectory is to generate the initial conditions , i.e. the initial position and velocity of the orbit.
In App.~\ref{app:initial conditions}, we present the explicit procedure to sample initial conditions, which we only summarize at this point.
The initial velocities of DM~particles which are about to enter the Sun follow the distribution proportional to the differential rate of particles entering the Sun.
Based on Eq.~\eqref{eq: total entering rate}, the distribution is given by
\begin{align}
    f_\mathrm{IC}(\vec{v}) = \mathcal{N}_\mathrm{IC}\, \left(v + {\vesc{R_\odot}^2 \over v} \right) f_\odot(\vec{v})\, , \label{eq: IC velocity distribution}
\end{align}
where $\mathcal{N}_\mathrm{IC} = \left[ \left(\langle v\rangle + \vesc{R_\odot}^2 \langle v^{-1}\rangle \right)\right]^{-1}$ is a normalization constant, and $f_\odot(\vec{v})$ is the velocity distribution of the SHM in the solar rest frame given in Eq.~\eqref{eq: SHM boost}.
The first term in Eq.~\eqref{eq: IC velocity distribution} accounts for faster particles entering the Sun with higher rate, whereas the second term reflects that slower particles get gravitationally focused toward the Sun from a larger volume.

However, due to the gravitational acceleration of the infalling particles, the SHM only applies asymptotically far away from the Sun, hence the initial position of the DM~particle needs to be chosen at a large distance.
The initial positions of the sampled DM~particles are distributed uniformly, but we sample them at a distance of at least 1000 AU, where the SHM surely applies.
However, at this distance, a DM~particle with a given velocity~$\mathbf{v}$ is very unlikely to be on a trajectory that intersects the solar surface.
Therefore, instead of sampling the initial position from all of space and rejecting the vast majority of initial conditions, we identify the sub-volume where particles with the velocity~$\mathbf{v}$ are bound to pass through the Sun.
By taking the initial positions to be uniformly distributed in that sub-volume, the initial conditions are guaranteed to result in particles entering the Sun.
In App.~\ref{app:initial conditions}, we derive a bound on the impact parameter~$b$ for the initial conditions, given by Eq.~\eqref{eq: hit condition}, which defines this sub-volume and also accounts for gravitational focussing.
This procedure is only applicable due to the assumed constant density of DM in the neighborhood of the solar system.
As a result, we obtain the initial location and velocity of a particle from the galactic halo on collision course with the Sun.

\vspace{10pt}\noindent\textbf{Free propagation: }
Neglecting scatterings for a moment, the orbits of DM~particles through the Sun and solar system are described by Newton's second law.
The specific equations of motion are listed in Eqs.~\eqref{eq: equations of motion} of App.~\ref{app: equations of motion}.
Furthermore, on their way towards the Sun, the gravitationally unbound DM~particles follow hyperbolic Keplerian orbits, whose analytic description is useful and therefore reviewed in some detail in App.~\ref{app: kepler}.
From their initial positions at large distances, we can propagate the particles into the Sun's direct vicinity by one analytic step saving time and resources.

Once the DM~particles arrived in the Sun's direct neighbourhood and are about to enter stellar bulk, we continue the trajectory simulation by solving the equations of motion  numerically.
We choose the Runge-Kutta-Fehlberg~(RKF) method to solve the Eqs.~\eqref{eq: 1st order eom}.
The RKF method is an explicit, iterative algorithm for the numerical integration of $1^\mathrm{st}$~order ordinary differential equations with adaptive time step size~\cite{Fehlberg1969}.
The method is reviewed in App.~\ref{app:runge kutta fehlberg}.
We further simplify the numerical integration by using the fact that the orbits lie in a two-dimensional plane due to conservation of angular momentum~$\vec{J}$.
Keeping track of the plane's orientation using the relations of App.~\ref{app: equations of motion} allows to switch back and forth between 2D and 3D, which is e.g. necessary when the simulated particle scatters on a solar target changing the direction of~$\vec{J}$.

\vspace{10pt}\noindent\textbf{Scatterings: }
Now that we can describe the motion of free DM~particles in- and outside the Sun, we turn our attention to the possibility of scatterings on nuclei or electrons of the solar plasma.
The probability for a DM~particle to scatter during the time interval~$\Delta t$ while following its orbit inside the Sun is given by
\begin{align}
    P(\Delta t) = 1 - \exp\left(-\int_0^{\Delta t}\frac{\dd t}{\tau(r(t),v(t))}\right)\, ,
\end{align}
where we defined the mean free time
\begin{align}
    \tau(r,v) = \Omega(r,v)^{-1} \label{eq: mean free time}
\end{align}
in terms of the scattering rate given in Eq.~\eqref{eq: total scattering rate}.
We find the location of the next scatterings by inverse transform sampling.
This involves sampling a random number~$\xi\in [0,1]$ from a uniform distribution~$\mathcal{U}_{[0,1]}$ and solving~$P(\Delta t) = \xi$ for~$\Delta t$.
Our knowledge of the shape and dynamics of the DM~particle's orbit (i.e.~$r(t)$ and~$v(t)$) in combination with the time of scattering provides the location of scattering.

In practice, we solve the equations of motions iteratively in small finite time steps~$\Delta t$, where~$\Delta t$ is adjusted adaptively by the RKF~method via Eqs.~\eqref{eq: adaptive delta t} and~\eqref{eq: adaptive delta t 2}.
For every integration step~$i$ at radius~$r_i$ and DM~speed~$v_i$, we add up~$\Delta t_i / \tau(r_i,v_i)$, until the sum surpasses the critical value,
\begin{align}
    \sum_i \frac{\Delta t_i}{\tau(r_i,v_i)} > -\ln (1-\xi)\, .
\end{align}
This is the condition for a scattering to occur.

At this point, we should make a remark regarding the definition of a mean free path~$\lambda$.
Since the DM~particle moves through a thermal plasma, where the targets' motion is crucial and may not be neglected, the total scattering rate~$\Omega(r,v_\chi)$, or equivalently the mean free time~$\tau(r,v_\chi)$, is a more meaningful physical quantity determining scattering probabilities than~$\lambda$.
Even a hypothetical particle at rest would eventually scatter on an incoming thermal electron or nucleus.
Nonetheless, if we insist on defining a mean free path, we can always do that by
\begin{subequations}
\label{eq: mean free path}
\begin{align}
    \lambda(r,v_\chi) &\equiv \tau(r,v_\chi) v_\chi =  \frac{v_\chi}{\Omega(r,v_\chi)}\\
    &=\left(\sum_i n_i(r) \sigma_i \frac{\langle|\vec{v}_\chi-\vec{v}_{T,i}|\rangle}{v_\chi}\right)^{-1}\, ,
\end{align}
\end{subequations}
which was applied in e.g.~\cite{An:2017ojc}.
\footnote{A number of previous works on DM~scatterings in the Sun have defined the mean free path as~$(\sum_i n_i \sigma_i)^{-1}$ or~$(\sum_i n_i \langle\sigma_i\rangle)^{-1}$, which will lead to an underestimation of the scattering rate~\cite{Taoso:2010tg,Lopes:2014aoa,Garani:2017jcj,Busoni:2017mhe}. While this might be a more acceptable approximation for nuclei where~$v_\chi$ is comparable to~$\langle|\vec{v}_\chi-\vec{v}_i|\rangle$ (deviations are less than 30\% for solar protons), the error becomes severe in the case of DM-electron scatterings, where the ratio of the two speeds can be as high as~$\sim 40$ in the Sun.}
The fact that the mean free path decreases for lower DM speeds reflects that the mean free time~$\tau$ is the underlying relevant physical quantity.
\footnote{In addition, using the mean free time also has computational advantages: 
$\tau(r,v_\chi)$ is a better-behaved function of~$v_\chi$ than~$\lambda(r,v_\chi)$ and its evaluation via two-dimensional interpolation is more efficient due to the need for fewer grid points in the $(r,v_\chi)$ plane.}

After this digression about mean free paths, we return to the MC~algorithm for the DM~scattering. After the determination of the location of the next scattering, we need to identify the target's particle species, i.e. we have to determine if the thermal target is an electron or one of the present nuclear isotopes.
If we label the different target by an index~$i$ and define the partial scattering rate of each target as~$\Omega_i(r,v_\chi) \equiv n_i(r) \langle \sigma_i |\mathbf{v}_\chi-\mathbf{v}_{T,i}|\rangle$, such that~$\Omega = \sum_i \Omega_i$, then the probability for the DM~particle to scatter on target~$i$ is
\begin{align}
    P(\text{scattering on target $i$}) = \frac{\Omega_i(r,v_\chi)}{\Omega(r,v_\chi)}\, .
\end{align}
Knowing the target's identity~$i$ and mass~$m_i$, we can formulate the density function of the target velocity~$\mathbf{v}_T$, which is given by the thermal, isotropic Maxwell-Boltzmann distribution in Eq.~\eqref{eq: maxwell boltzmann} weighted by the target's contribution to the scattering probability for a DM~particle of velocity~$\mathbf{v}_\chi$~\cite{Romano2018}.
In particular, faster particles moving in an opposite direction to~$\vec{v}_\chi$ are more likely to scatter.
In App.~\ref{app: target velocity sampling}, we derive that the PDF of~$\vec{v}_T$ is
\begin{align}
    f(\mathbf{v}_T) = \frac{|\vec{v}_\chi - \vec{v}_T|}{\langle|\vec{v}_\chi - \vec{v}_T|\rangle}f_i(\mathbf{v}_T,T)\, , \label{eq: target velocity pdf 3}
\end{align}
where we used the assumption that the total cross section does not depend on the target speed.
App.~\ref{app: target velocity sampling} also contains a detailed description of the target velocity sampling algorithm.
This procedure enables us to draw velocity vectors~$\vec{v}_T$ from the distribution corresponding to Eq.~\eqref{eq: target velocity pdf 3}.

In the final step of the scattering process, we compute the DM~particle's new velocity~$\mathbf{v}_\chi^\prime$, which is given by Eq.~\eqref{eq: kinematics}.
The determination of the new velocity involves sampling the scattering angle~$\alpha$.
In this paper, we limit ourselves to isotropic contact interactions, and the vector~$\mathbf{n}$ in Eq.~\eqref{eq: kinematics} is an isotropically distributed unit vector.
We note in passing that modifying the fundamental DM-target interaction can have crucial consequences for the distribution of~$\alpha$. 
For example, if the interactions were mediated by a light dark photon mediator, the resulting suppression of large momentum transfers would strongly favour small values of~$\alpha$, i.e. forward scatterings, and it will be interesting to see the implications for solar reflection~\cite{SolarReflectionLightMediator}.

Given a new velocity of the DM~particle, its free propagation along its orbit continues until the particles either re-scatters or escapes the star.
The combination of MC simulating the initial conditions, the free propagation of DM~particles in the Sun, and their scattering events with solar electrons and nuclei result in DM~trajectories of vast variety of orbital shapes, one example was shown in Fig.~\ref{fig: trajectory} of the introduction.
Finally, we need to define certain termination conditions under which the trajectory simulation of a DM~trajectory concludes.

\vspace{10pt}\noindent\textbf{Termination conditions: }
The simulation of a DM~particle's orbit continues until one of two termination events occurs:
\begin{enumerate}
    \item The particle escapes the Sun, i.e. it leaves the solar bulk with~$v>v_\mathrm{esc}(R_\odot)$.
    \item The particle gets gravitationally captured.
\end{enumerate}
In the case of the first event, we consider the particle as \textit{reflected} contributing to the SRDM~flux if it has scattered at least once.
Otherwise, it is categorized as \textit{un-scattered} or \textit{free}, and is not relevant for our purposes as its energy has not changed.
Regarding the second termination event, the definition of gravitational capture in our MC simulations is not straightforward, as it is at least in principle always possible to gain energy via a scattering and escape the Sun.
We define a DM~particle as \textit{captured}, if it either has scattered more than 1000 times, or if it propagated through the Sun for a very long time along a bound orbit freely, i.e. without scatterings.
The first one is supposed to apply to cases where especially heavier DM~particles get captured and thermalize with the solar plasma without a significant chance to escape.
For the second capture condition, rather than defining a maximum time of free propagation, we terminate the simulation after~$10^7$ time steps~$\delta t$ of the RKF procedure.

Both criteria are an arbitrary choice, and it is of course not impossible for a DM~particle to get reflected with more than 1000 scatterings or after having propagated freely over a very long duration.
The captured particles (as defined above) will build up a gravitationally bound population eventually reaching an equilibrium where their reflection or evaporation rate equals the capture rate.
As we will see later in Fig.~\ref{fig: free vs capture vs reflection}, for low DM~masses the amount of captured particles falls significantly below the number of reflected particles.
This is why the contributions of such particles to the total SRDM~particle flux are sub-dominant in the parameter space relevant for solar reflection, while their MC description is resource expensive.
Their omission saves a great amount of computational time and renders the resulting MC~estimates as slightly conservative only.

\subsection{The MC estimate of the SRDM flux}
Once a DM~particle gets reflected and leaves the Sun with~$v>v_\mathrm{esc}(R_\odot)$, we propagate it away from the Sun to the Earth's distance of 1~AU.
Just like in the case of the initial conditions, we use the analytic description of hyperbolic Kepler orbits to save time, see App.~\ref{app: kepler}.
The speed~$v_\chi$ of the DM~particle as it passes a distance of 1~AU is saved and constitutes one data point.

By counting the number of reflected particles passing through a sphere of 1~AU radius and recording their speed, we can obtain solid estimates of the total and differential rate of solar reflection.
Assuming the simulation of~$N_\mathrm{sim}$ trajectories resulting in~$N_\mathrm{refl}$ reflected particles or data points, we estimate the reflection probability~$p_\mathrm{refl}$ introduced in Eq.~\eqref{eq: srdm rate estimate} by the ratio of the two numbers, $N_\mathrm{refl}/N_\mathrm{sim}$.
Hence, the total SRDM rate~$\mathcal{R}_\odot$ is given by
\begin{align}
    \mathcal{R}^\mathrm{MC}_\odot = \frac{N_\mathrm{refl}}{N_\mathrm{sim}} \Gamma(m_\chi)\, , \label{eq: total srdm rate MC}
\end{align}
where we used the fact that the initial conditions guarantee that all simulated particles will enter the Sun's interior (see App.~\ref{app:initial conditions}).
The total rate of infalling particles was given previously by Eq.~\eqref{eq: total entering rate}.
Consequently, the total SRDM~flux through the Earth is
\begin{align}
    \Phi_\odot^\mathrm{MC} &= \frac{ \mathcal{R}^\mathrm{MC}_\odot }{4\pi\ell^2}\, ,\label{eq: total srdm flux MC}
\end{align}
with~$\ell = 1\text{ AU}$.

In order to get statistically stable results for the direct detection rates, we only count data points to $N_\mathrm{refl}$ and only record the speed of a DM~particle if its kinetic energy is sufficient to trigger the detector, i.e. if~$v_\chi$ exceeds an experiment-specific speed threshold.
As a consistency check, we confirmed that the MC~estimate of the SRDM~spectrum at high speeds does not depend on the choice of the threshold.
We continue the data collection until the relevant speed interval is populated with a sufficient amount of data points, such that the statistical fluctuations of the derived direct detection rates from Eq.~\eqref{eq: nuclear recoil spectrum halo + SRDM},~\eqref{eq: energy spectrum ionization}, or~\eqref{eq: energy spectrum semiconductor} are negligible.

The recorded speed data points~$v_i$ allow us to estimate the speed distribution of the SRDM particles.
Here, we prefer a smooth distribution function over histograms.
One powerful, non-parametric method is kernel density estimation~(KDE)~\cite{rosenblatt1956,parzen1962}.
In combination with the pseudo-data method by Cowling and Hall for boundary bias reduction~\cite{CowlingHall,Karunamuni2005}, KDE provides a robust estimator of the SRDM~distribution~$f^\mathrm{KDE}_\odot(v_\chi)$.\footnote{For a brief review of KDE, we refer to App. A of~\cite{Emken:2018run}. The only free parameter of KDE is the kernel's bandwidth. It is chosen following Silverman's rule of thumb~\cite{silverman1986density}.}
The full MC~estimates of the differential SRDM rate and flux are then given by
\begin{align}
\frac{\dd \mathcal{R}_\odot}{\dd v_\chi} &= \mathcal{R}^\mathrm{MC}_\odot f^\mathrm{KDE}_\odot(v_\chi)\, ,\label{eq: SRDM spectrum}\\
    \frac{\dd \Phi^\mathrm{MC}_\odot}{\dd v_\chi} &= \frac{1}{4\pi\ell^2} \frac{\dd \mathcal{R}_\odot}{\dd v_\chi} \, . \label{eq: differential SRDM flux}
\end{align}

In Eqs.~\eqref{eq: total srdm flux MC} and~\eqref{eq: differential SRDM flux}, we did not consider the possibility of an anisotropic flux and implicitly assumed that the SRDM flux through the Earth equals the flux averaged over all isoreflection angles.
In Sec.~\ref{ss: anisotropy}, we introduced the isoreflection angle~$\theta$ as a natural parameter for possible anisotropies of the solar reflection flux.
Taking into account the possibility of anisotropies, or in other words a strong dependence of~$\Phi_\odot$ on~$\theta$, a more precise estimate of the average flux through Earth is given by
\begin{align}
    \left\langle \Phi_\odot^\mathrm{MC} \right\rangle &= \frac{1}{\Delta t}\int_0^{\Delta t}\dd t\; \Phi_\odot^\mathrm{MC}(\theta(t))\, ,\label{eq: total srdm flux MC averaged}\\
    \left\langle \frac{\dd \Phi^\mathrm{MC}_\odot}{\dd v_\chi}\right\rangle &= \frac{1}{\Delta t}\int_0^{\Delta t}\dd t\; \frac{\dd \Phi^\mathrm{MC}_\odot}{\dd v_\chi}(\theta(t))\, ,\label{eq: differential SRDM flux averaged}
\end{align}
where $\Delta t = \text{1 yr}$ and $\theta(t)$ gives the time evolution of the Earth's isoreflection angle.
However, due to the Earth's particular isoreflection coverage, the average over all isoreflection angles is a good estimate of Eq.~\eqref{eq: total srdm flux MC averaged}, i.e. $\left\langle \Phi_\odot^\mathrm{MC} \right\rangle\approx \Phi_\odot^\mathrm{MC}$, which can be seen in Fig.~\ref{fig: anisotropy 1}.
This justifies the use of Eq.~\eqref{eq: total srdm flux MC} and~\eqref{eq: differential SRDM flux} to estimate the SRDM~flux through the Earth in practice.

\subsection{Anisotropies and isoreflection rings}
\label{ss: isoreflection rings}
\begin{figure}[t!]
    \centering
    \includegraphics[width=0.3\textwidth]{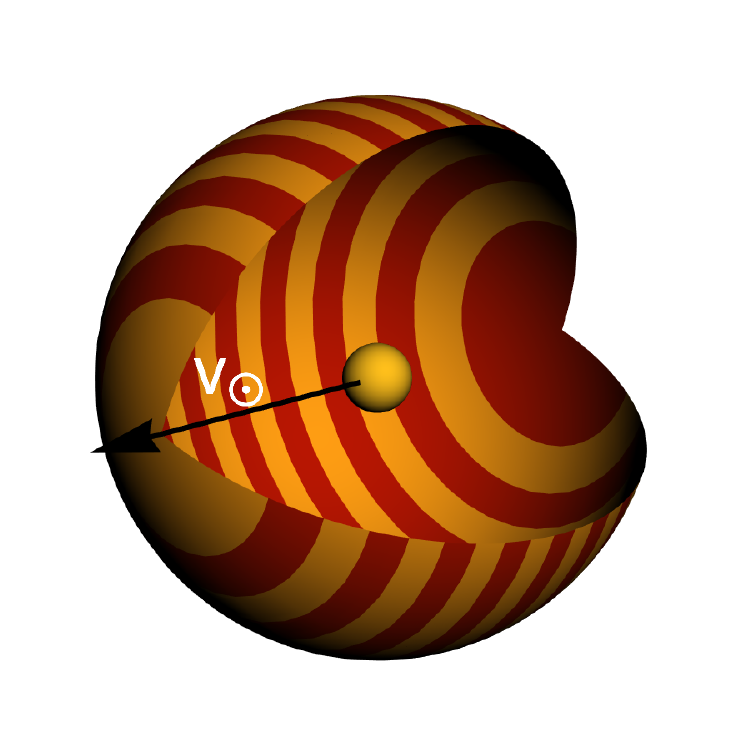}
    \caption{Isoreflection rings of equal area~($N_\mathrm{rings}=20$).}
    \label{fig: isoreflection rings}
\end{figure}

Equation~\eqref{eq: isoreflection angle} of Sec.~\ref{ss: anisotropy} introduced the isoreflection angle~$\theta$ as a crucial quantity in the study of anisotropies of the SRDM~flux.
For the MC~simulations, we define a number of \textit{isoreflection rings} of finite angular size, for each of which we collect data and perform the analysis independently.
This way we obtain the MC estimates of the SRDM~flux, event spectra, and signal rates as a function of~$\theta$ allowing us to measure the anisotropy as well as the resulting annual signal modulation discussed in Sec.~\ref{sss: expected signal modulation}.
Instead of a fixed angular size~$\Delta \theta$, we define rings of equal surface area, i.e. for~$N_\mathrm{rings}$ isoreflection rings, their boundary angles are given by
\begin{align}
    \theta_i = \arccos\left(\cos\theta_{i-1}-{2\over N_\mathrm{rings}}\right)\, , \label{eq: isoreflection rings}
\end{align}
where~$\theta_0=0$.
We illustrate the example of~$N_\mathrm{rings}=20$ in Fig.~\ref{fig: isoreflection rings}.

Due to the isoreflection rings being of equal area, we lose angular resolution close to~$\theta=0^\circ$ and~$\theta = 180^\circ$, and obtain rings of smaller angular size, i.e. better angular resolution around~$\theta = 90^\circ$.
This has great advantages over equal-angle rings as the Earth only covers the interval~$\theta\in[60^\circ,120^\circ]$ over a full year, and we do not require good resolution close to the poles.
Furthermore, the time required to collect speed data through MC~simulations is more uniform among equal-area rings.

\subsection{Software and computational details}
The \textit{Dark Matter Simulation Code for Underground Scatterings - Sun Edition} (\texttt{DaMaSCUS-SUN}) underlying all results of this paper is publicly available~\cite{Emken2021}.
During the development of \texttt{DaMaSCUS-SUN}, it was our goal to follow best software development practices to facilitate reproducibility and usability and to ensure reliable results. 
The code is written in C++ and built with CMake.
It is maintained and developed on \href{https://github.com/temken/DaMaSCUS-SUN}{Github} and is set up with a Continuous Integration (CI) pipeline including build testing, code coverage, and unit testing.
Furthermore, version 0.1.0 of the code has been archived under \href{https://doi.org/10.5281/zenodo.4559874}{[DOI:10.5281/zenodo.4559874]}.
\texttt{DaMaSCUS-SUN} is parallelized with the Message Passing Interface (MPI) such that the MC~simulations can run in parallel on HPC clusters.
For the more general functionality related to DM~physics and direct detection, \texttt{DaMaSCUS-SUN} relies on the C++~library \textit{obscura}~\cite{Emken2021-2}.
Most results were obtained with the HPC clusters \textit{Tetralith} at the National Supercomputer Centre (NSC) in Link\"oping, Sweden, and \textit{Vera} at the Chalmers Centre for Computational Science and Engineering (C3SE) in G\"oteborg, Sweden, both of which are funded by the Swedish National Infrastructure for Computing (SNIC).

\section{Models of DM-matter interactions}
\label{sec:DM models}

So far, we have not specified our assumptions on how to model the interactions between DM~particles and ordinary matter, either in the solar plasma or terrestrial detector.
In this paper, we will study solar reflection of light~DM upscattered by either nuclei or electrons, or a combination of both.
The chosen models are the ``standard choices'' of the direct detection literature.
For nuclear interactions, we consider spin-independent (SI) and spin-dependent~(SD) interactions.
We also consider the possibility of ``leptophilic''~DM, which interacts only with electrons.
Lastly, we would like to study the scenario where~DM can be reflected by both electrons and nuclei, which is the case for the ``dark photon'' model.
Throughout this work, we assume the interactions' mediators to be heavy, such that the scattering processes are isotropic contact interactions.
Another interesting scenario is the presence of light mediators, which we will consider in a separate work~\cite{SolarReflectionLightMediator}.
In this section, we summarize the commonly used DM~interaction models and review the expressions for the cross sections needed for the description of solar reflection and detection.

\subsection{DM-nucleus interactions}
\label{ss: Dm nucleus interactions}
Starting with elastic DM-nucleus scatterings through SI interactions, the differential cross section is given by~\cite{Jungman:1995df},
\begin{align}
	\frac{\dd\sigma^{\rm SI}_{N}}{\dd E_R} &=\frac{m_N\,\sigma_p^{\rm SI}}{2\mu_{\chi p}^2 v_\chi^2}\left[Z+\frac{f_n}{f_p}(A-Z) \right]^2 \left|F_N\left(q\right)\right|^2\, ,\label{eq: differential cross section SI}
\end{align}
where~$q=\sqrt{2m_N E_R}$ is the momentum transfer, and~$Z$, $A$, and $F_N(q)$ are the nuclear charge, mass number, and form factor respectively. For low-mass DM~particles, the momentum transfers are sufficiently small such that~$F_N(q)\approx 1$.
While the ratio~$f_n/f_p$ of the effective DM-neutron and DM-proton couplings is a free parameter, we assume isospin-conserving interactions whenever considering nuclear scatterings only ($f_n=f_p$).
In the context of the dark photon model, the DM couples to electric charge and~$f_n=0$ applies, as we will discuss below in more detail.

\begin{figure*}
\centering
\includegraphics[width=0.329\textwidth]{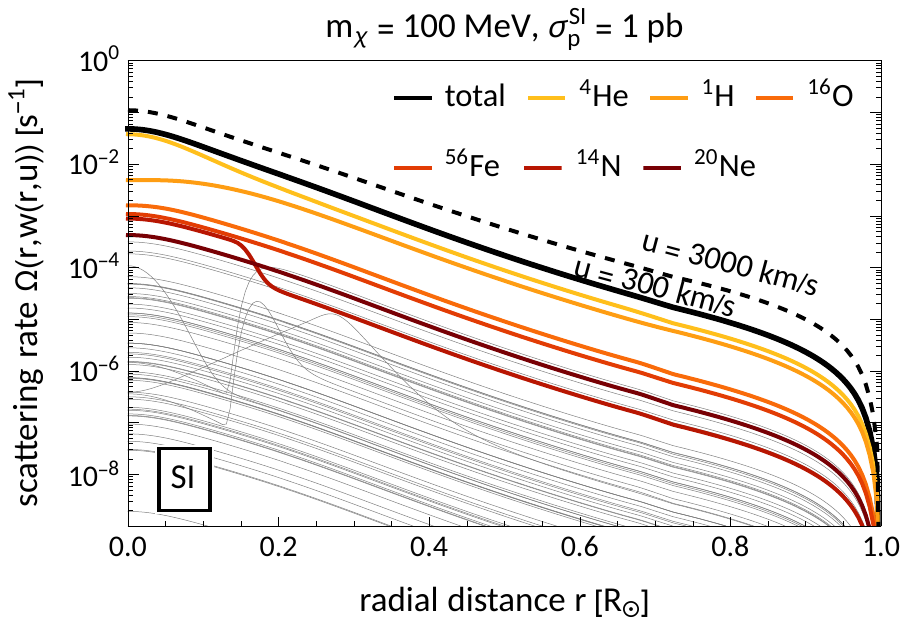}
\includegraphics[width=0.329\textwidth]{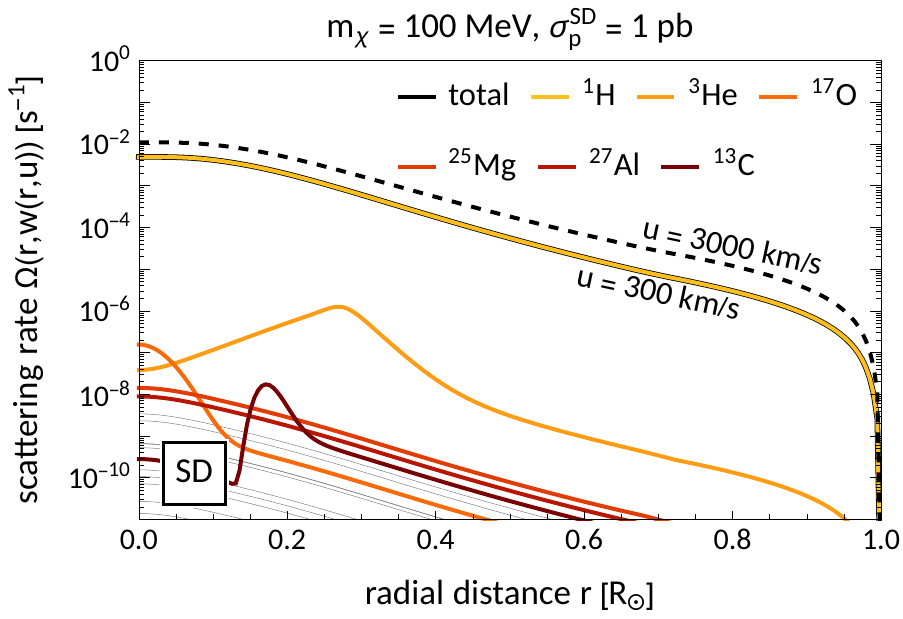}
\includegraphics[width=0.329\textwidth]{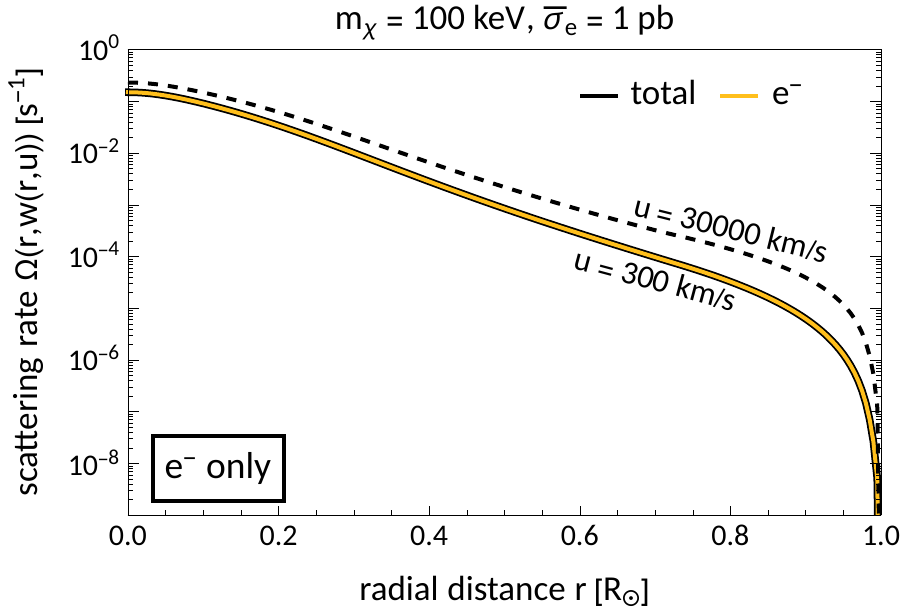}
\caption{The scattering rates for SI/SD nuclear interactions (left, middle) and electron interactions (right) between an incoming DM~particle with asymptotic speed~$u=300\text{ km/s}$ and different solar targets as a function of the radial distance~$r$ defined in Eq.~\eqref{eq: total scattering rate}. In the case of nuclear interactions, the six most important targets are highlighted in color, whereas the total scattering rate is given as a black, solid line. In addition, the dashed, black line depicts the scattering rate of a boosted DM~particle, in particular~$u=3000\text{ km/s}$ for SI/SD interactions, and $u=30000\text{ km/s}$ for electron interactions.}
\label{fig: scattering rate}
\end{figure*}

Alternatively, we also consider spin-dependent (SD) interactions~\cite{Engel:1992bf,Bednyakov:2004xq,Bednyakov:2006ux}, for which the differential cross section is given by
\begin{align} 
\frac{\dd \sigma_N^{\rm SD}}{\dd E_R}&=\frac{2m_N\sigma_p^{\rm SD}}{3\mu_{\chi p}^2v_\chi^2}\frac{J+1}{J}\left[\langle S_p\rangle +\frac{f_n}{f_p} \langle S_n\rangle\right]^2 \left|F_N^{\rm SD}(q)\right|^2\, . \label{eq: differential cross section SD}
\end{align}
Here, the DM~particle couples to the nuclear spin~$J$, and~$\langle S_p\rangle$ and~$\langle S_n\rangle$ are the average spin contribution of the protons and neutrons respectively.
As in the case of SI interactions, we may set the nuclear form factor~$F_N^{\rm SD}(q)$ to unity.
We use the isotope-specific values for the average spin contributions given in~\cite{Bednyakov:2004xq,Klos:2013rwa}.

Based on the differential DM-nucleus scattering cross sections given by Eqs.~\eqref{eq: differential cross section SI} and~\eqref{eq: differential cross section SD}, we can compute the total cross section by integrating over all kinematically allowed recoil energies,
\begin{align}
    \sigma_N &= \int_0^{E_R^\mathrm{max}}\dd E_R\; \frac{\dd \sigma_N}{\dd E_R}\nonumber\\
    &=\begin{cases} 
    \sigma_p^{\rm SI}\left(\frac{\mu_{\chi N}}{\mu_{\chi p}}\right)^2\left[Z+\frac{f_n}{f_p}(A-Z) \right]^2\, ,\;\text{ for SI,}\\
   \frac{4}{3}\sigma_p^{\rm SD}\left(\frac{\mu_{\chi N}}{\mu_{\chi p}}\right)^2\frac{J+1}{J}\left[\langle S_p\rangle +\frac{f_n}{f_p} \langle S_n\rangle\right]^2 \, ,\;\text{for SD,}
    \end{cases}
\end{align}
where we used~$E_R^\mathrm{max}={2\mu_{\chi N}^2 v_\chi^2\over m_N}$.
Substituting the total cross section into Eq.~\eqref{eq: total scattering rate}, we obtain the total rate of DM-nucleus scatterings in the Sun as a function of the distance to the solar center and the DM~speed.

\subsection{DM-electron interactions}
\label{ss: electron only}
Another scenario is solar reflection by electron scattering, which was previously studied in~\cite{An:2017ojc}. 
A straight-forward realization is to assume that the incoming DM exclusively interacts with the solar electrons and not the various nuclei.
While DM-nucleus interactions get generated at the loop level even in leptophilic models, this can be avoided by assuming pseudoscalar or axial vector mediators~\cite{Kopp:2009et}.

The differential cross section of elastic DM-electron scatterings can be parameterized in terms of a reference cross section~$\bar{\sigma}_e$,
\begin{align}
    \frac{\dd \sigma_{e}}{\dd q^2} &=\frac{\bar{\sigma}_e}{4\mu_{\chi e}^2v_\chi^2}F_{\rm DM}(q)^2\, . \label{eq: dSigma dq2 electron}
\end{align}
In the case of contact interactions, where~$F_\mathrm{DM}(q)=1$, the reference cross section yields the total scattering cross section, and~$\sigma_e = \bar{\sigma}_e$.
It should however be noted that this is no longer accurate e.g. for light mediators, where the DM~form factor scales as~$F_\mathrm{DM}(q)\sim q^{-2}$ and~$\sigma_e \neq \bar{\sigma}_e$.
We will consider solar reflection through light mediators in a separate work~\cite{SolarReflectionLightMediator}.

\subsection{Dark photon model}
\label{ss: dark photon}

Finally, we also want to consider the possibility of solar reflection of electrons and nuclei simultaneously.
Arguably the most considered simplified Beyond the Standard Model (BSM) model in the context of sub-GeV DM searches is the dark photon model, which leads to a simple relation between DM-nucleus and DM-electron interactions.

In this model, the SM is extended by a fermion~$\chi$, acting as DM, and an additional~$U(1)$ gauge group, which is spontaneously broken.
The corresponding gauge boson, the dark photon~$A^\prime$ of mass~$m_{A^\prime}$ can mix with the photon via the kinetic mixing term~\cite{Galison:1983pa,Holdom:1985ag}.
This simple dark sector is described by the effective Lagrangian
\begin{align}
    \mathscr{L}_D &= \bar{\chi} (i\gamma^{\mu}D_\mu-m_{\chi}) \chi  + \frac{1}{4}F'_{\mu \nu}F'^{\mu \nu}\nonumber\\
    &+ m_{A^\prime}^{2} A'_{\mu}A'^{\mu}+ \frac{\varepsilon}{2} F_{\mu \nu}F'^{\mu \nu}\, .
\end{align}
Through the mixing term, the dark photon couples to all charged fermions of the~SM, generating the effective DM-proton and DM-electron interaction terms~\cite{Kaplinghat:2013yxa}
\begin{align}
   	\mathscr{L}_{\rm int} = e \varepsilon A^\prime_\mu \left(\bar{p}\gamma^\mu p- \bar{e}\gamma^\mu e\right)\, .
\end{align}
The resulting differential scattering cross section for DM-electron scatterings is identical to Eq.~\eqref{eq: dSigma dq2 electron} where the reference cross section is related to the model parameter via
\begin{align}
    \bar{\sigma}_e &\equiv \frac{16\pi \alpha\alpha_D\epsilon^2 \mu_{\chi e}^2}{(q_{\rm ref}^2+m_{A^\prime}^2)^2}\, . \label{eq: reference cs electron}
\end{align}
The corresponding cross section for nuclear interactions is given by the SI cross section of Eq.~\eqref{eq: differential cross section SI} with~$f_n=0$ and the reference DM-proton cross section
\begin{align}
    \bar{\sigma}_p&\equiv \frac{16\pi \alpha\alpha_D\epsilon^2 \mu_{\chi p}^2}{(q_{\rm ref}^2+m_{A^\prime}^2)^2}\, .\label{eq: reference cs proton}
\end{align}
Comparing Eqs.~\eqref{eq: reference cs electron} and~\eqref{eq: reference cs proton}, we find the following relations between the two reference cross sections,
\begin{align}
\frac{\bar{\sigma}_p}{\bar{\sigma}_e} = \left(\frac{\mu_{\chi p}}{\mu_{\chi e}}\right)^2\, . \label{eq: dark photon cross section ratio}
\end{align}
In the case of contact interactions of low-mass DM~particles, the two reference cross sections correspond to the total scattering cross sections.
We find that the two cross sections are of comparable size for DM~masses below the electron mass, while for larger masses, the DM-proton scattering cross section satisfies~$\bar{\sigma}_p\gg \bar{\sigma}_e$.
In the next section, we will take a look at the implications of relation~\eqref{eq: dark photon cross section ratio} for the relative contributions to the scattering rate inside the Sun.

\begin{figure*}
\centering
\includegraphics[width=0.329\textwidth]{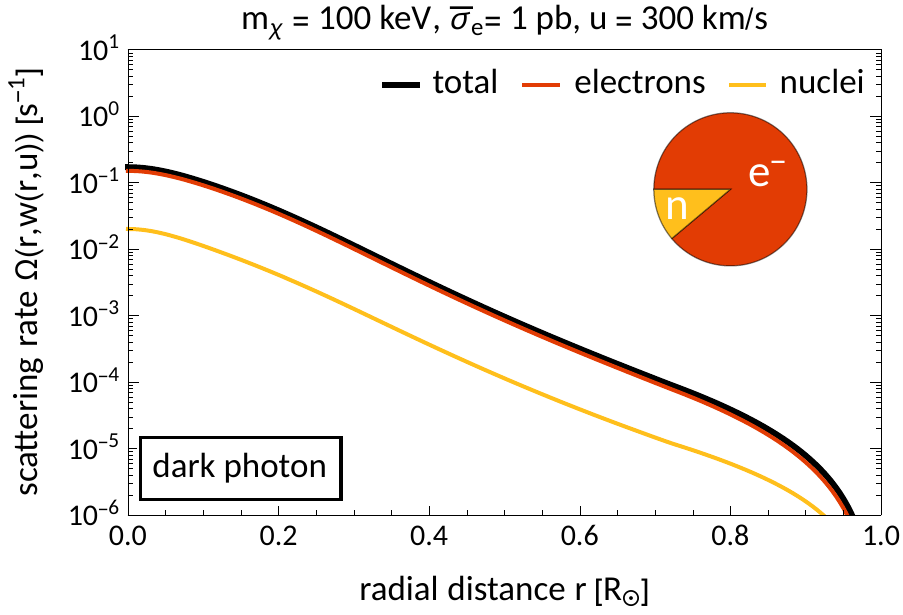}
\includegraphics[width=0.329\textwidth]{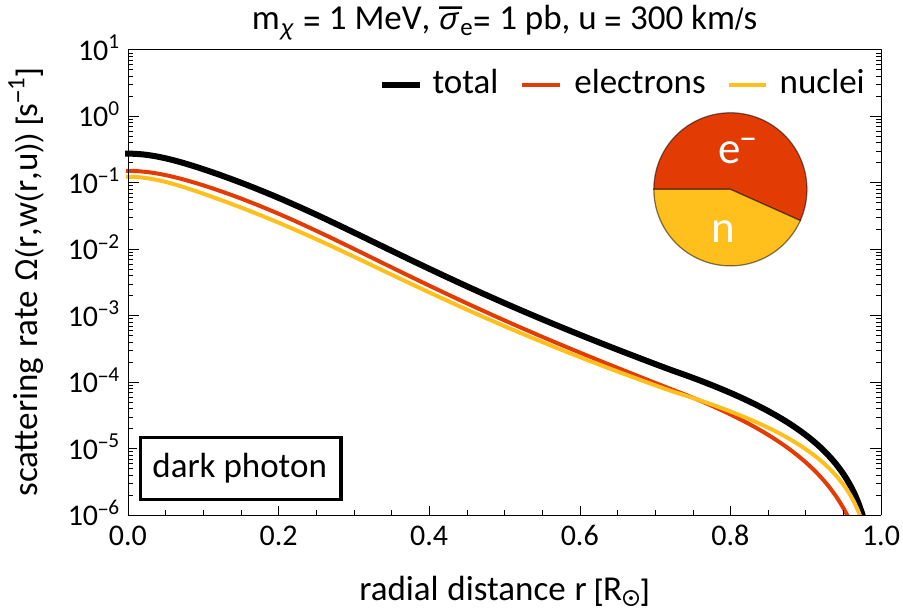}
\includegraphics[width=0.329\textwidth]{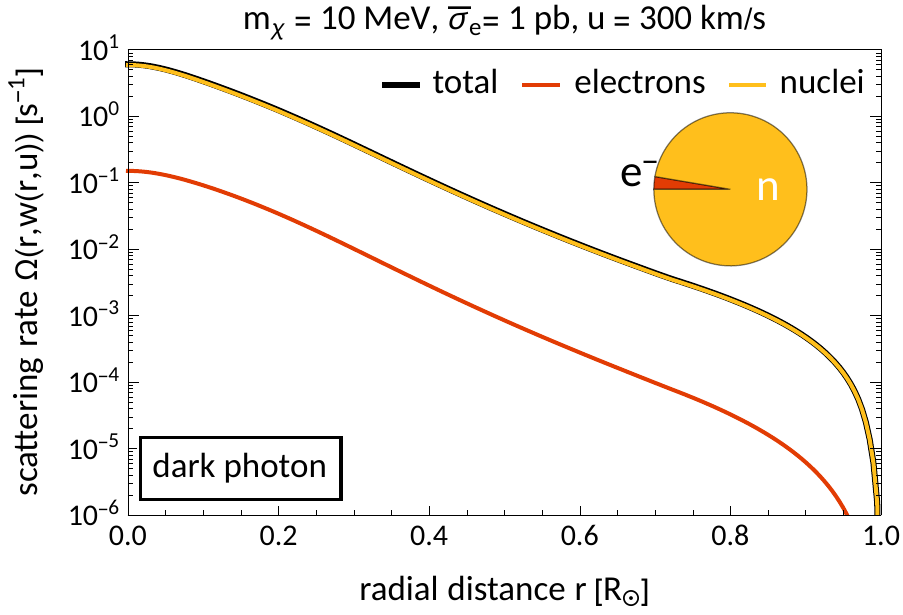}
\caption{Similarly to Fig.~\ref{fig: scattering rate}, these plots compare the contributions of DM-electron and DM-proton interactions to the total scattering rate in the Sun for the dark photon model. The DM~mass is increasingly set to 100~keV (left), 1~MeV (middle), and 10~MeV (right) to illustrate the different regimes. The pie-charts show the relative contributions of the two processes.}
\label{fig: scattering rate dark photon}
\end{figure*}

\subsection{Scattering rates}

Now that we specified the DM model and interactions we would like to study, we can evaluate the scattering rate inside the Sun given by Eq.~\eqref{eq: total scattering rate}.
In Fig.~\ref{fig: scattering rate}, we show the total scattering rate as a function of the radial distance of the solar center~$r$ for SI/SD nuclear interactions as well as electron interactions only.
We assume a DM~particle of asymptotic speed~$u=300\text{ km/s}$, i.e. the gravitational acceleration is taken into account.
The different colors depict the contributions of the most important targets.
The dashed lines correspond to an asymptotic speed of~$u=3000\text{ km/s}$ ($u=30000\text{ km/s}$), a possible asymptotic speed of a DM~particle reflected by scatterings on nuclei (electrons).
While multiple nuclear targets contribute significantly to the total scattering rate in the case of SI interactions, for SD interactions the interactions with protons dominates the total rate completely.

In the dark photon model, both DM-nucleus and DM-electron interactions are present, and DM~particles may scatter on solar nuclei and electrons.
It is therefore interesting to consider the relative contribution and therefore the relevance for the reflection process of the two processes.
Let us consider the ratio of the scattering rate contributions of DM-electron and DM-proton scatterings,
\begin{align}
    \frac{\Omega_e}{\Omega_p}&\propto \frac{\bar{\sigma}_e}{\bar{\sigma}_p} \frac{\langle |\vec{v}_\chi-\vec{v}_e|\rangle}{\langle |\vec{v}_\chi-\vec{v}_p|\rangle}\, ,
\end{align}
where the thermal average of the relative speed can be found in Eq.~\eqref{eq: thermal average of relative speed}. Depending on the temperature~$T(r)$, the ratio of the average relative speeds is a number of~$\mathcal{O}(10)$, and the ratio of the cross sections is given by the squared ratio of the reduced masses following Eq.~\eqref{eq: dark photon cross section ratio},
\begin{align}
   \frac{\Omega_e}{\Omega_p}&\propto   \left(\frac{\mu_{\chi p}}{\mu_{\chi e}}\right)^2 \times\mathcal{O}(10)\, .
\end{align}
Based on this rough estimate, we can expect electron scatterings to be sub-dominant for DM~masses above a few MeV.
We also expect a regime around~$m_\chi\approx 1\text{ MeV}$, where DM-nucleus and DM-electron scatterings contribute to the total scattering rate with comparable amounts.
For masses well below the electron mass, the ratio of the reduced masses approaches one, and DM-electron interactions will dominate the collision rate.

These three cases are illustrated in Fig.~\ref{fig: scattering rate dark photon}, which shows the total scattering rate similarly to Fig.~\ref{fig: scattering rate}, but for the dark photon model and three exemplary DM~masses.
The embedded pie-charts depict the relative contributions of nuclear and electron interactions respectively.
We conclude that in the heavy-mediator regime of the dark-photon model, either nuclear or electron interactions dominate depending on the DM~mass, with the exception of a small intermediate mass interval around 1~MeV.
In particular, the results of~\cite{An:2017ojc} and \cite{Emken:2017qmp} which only considered DM-electron and DM-nucleus interactions respectively will also apply to the dark photon model to good approximation, since the masses considered therein are of order~$\mathcal{O}(100\text{ keV})$  and $\mathcal{O}(100\text{ MeV})$, such that either electron or nucleus scatterings dominate.

\begin{figure*}
    \centering
    \includegraphics[width=0.32\textwidth]{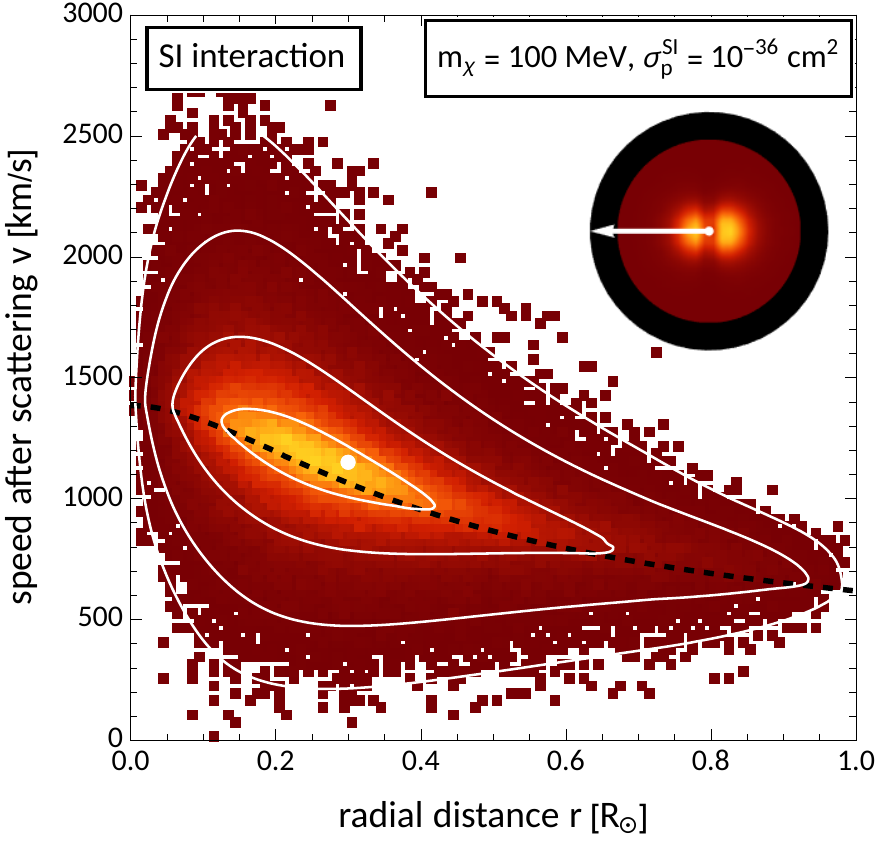}
    \includegraphics[width=0.32\textwidth]{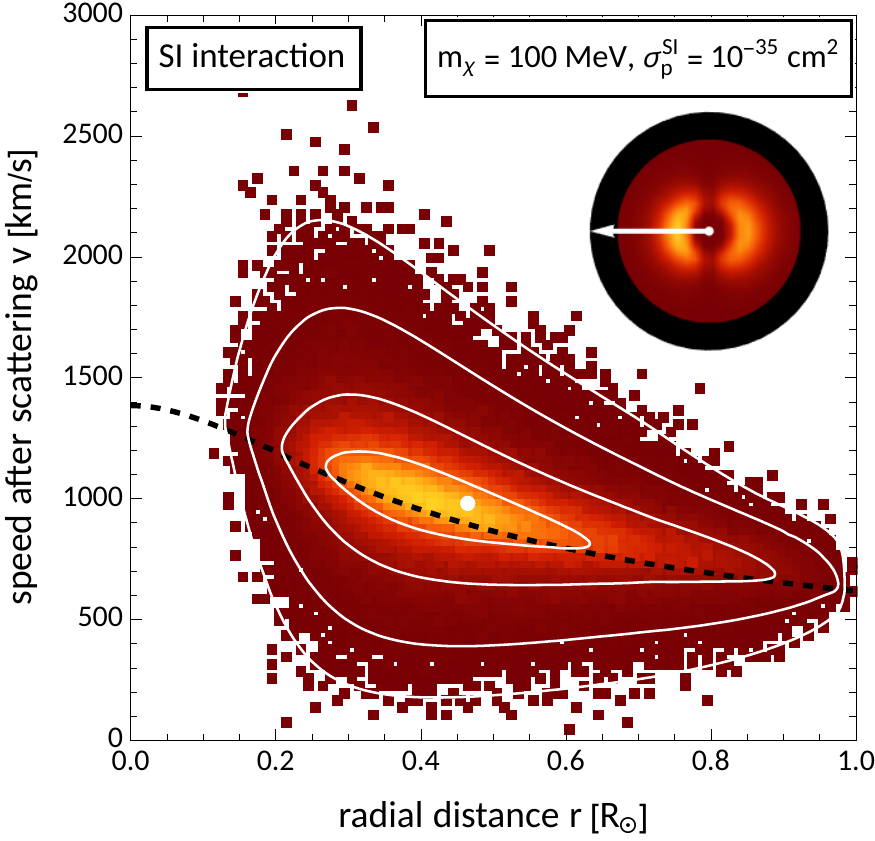}
    \includegraphics[width=0.32\textwidth]{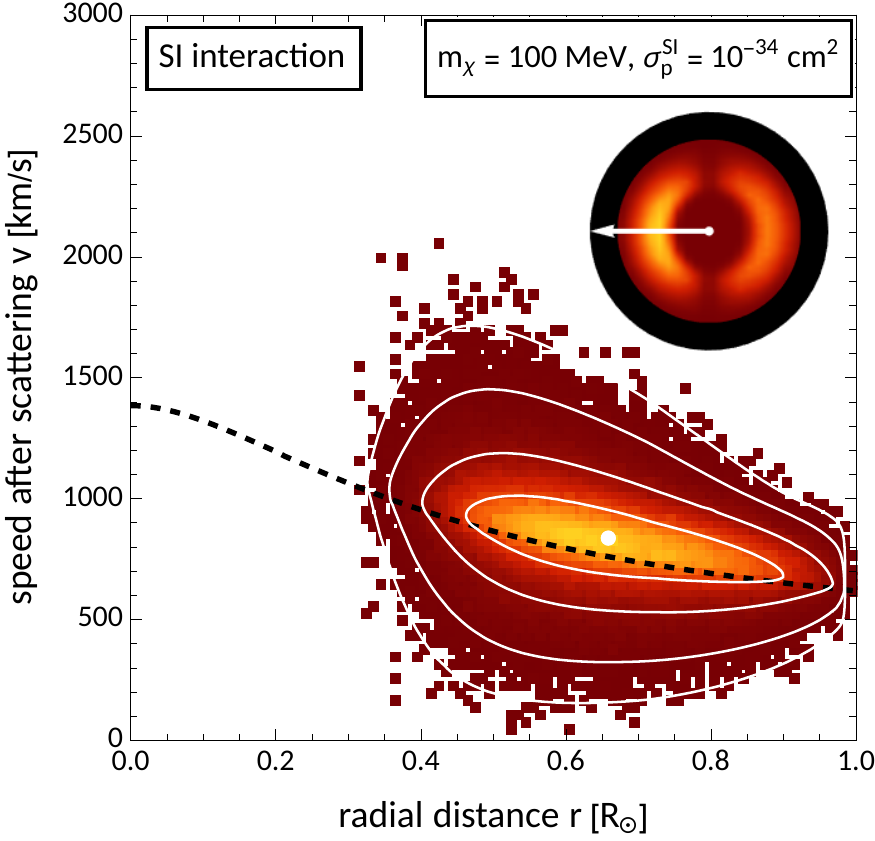}
    \includegraphics[width=0.32\textwidth]{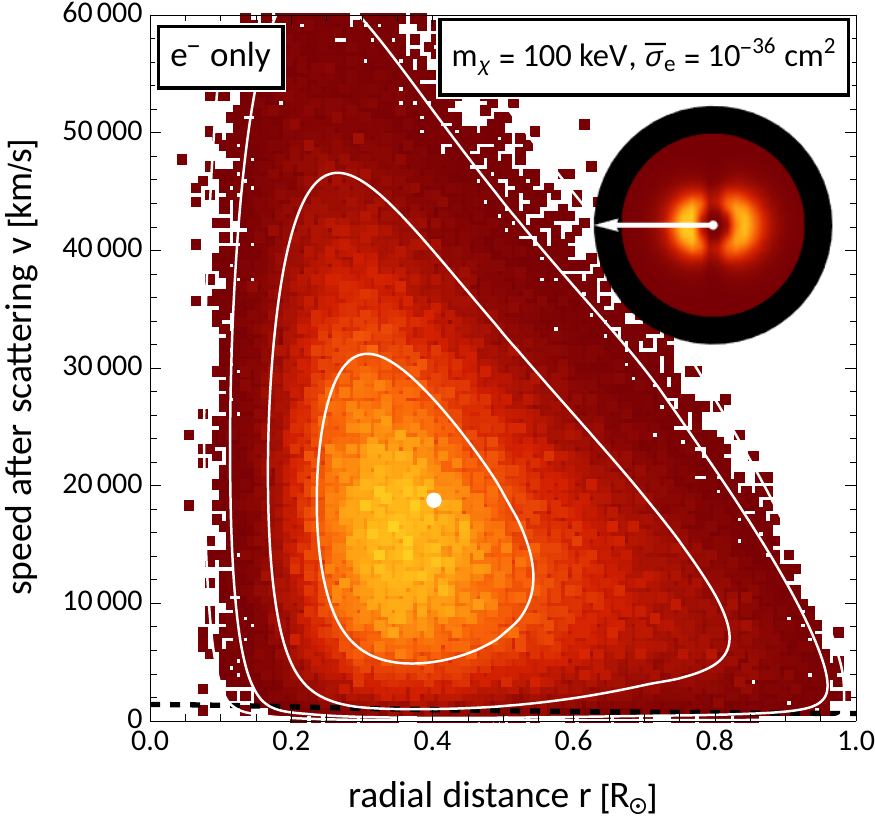}
    \includegraphics[width=0.32\textwidth]{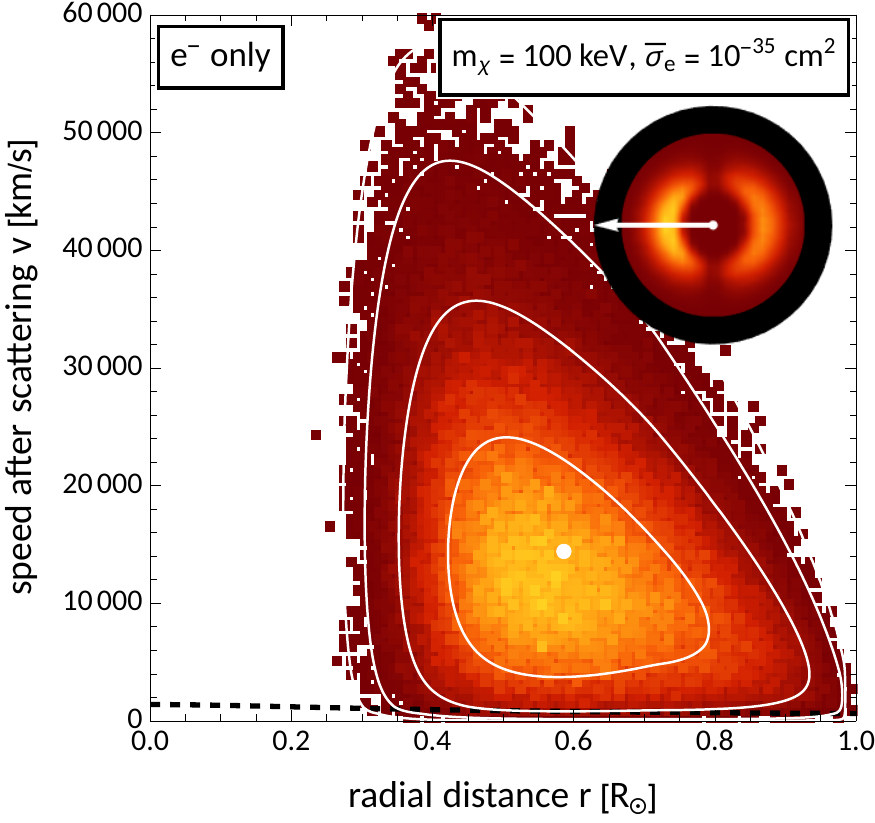}
    \includegraphics[width=0.32\textwidth]{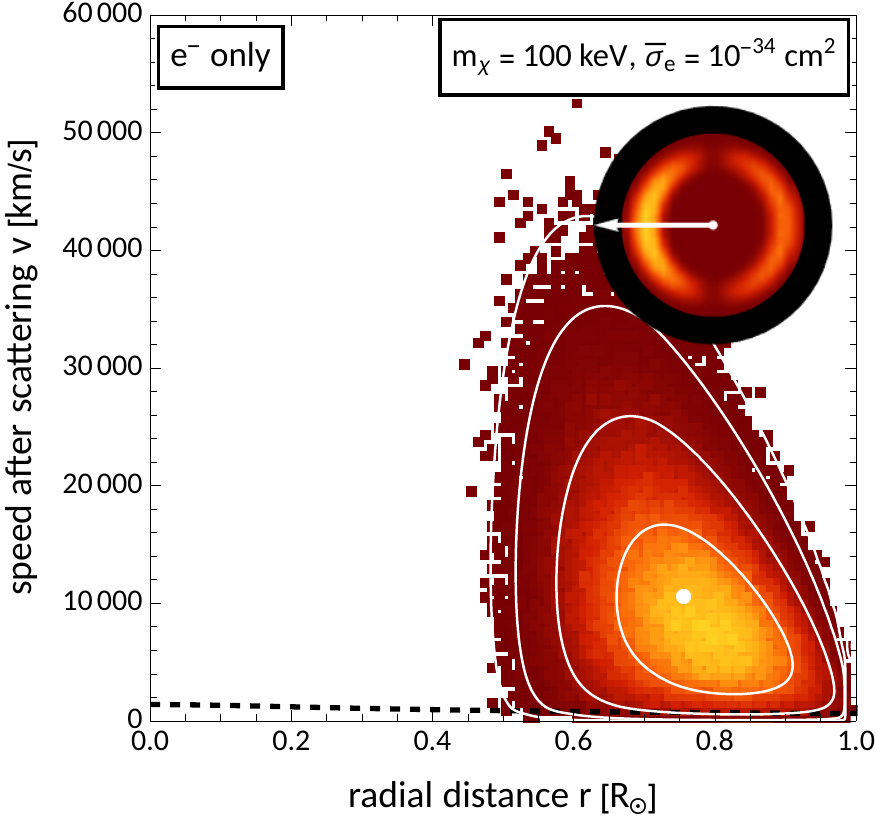}
    \caption{Two-dimensional histograms of the radial distance~$r$ and final DM~speed~$v$ of the first scattering ($N=2\times 10^5$). The white lines show the analytic differential scattering rate~$\frac{\dd S}{\dd r \dd v}$ derived in~\cite{Emken:2017hnp} for comparison. The white dot indicates the maximum, and the contour lines show the differential rate's decline to 50\%/10\%/1\% of the peak's value. The black dashed line gives the local escape velocity, see Eq.~\eqref{eq: sun escape velocity} (barely visible in the lower panels). The small inlay figures are an alternative illustration of the distribution of the first scattering's radius, where the white arrow indicates the direction of the Sun's velocity.}
    \label{fig: first scattering}
\end{figure*}

\section{Results}
\label{sec:results}

This section contains the main results of this study.
It is divided into two parts:
We start by investigating the general results of the MC~simulations and particularly more general properties of the SRDM~flux in Sec.~\ref{ss: results spectrum}.
In the second part, Sec.~\ref{ss: results direct detection}, we investigate the prospects of observing the SRDM~flux in terrestrial direct detection experiments.
We set exclusion limits based on existing experiments, and present projections for next-generation detectors.
Furthermore, we study the rich modulation signature of SRDM as a potential key to identify the solar reflection as the source of a hypothetical DM~discovery.

\subsection{Reflection spectrum and flux}
\label{ss: results spectrum}

\noindent\textbf{The first scattering: }
In a previous work, we found analytic expressions to describe solar reflection by a single scattering inside the Sun~\cite{Emken:2017hnp}.
Among other things, the analytic equations describe the location and final DM~speed of the first scattering of an infalling halo DM~particle along its trajectory inside the Sun.
Comparing these results to the fundamentally different approach of our MC~simulations provides an invaluable consistency check.

In Eq.~(11) of~\cite{Emken:2017hnp}, we derived an expression for the differential scattering rate, $\frac{\dd S}{\dd v \dd r}$, which extended the Gould's analytic theory of solar DM~capture and evaporation to include the Sun's opacity~\cite{Gould:1987ju}.
It captures the distribution of the first scattering's radial distance~$r$ to the solar center and the DM~particle's speed~$v$ after the scattering.
By running a number of MC~simulations where we record $r$ and $v$ of the trajectories' first scatterings, we obtain a two-dimensional histogram in the $(r,v)$-plane, which are shown in Fig.~\ref{fig: first scattering}.

In the first and second rows, we focus on SI~nuclear interactions, and electron interactions respectively, whereas the columns correspond to increasing scattering cross-sections.
For larger cross sections the first scattering is more likely to occur closer to the solar surface (as additionally illustrated by the inlays) and to result in a slower final DM~speed.
This is not surprising since the Sun's outer layers are cooler than the core.
The histograms are shown in combination with the contour lines of the differential scattering rate~$\frac{\dd S}{\dd v \dd r}$, which demonstrate a convincing agreement between the analytic expressions and the MC~simulations.

Furthermore, Fig.~\ref{fig: first scattering} also shows the local escape velocity of Eq.~\eqref{eq: sun escape velocity} as a black dashed line (barely visible in the second row).
In the case of electron scatterings, the first scattering accelerates a DM~particle of 100~keV mass so efficiently that only very few lose enough energy to get gravitationally captured.
We therefore expect many of those particles to get reflected by a single scattering.
In contrast, about half of the DM~particle of 100~MeV mass get decelerated below the local escape velocity by the first scattering.
Here, a ``single scattering regime'' as in the case of electron scatterings does not exist since the captured particles are bound to eventually scatter \textit{at least} a second time.
We can thereby suspect that the contribution of multiple scatterings to the SRDM~flux is more significant in the case of nuclear collisions.
At this point, it makes sense to consider the contribution of multiple scatterings under the different DM~scenarios.

\vspace{10pt}\noindent\textbf{Multiple scatterings: }
In order to quantify the relative contribution of multiple scatterings to the solar reflection particle flux, we compute the MC~estimate of the reflection spectrum~$\frac{\dd \mathcal{R}_\odot}{\dd v_\chi}$ using Eq.~\eqref{eq: SRDM spectrum}.
For each data point, i.e. reflected DM~particle contributing to our estimates, we keep track of how often that particle had scattered.

\begin{figure*}
    \centering
    \includegraphics[width=0.48\textwidth]{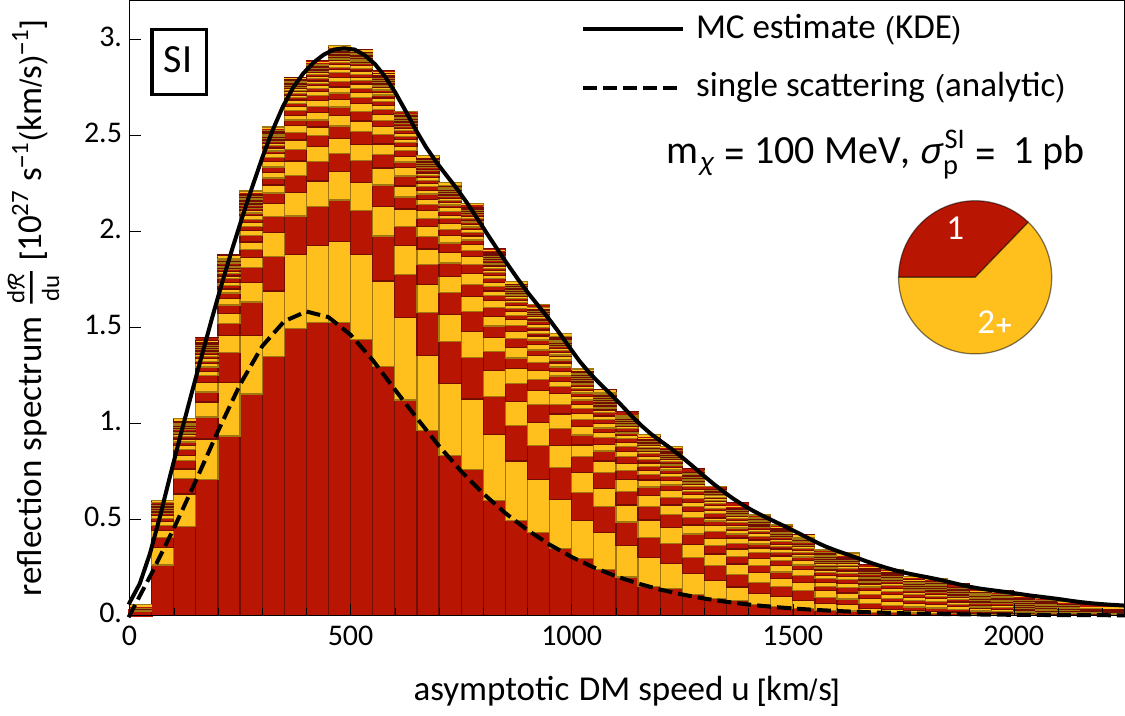}
    \includegraphics[width=0.48\textwidth]{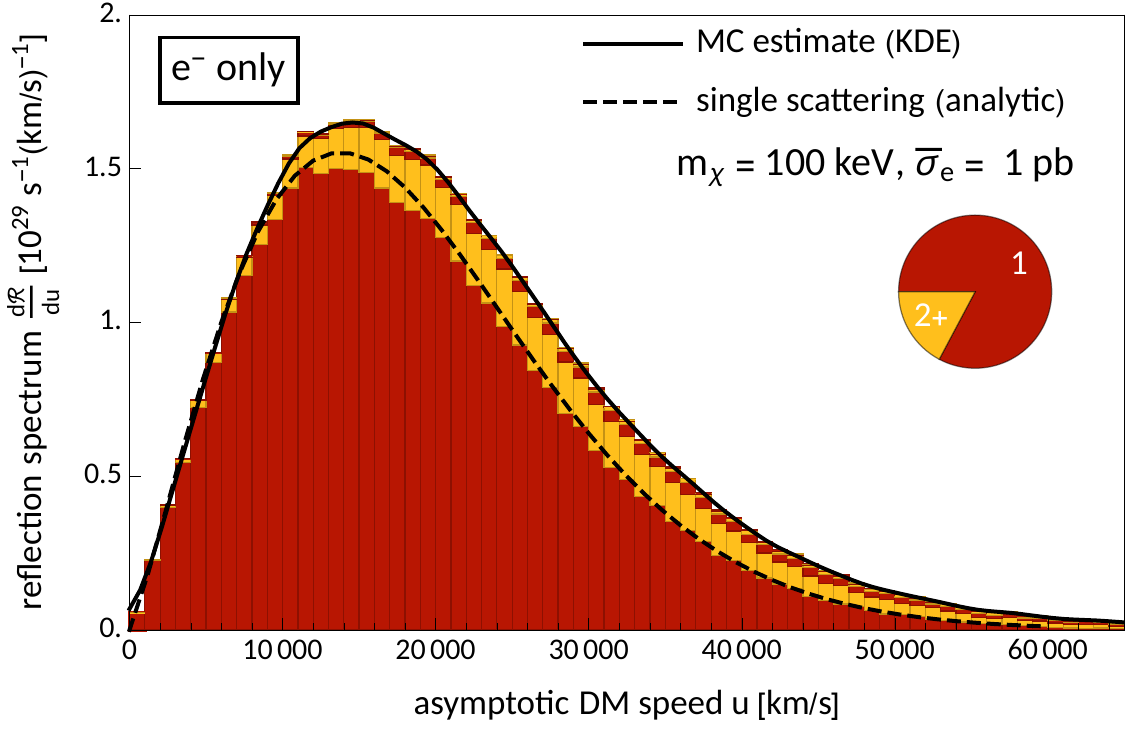}
    \caption{Histogram estimate of the solar reflection spectrum~$\frac{\dd\mathcal{R}_\odot}{\dd u}$ for SI nucleus interactions (left) and interactions with electrons only (right). Going up, the alternating colors depict the contributions of 1, 2, ... scatterings. The black solid line is the kernel density estimate of the spectrum given by Eq.~\eqref{eq: SRDM spectrum} and based on the same data set as the histogram. The black dashed line depicts the analytic result for the single scattering rate~\cite{Emken:2017hnp}. The pie-charts show the relative contributions of single (`1') and multiple (`2+') collisions to the total SRDM population.}
    \label{fig: multiple scatterings}
\end{figure*}

The resulting stacked histograms in Fig.~\ref{fig: multiple scatterings} verify our previous conjecture.
In the case of SI~nuclear interactions, the contributions of multiple scatterings exceed those of single collision reflection.
The relative contribution of single and multiple scatterings are depicted in the pie-chart.
For interactions with electrons only, the reflection spectrum is dominated by DM~particles scattered only once inside the Sun.
\footnote{
With this in mind, it is ironic that solar reflection of light~DM was established using analytic methods for single nucleus~\cite{Emken:2017hnp} and MC~techniques for multiple electron interactions~\cite{An:2017ojc} respectively, whereas the respective other approach might have been more appropriate.}
Collisions with solar nuclei are much more likely to reduce the DM~particle's energy than electron scatterings.
Therefore, there is a higher probability to get gravitationally captured by a nuclear collision such that multiple scatterings are inevitable.

At this point, we can perform another consistency check and apply again our analytic theory of single scattering reflection~\cite{Emken:2017hnp}.
We compare the analytic reflection spectrum given by Eq.~(16) of~\cite{Emken:2017hnp} (black, dashed line) to the one-scattering histogram in Fig.~\ref{fig: multiple scatterings} for both nuclear and electron scatterings.
We find excellent agreement between the theory and simulation, further solidifying our confidence in the simulations' accuracy.

\vspace{10pt}\noindent\textbf{Gravitational capture of DM: }
While our simulation code was developed to study solar reflection, it naturally describes the gravitational DM~capture as well.
In Sec.~\ref{sec:monte carlo}, we defined a DM~particle as \textit{captured}, if it propagates through the Sun's interior along a bound orbit for a long time without scatterings, or if it scatters many times without getting reflected.
We set the arbitrary limit at 1000~scatterings.
As discussed previously, these choices might render our SRDM flux estimates as marginally conservative, and in turn slightly overestimate the capture rate.

\begin{figure*}
    \centering
    \includegraphics[width=0.33\textwidth]{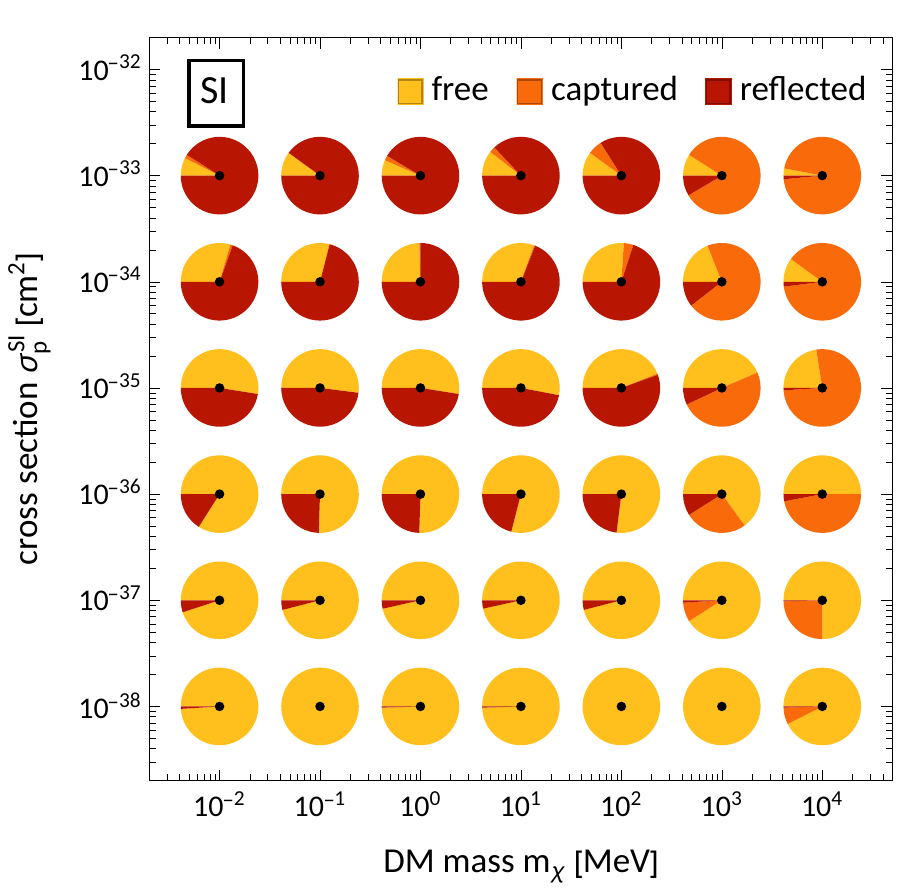}
        \includegraphics[width=0.33\textwidth]{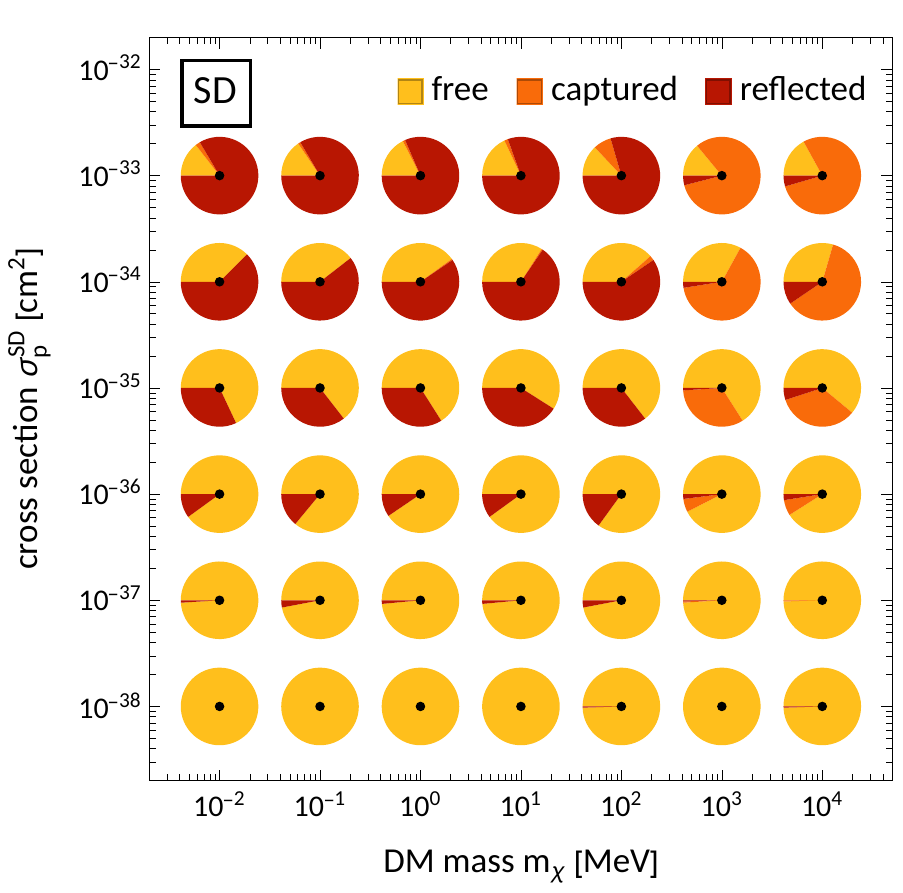}
    \includegraphics[width=0.33\textwidth]{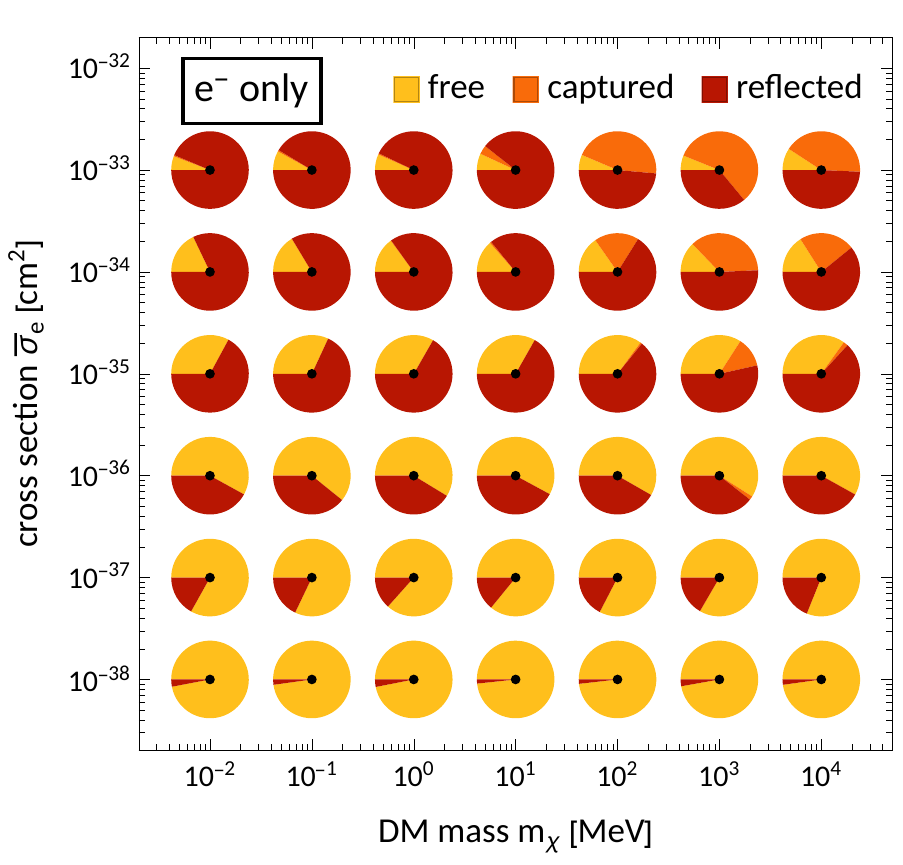}
     \includegraphics[width=0.33\textwidth]{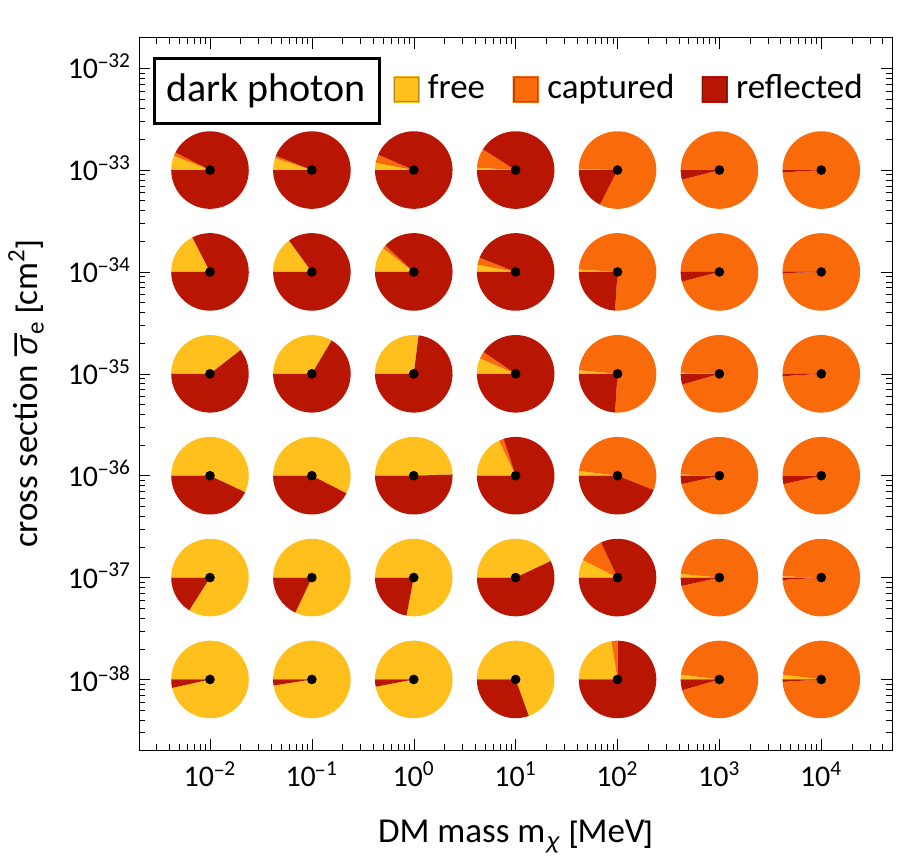}
    \caption{Pie-charts depicting the relative proportions of the DM~particles entering the Sun which do not scatter at all (yellow), get gravitationally captured (orange), or get reflected (red), as a function of DM~mass~$m_\chi$ and interaction cross section.}
    \label{fig: free vs capture vs reflection}
\end{figure*}

In Fig.~\ref{fig: free vs capture vs reflection}, we show the relative proportions of free, captured, and reflected particles depending on the DM~mass and cross section for SI/SD nuclear interactions (top row), electron interactions (bottom left panel), and for the dark photon model~(bottom right panel).
Since we are focussing on low-mass DM, gravitational capture of DM~particles plays a sub-dominant role and most particles either pass the Sun without scatterings or get reflected.
Only for large cross sections and masses do we find particles getting captured.
In particular for large DM~masses in the dark photon model, the majority particles get captured.
In this model, even small values of the electron scattering reference cross section~$\bar{\sigma}_e$ are accompanied by large nuclear cross sections as explained by Eq.~\eqref{eq: dark photon cross section ratio}.
Most DM~particles lose their energy through nuclear scatterings and enter bound orbits, and eventually thermalizing with the plasma.
In conclusion, even though it is not the purpose of this study, the simulations can indeed be used to describe DM~capture and captured particles' properties as well as the process of thermalization inside the Sun.

\begin{figure}[t!]
    \centering
    \includegraphics[width=0.5\textwidth]{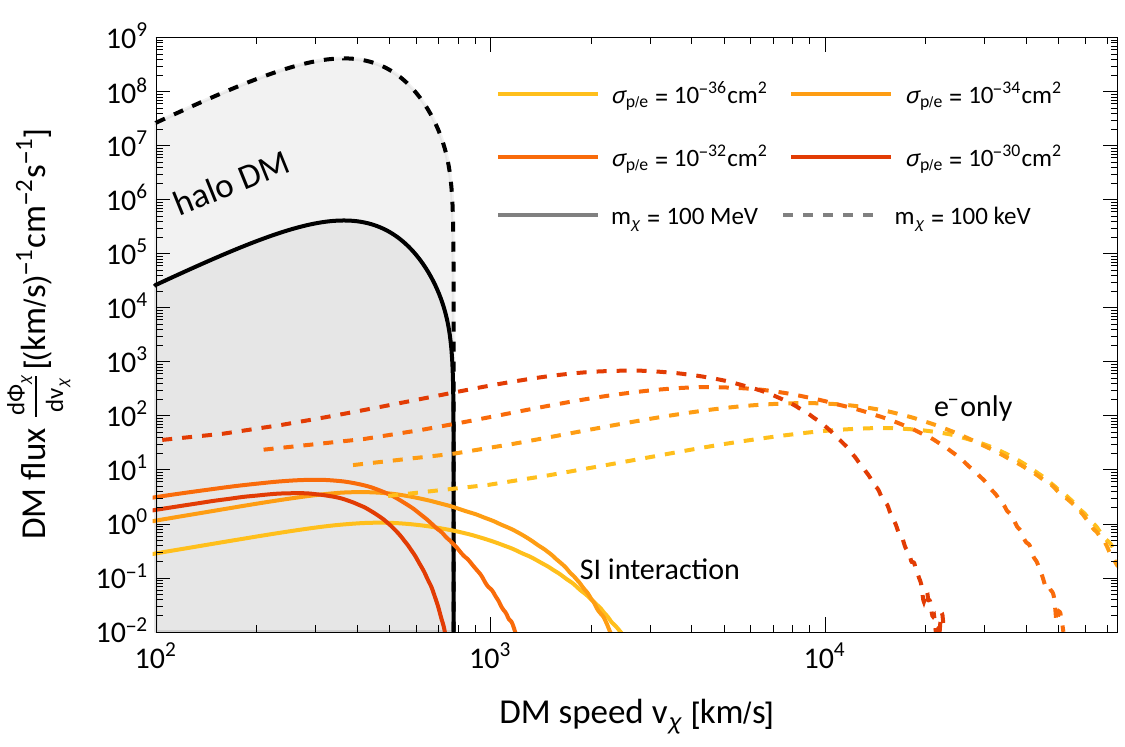}
    \caption{The differential SRDM~flux defined by Eq.~\eqref{eq: differential cross section SI} for nuclear (solid, colored lines) and electron interactions (dashed, colored lines). The black lines show the SHM flux for comparison.}
    \label{fig: SRDM spectra}
\end{figure}

\vspace{10pt}\noindent\textbf{Solar reflection spectra:}
In Eq.~\eqref{eq: differential SRDM flux}, we introduced the MC~estimate for the differential SRDM~flux~$\frac{\dd \Phi_\odot}{\dd v_\chi}$, where~$v_\chi$ is the DM~speed at a distance of 1~AU from the Sun.
We compare the SRDM spectrum to the particle flux of halo~DM in Fig.~\ref{fig: SRDM spectra}.
Here, we assume either SI~nuclear interactions for a DM~mass of 100~MeV (solid lines), or electron interactions for a mass of 100~keV (dashed lines) for cross sections between~$10^{-36}$ and~$10^{-30}\text{ cm}^2$.

Comparing the total fluxes, the contribution of solar reflection is suppressed by many orders of magnitude when comparing to the standard galactic DM~population.
However, in contrast to the sharp cut-off of the SHM~flux, the differential SRDM spectrum extends far beyond~$(v_\mathrm{gal}+v_\odot)$.
Especially for electron interactions, the reflected DM~particles ejected from the Sun can be boosted significantly.

In both cases, we observe that the SRDM~flux increases for stronger interactions as we might have expected.
Simultaneously, the high-energy tail gets suppressed due to the hot solar core being shielded off by the outer, cooler layers.

\vspace{10pt}\noindent\textbf{Anisotropy of solar reflection: }
The Sun's motion in the galactic rest frame causes the apparent ``DM~wind'' which entails that the incoming particles approach and enter the Sun in an anisotropic way.
The original direction of an DM~particle will be deflected by the Sun's gravitational pull and by collisions on nuclei and electrons.
We expect in particular the scatterings to ``wash out'' the anisotropy, and we may ask if any trace of the DM~wind survives the process of solar reflection.

\begin{figure*}
    \centering
    \includegraphics[width=0.45\textwidth]{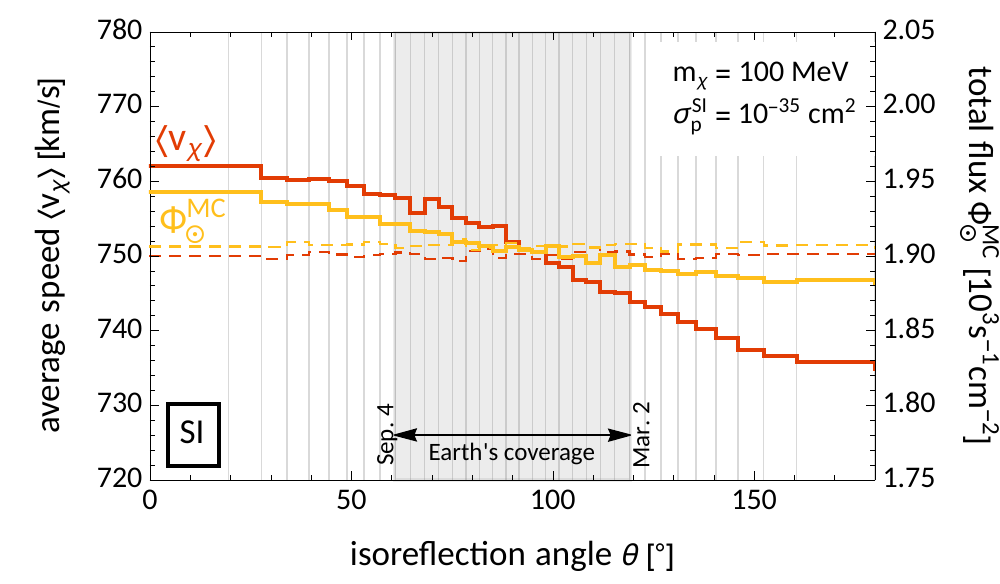}
    \qquad
    \includegraphics[width=0.45\textwidth]{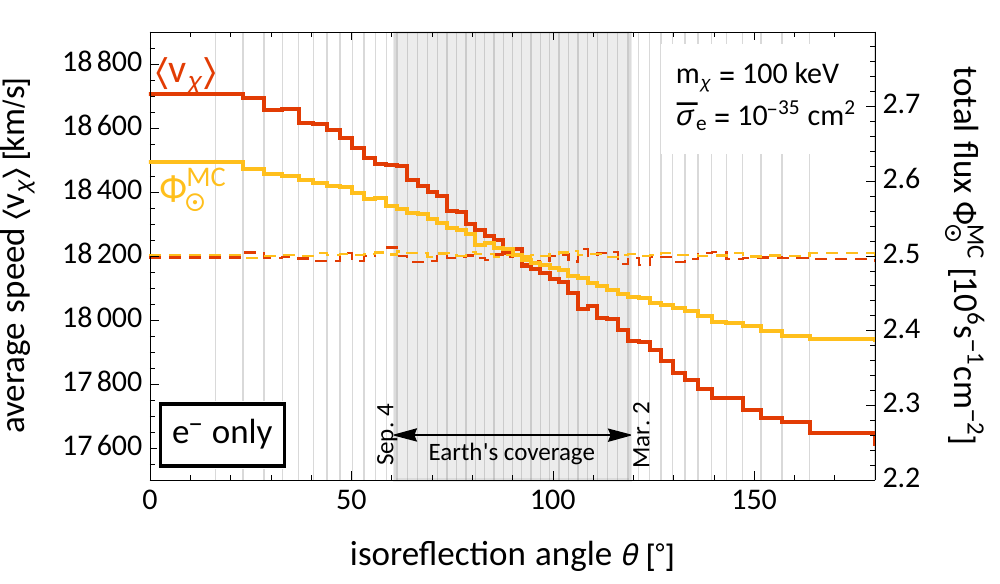}
    \caption{Anisotropies of the SRDM~flux. The figure depicts the total SRDM~flux~$\Phi_\odot^\mathrm{MC}$ through the Earth (yellow) and the mean speed~$\langle v_\chi\rangle$ (red) of reflected particles as a function of the isoreflection angle~$\theta$ for SI interactions with nuclei (left) and interactions with electrons only (right). The gray area shows the $\theta$ interval covered by the Earth's elliptical orbit through the solar system. The dashed lines are the result of MC~simulations using isotropic initial conditions, i.e. without the preferred direction of the `DM~wind', and serve as a consistency check.}
    \label{fig: anisotropy 1}
\end{figure*}

The isoreflection angle introduced in Sec.~\ref{ss: anisotropy} is the natural angle to parametrize the anisotropies.
We compute the SRDM~flux defined in Sec.~\ref{ss: isoreflection rings} for SI~nuclear (electron only) interactions using 35~(50)~isoreflection rings, which reveal its directional dependence.
For nuclear (electron) interactions, we assume a DM~particle of 100~MeV (keV) mass and an interaction cross section of~$\sigma^\mathrm{SI}_p=10^{-35}\text{cm}^2$ ($\bar{\sigma}_e=10^{-35}\text{cm}^2$).
In Fig.~\ref{fig: anisotropy 1}, we show the total flux~$\Phi_\odot$ (yellow) and the average speed~$\langle v_\chi\rangle$ (red) as a function of the isoreflection angle~$\theta$.
The area shaded in gray highlights the $\theta$~interval covered by the Earth over the course of a year, as illustrated previously in Fig.~\ref{fig: isoreflection angle}.

Furthermore, the dashed lines are the corresponding results of a consistency check where we ``removed'' the DM~wind by sampling initial conditions with isotropically distributed initial velocities.
In this case, there is no preferred direction in the simulated system, and we do not expect the average speed and total flux of SRDM to show any dependence on~$\theta$.
Indeed, this is what we find.

We demonstrate in Fig.~\ref{fig: anisotropy 1} that the SRDM~flux is generally anisotropic.
Both the total flux and the average speed of the SRDM spectrum are decreasing functions of~$\theta$, which can be understood at least in part by the fact that DM~particles reflected by single scatterings are most accelerated by hard, and thereby backward scatterings.
For SI nuclear interactions, both quantities deviate from the mean values by up to 2\% across the solar system.
We find larger anisotropies for electron interactions, where the flux~$\Phi_\odot$ ejected into the solar system by the Sun varies around its mean value by around~5\%.
The variation of the mean speed~$\langle v_\chi\rangle$ in this case is around 3\%.
One explanation for these larger anisotropies could be that the majority of particles gets reflected by a single scattering as we saw in Fig.~\ref{fig: multiple scatterings}.
For nuclear interactions, we found that most reflected particles scattered more than once which could wash out the initial anisotropies more efficiently.

The directional dependence of the total and differential reflection flux, i.e. their non-trivial dependence on the isoreflection angle, has important implications for the detection of SRDM~particles.
It sources the annual anisotropy modulation introduced in Sec.~\ref{sss: expected signal modulation}, which has to be considered alongside the orbital modulation due to the eccentricity of the Earth's orbit.
In Ch.~\ref{ss: results direct detection}, we will investigate the total modulation signature in more detail.

\vspace{10pt}\noindent\textbf{Halo model dependence: }
In the previous work on solar reflection, we conjectured that the spectrum of SRDM particles is insensitive to the details of the halo model.
After falling into the gravitational well of the Sun, the DM~particles' speed is determined mostly by the gravitational acceleration by the Sun's gravity and less so by its original, asymptotic halo speed.
In addition, an isotropic scattering of such an accelerated particle is expected to `wash out' all remaining information from the halo model.
In this section, we want to put this claim to the test and quantify the SRDM flux's dependence on the halo model.

We run a number of MC~simulations $(N=100)$, where we use the SHM model, given in Eq.~\eqref{eq: SHM boost}, to sample the initial velocities from the distribution in Eq.~\eqref{eq: IC velocity distribution}.
Here, we allow Gaussian variations of the parameters~$v_0$ and~$v_\mathrm{gal}$ of the model,
\begin{subequations}
\label{eq: shm variations}
\begin{align}
   v_0&=(220\pm 20)\text{ km/s} \, , \\
   v_\mathrm{gal}&=(544\pm 50)\text{ km/s}\, ,
\end{align}
\end{subequations}
and study the impact of these variations on the reflection spectra.
Note that $v_0$ also enters the Sun's velocity in the galactic rest frame, see Eq.~\eqref{eq: Sun velocity}.

\begin{figure*}
    \centering
    \includegraphics[width=0.32\textwidth]{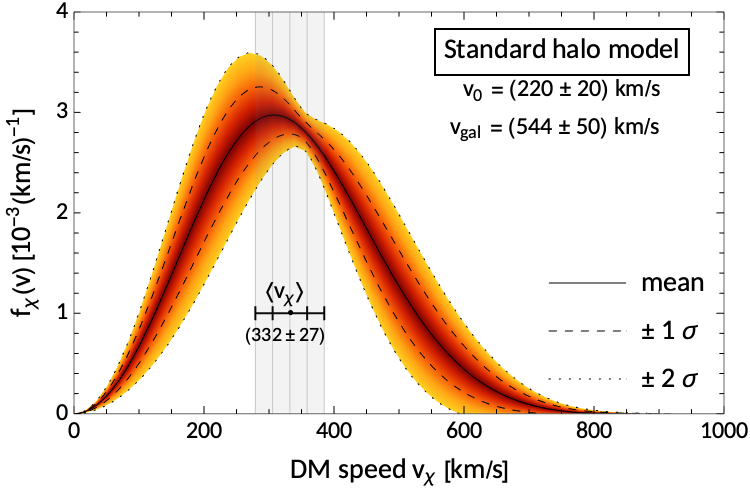}
    \includegraphics[width=0.32\textwidth]{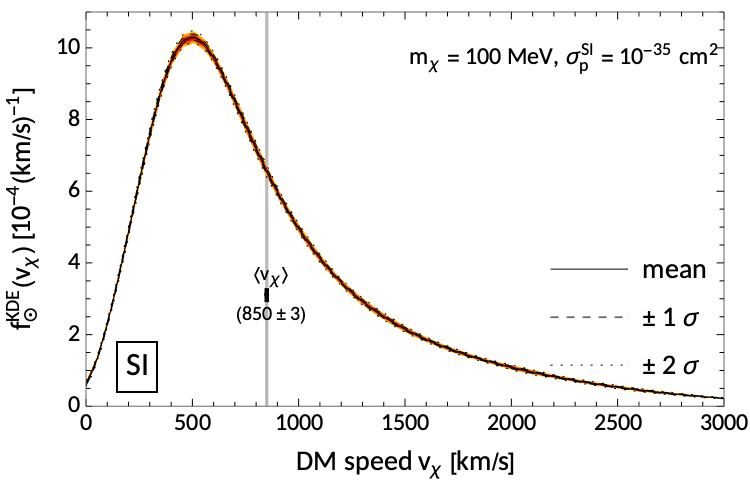}
    \includegraphics[width=0.32\textwidth]{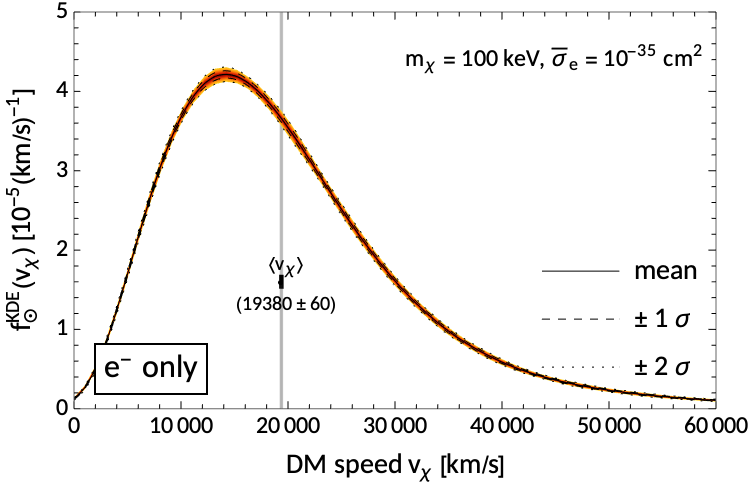}
    \caption{Halo model dependence of the reflection spectrum: The left panel shows variations of the Standard Halo Model~(SHM) with $v_0=(220\pm 20)\text{ km/s}$ and $v_\mathrm{gal}=(544\pm 50)\text{ km/s}$, whereas the middle (right) panel shows the corresponding impact on the speed distribution of the resulting reflected particles with SI-nuclear (electron only) interactions. In each panel, the dashed (dotted) line indicate the vertical $1\sigma(2\sigma)$ variations around the mean. The shaded speed range indicate the $1\sigma(2\sigma)$ variations of the mean speed~$\langle v_\chi\rangle$. (Based on 100 simulations with a sample size of 15000 reflected particles each.)}
    \label{fig: halo dependence}
\end{figure*}

The left panel of Fig.~\ref{fig: halo dependence} shows the resulting variation of the SHM speed distribution, whereas the middle (left) panel show the corresponding speed distributions of the SRDM particles and their variations.
The distribution of solar reflected particles is remarkably stable under the variations of the initial conditions. 

But while the energy distribution is shown to be largely independent of the halo model, varying the parameters of the SHM following Eq.~\eqref{eq: shm variations} may still impact the total solar reflection rate.
The rate of DM~particles entering the Sun, given by Eq.~\eqref{eq: total entering rate}, changes under these variations,
\begin{align}
\Gamma_\odot = (1.1\pm 0.1) \times 10^{33} \left(\frac{m_\chi}{\text{MeV}}\right)^{-1} \text{ s}^{-1}\, .
\end{align}
As a consequence, the total solar reflection rate also varies by the same degree under variations of the SHM~parameters.

\subsection{Direct detection results}
\label{ss: results direct detection}
Our studies of the SRDM~flux in the previous chapter were mainly motivated by the hope that direct detection experiments might observe signals from fast low-mass DM~particles boosted inside the Sun.
In this section, we derive solar reflection exclusion limits for existing experiments as well as as projections for next-generation detectors.
Secondly, we quantify the annual modulation of a potential SRDM~signal as a superposition of two independent sources of modulation.

\begin{figure*}
    \centering
    \includegraphics[width=0.48\textwidth]{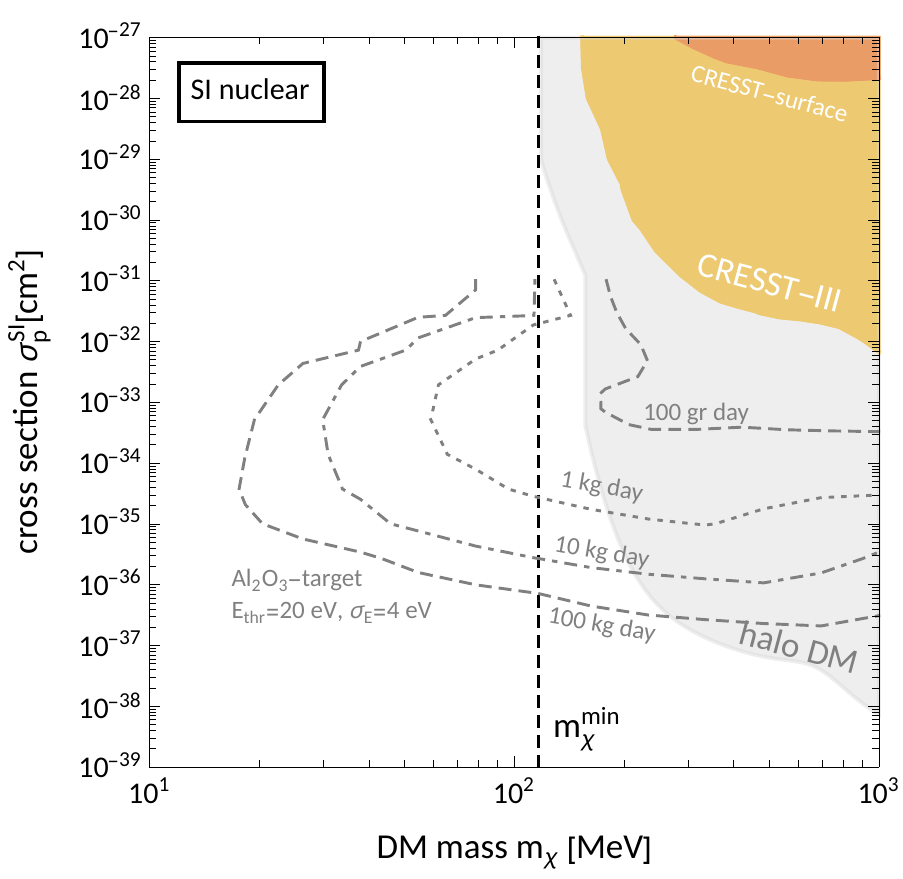}
    \includegraphics[width=0.48\textwidth]{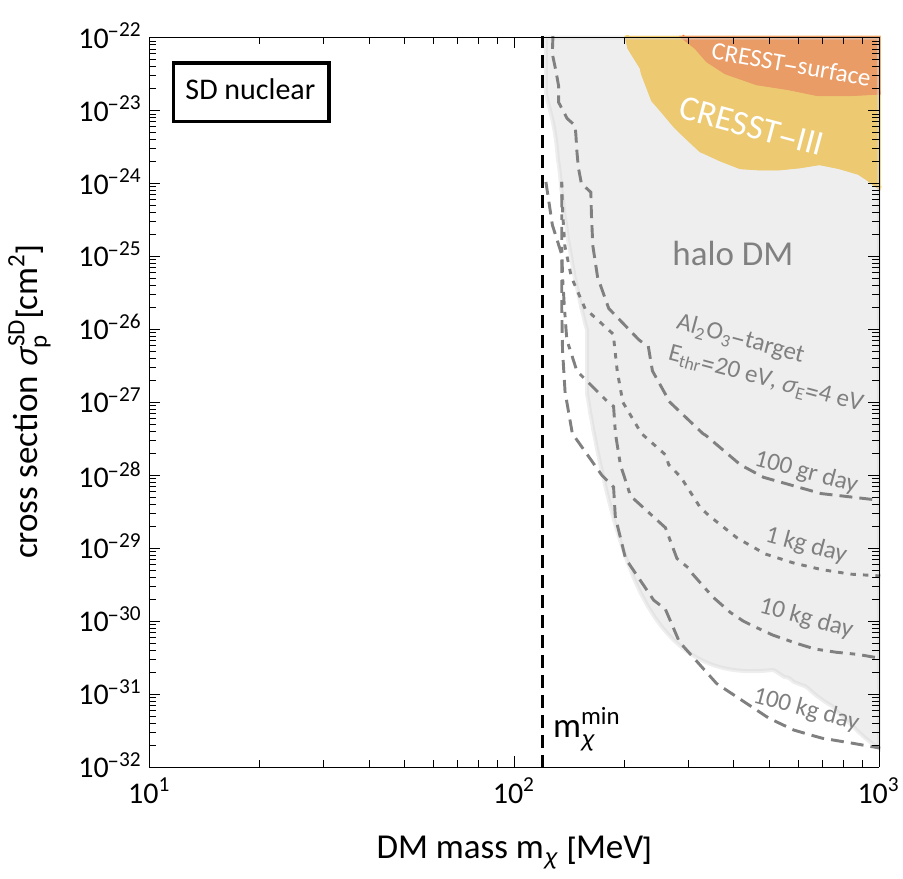}
    \includegraphics[width=0.48\textwidth]{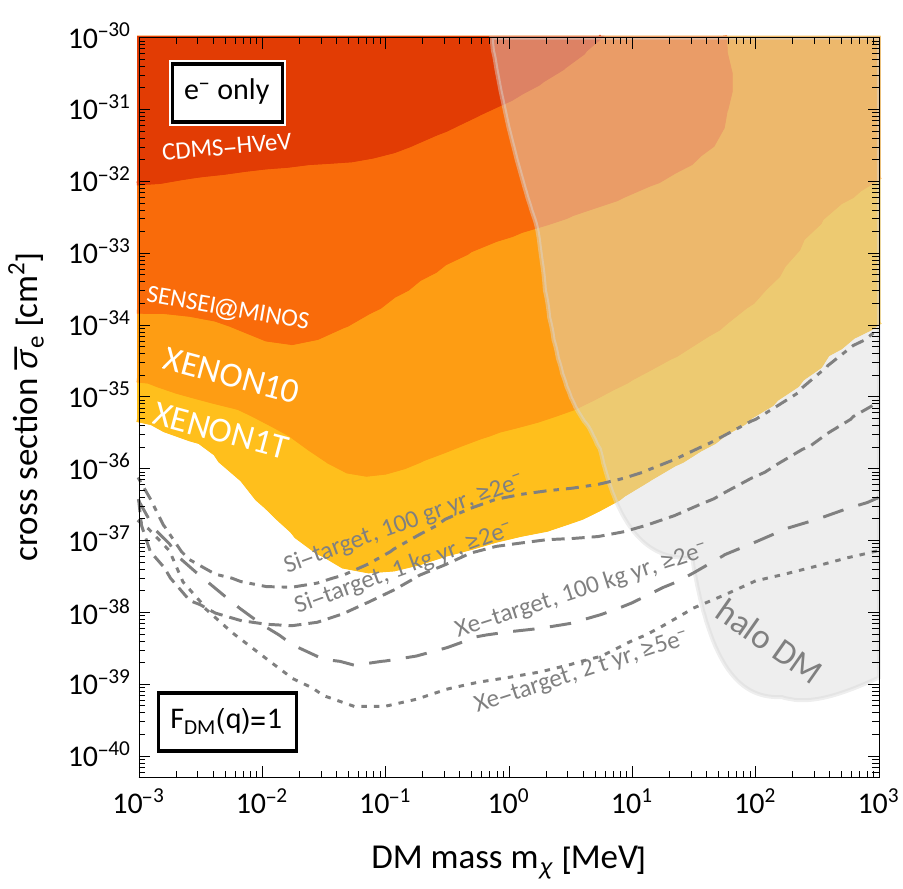}
    \includegraphics[width=0.48\textwidth]{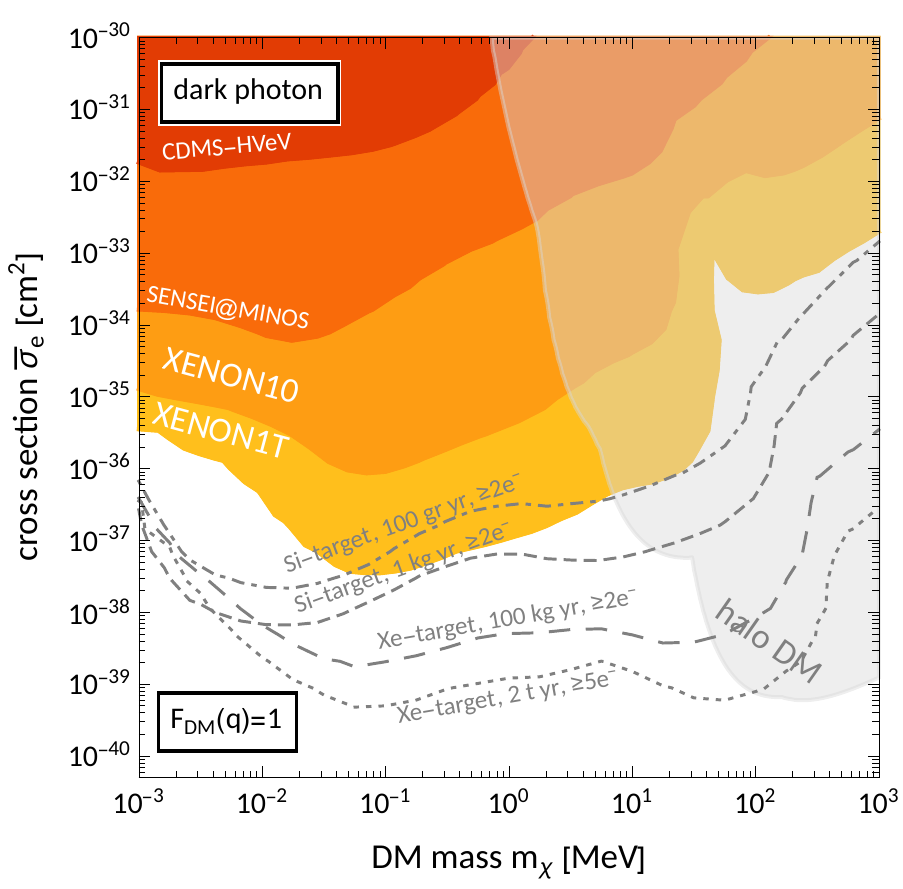}
    \caption{Solar reflection exclusion limits (95\%~CL) from direct detection experiments for SI (top left) and SD (top right) nuclear interactions, as well es electron interactions (bottom left) and the dark photon model (bottom right). The constraints are based on CRESST-III~\cite{Abdelhameed:2019hmk,Abdelhameed:2019mac}, and CRESST-surface~\cite{Angloher:2017sxg} for nuclear, and CDMS-HVeV~\cite{Agnese:2018col}, XENON10~\cite{Angle:2011th,Essig:2012yx,Essig:2017kqs}, and XENON1T~\cite{Aprile:2019xxb} for electron scattering experiments. The areas shaded in gray are already excluded by the same experiments based on halo~DM. The gray lines show projected exclusion limits for upcoming experiments. For the nuclear case, we assume a direct detection experiment with a sapphire target and an energy threshold of~$E_\mathrm{thr}=20$ eV (similarly to CRESST-surface or~$\nu$-cleus) and vary the exposure between 0.1 and 100~kg~days. The vertical black dashed line marks the minimal DM~mass that can be probed with halo DM alone. For the leptophilic and dark photon model, the projections with Si-semiconductor targets are inspired by SENSEI@SNOLAB~\cite{Barak:2020fql}, and DAMIC-M~\cite{Castello-Mor:2020jhd}, whereas the xenon target experiments are set up in anticipation of XENONnT~\cite{Aprile:2020vtw} and LBECA~\cite{Bernstein:2020cpc}. All projected limits represent an idealized scenario as we assume zero-background. For more realistic background assumptions, the projected S2-only search limits are expected to weaken by about 1-2 orders of magnitude.}
    \label{fig: direct detection constraints}
\end{figure*}

\vspace{10pt}\noindent\textbf{Exclusion limits and projections: }
Setting new exclusion limits for low-mass DM is one of the main motivation to study solar reflection.
For this purpose, we perform a parameter scan in the~$(m_\chi,\sigma)$ plane and compute the $p$-value of each parameter point.
For a given confidence level (CL), the excluded regions' boundaries are defined by $p=1-\text{CL}$.
\footnote{We can in principle find these contours by scanning the parameter space with equal-sized steps in log-space.
A more resourceful alternative is to use a contour tracing algorithm to find the excluded parameter regions.
In \texttt{DaMaSCUS-SUN}, we implemented the Square Tracing Algorithm (STA) for this purpose~\cite{STAwebsite}. Using the STA, only the subset of parameter points along the~$p=1-\text{CL}$ boundary need to be evaluated with MC~simulations.}
All of the obtained limits, both from existing and future detectors, can be found in Fig.~\ref{fig: direct detection constraints}.

The top row of Fig.~\ref{fig: direct detection constraints} shows the SRDM~limits~(95\%~CL) for SI (SD) nuclear interactions on the left (right) panel.
We derive the exclusion limits for the direct detection experiments CRESST-III~\cite{Abdelhameed:2019hmk,Abdelhameed:2019mac} and CRESST-surface~\cite{Angloher:2017sxg}.
These figures also depict the usual constraints from halo~DM by the same experiments in gray.
While we do find a regions in parameter space with large SI and SD cross sections that the CRESST experiments can exclude based on solar reflection, their exposures are insufficient to extend their sensitivity toward lower masses.
Unlike for halo~DM, an increased exposure not only allows to probe weaker interactions but potentially also lower masses when considering SRDM.
Unfortunately, the excluded masses fall above the mass threshold of halo~DM, and therefore the SRDM constraints cannot compete with the standard limits and ``drown'' in the halo limit.

In the absence of newly excluded parameter space for nuclear interactions, we want to specify what kind of nuclear recoil detector would be able to exploit solar reflection to probe lower DM~masses.
Inspired by gram-scale cryogenic calorimeters~\cite{Strauss:2017cam,Angloher:2017sxg}, we assume a sapphire target~($\text{Al}_2\text{O}_3$) and fix the energy threshold and resolution to~$E_\mathrm{thr}=20\text{ eV}$ and~$\sigma_E=4\text{ eV}$ respectively.
The gray lines in the top panels of Fig.~\ref{fig: direct detection constraints} depict the projection results for exposure between 0.1 and 100~kg~days.
Furthermore, we assume an idealized zero-background run for the sapphire target experiments.
In the case of SI~interactions, we indeed find that higher exposures probe lower masses.
In this scenario, the exposure necessary to exclude DM~masses below the minimum~$m_\chi^\mathrm{min}\approx 100\text{ MeV}$ detectable with halo~DM (shown as a vertical line) is of order~$\mathcal{O}(100~g~day)$.
For any larger exposure, the inclusion of solar reflection extends the exclusion limits.
In particular, with an exposure of 0.1, 1, 10, and 100 kg~days the respective limits reach down to about 170, 60, 30, and 15~MeV.
Therefore, we can expect future experiments such as the $\nu$-cleus experiment to be able to extend their sensitivity to lower masses by taking solar reflection into account~\cite{Strauss:2017cuu}.
This experiment might also realize an even lower energy threshold which would improve these prospects further.

For SD~interactions, the situation is generally less promising.
Even for the high-exposure projections, the SRDM limits never reach below the minimum mass probed in halo~DM searches.
Comparing SI and SD interactions, the process of solar reflection itself works very similar and the SRDM fluxes are comparable.
Assuming isospin-conserving interactions, a DM~mass of 100~MeV and a DM-proton scattering cross section of~$\sigma_p=10^{-35}\text{ cm}^2$, we find a total reflection flux on Earth of $\Phi_\odot^\text{SI}\approx2000\text{s}^{-1}\text{cm}^{-2}$ for SI, and $\Phi_\odot^\text{SD}\approx 1500\text{s}^{-1}\text{cm}^{-2}$ for SD interactions.
Also their spectra are similar, for the mean speed of the reflected particles we find~$\langle v_\chi \rangle^\text{SI} \approx 760 \text{km s}^{-1}$ and $\langle v_\chi \rangle^\text{SD} \approx 900 \text{km s}^{-1}$ respectively.
But while the SRDM flux might be comparable, the detection of SD~interactions suffer from the low number of target nuclei with non-vanishing spins and the lack of coherent scatterings. 
These factors did not affect the DM~particles' reflection in the Sun, since the most important solar target are hot protons.
For the above examples of DM~parameters and the same sapphire-target experiment, we expect a SRDM event rate of~$R^\mathrm{SI} \approx 1.1\text{kg}^{-1}\text{day}^{-1}$ for SI, which needs to be compared to a signal rate of $R^\mathrm{SD} \approx 1.9\times 10^{-4}\text{kg}^{-1}\text{day}^{-1}$ for SD cross sections.

The suppressed signal rates of SD~interactions has the consequence that only very high cross sections can be excluded even by our projected bounds.
For high cross section however, the reflection occurs in the cool outer layers of the Sun such that the reflected particles do not get boosted but are instead losing kinetic energy.
Therefore, the projections are limited by the same minimum DM mass as halo~DM constraints.
There exists significant amounts of parameter space for nuclear SD~interactions, where solar reflection is very efficient in accelerating infalling DM~particles.
For lower cross sections, i.e. values for $\sigma_p^\mathrm{SD}$ between $\sim 10^{-35}$ and $\sim10^{-33}\mathrm{cm}^2$, the DM~particles can get boosted by fast protons from the solar core and escape the Sun with higher kinetic energies.
Unfortunately, these low cross sections are not accessible to terrestrial searches at this point.

\begin{figure*}
    \centering
    \includegraphics[width=0.45\textwidth]{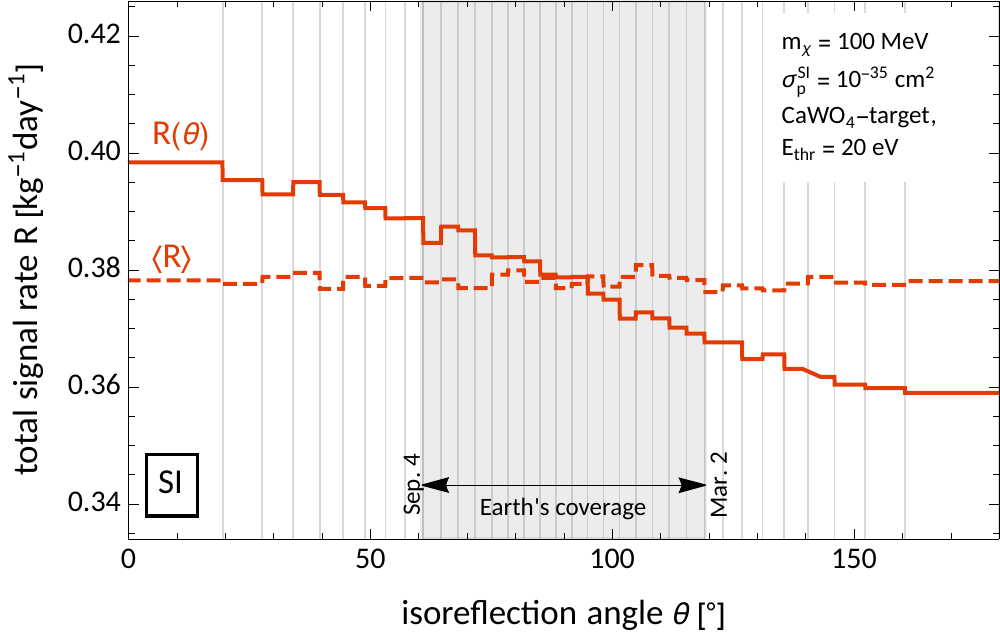}
    \includegraphics[width=0.44\textwidth]{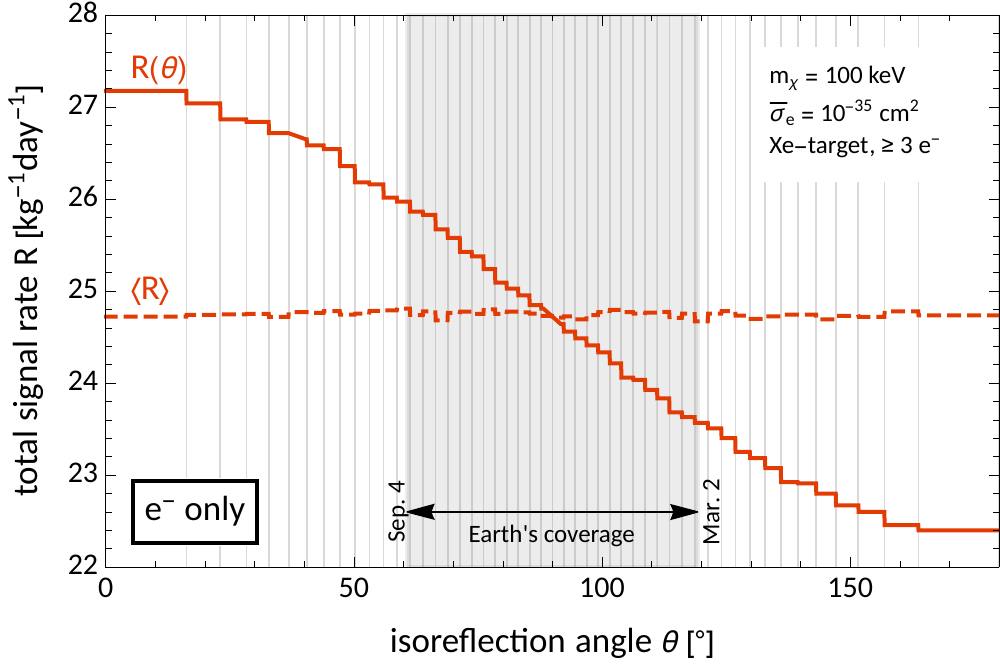}
    \includegraphics[width=0.43\textwidth]{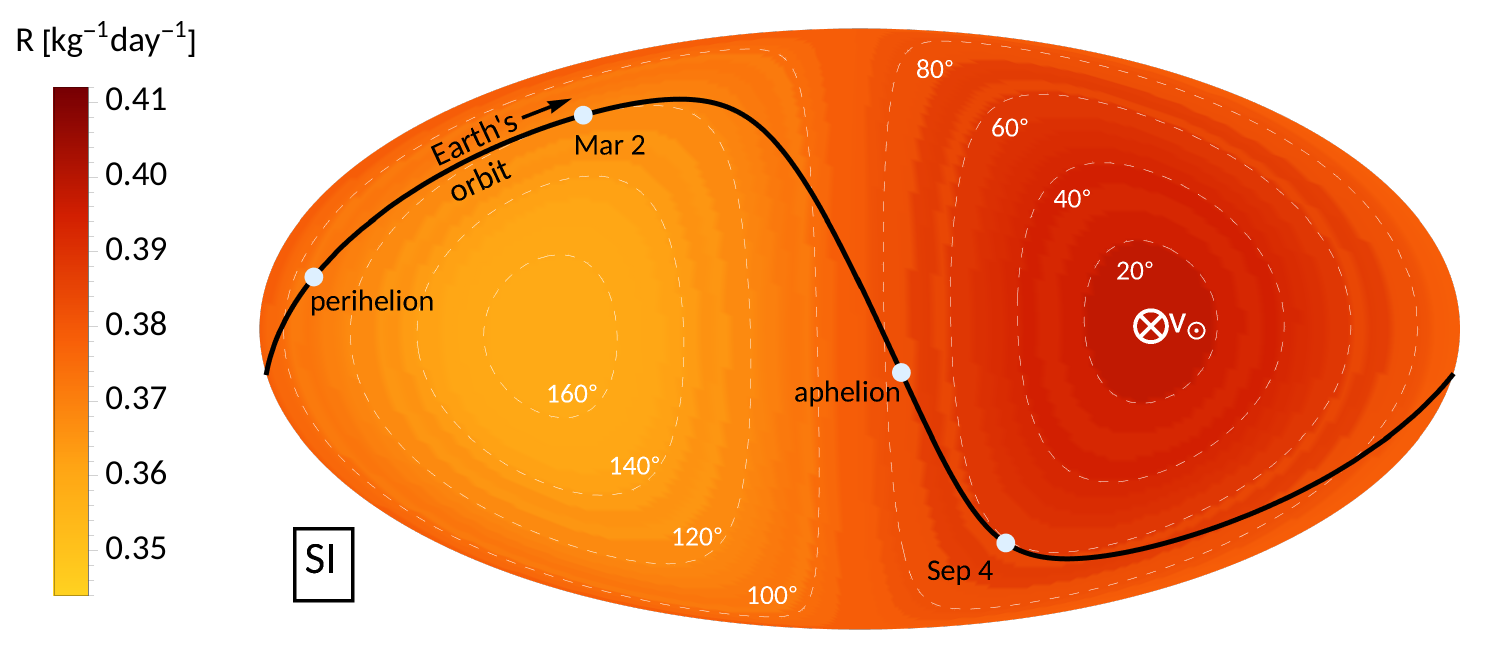}\qquad\qquad
    \includegraphics[width=0.43\textwidth]{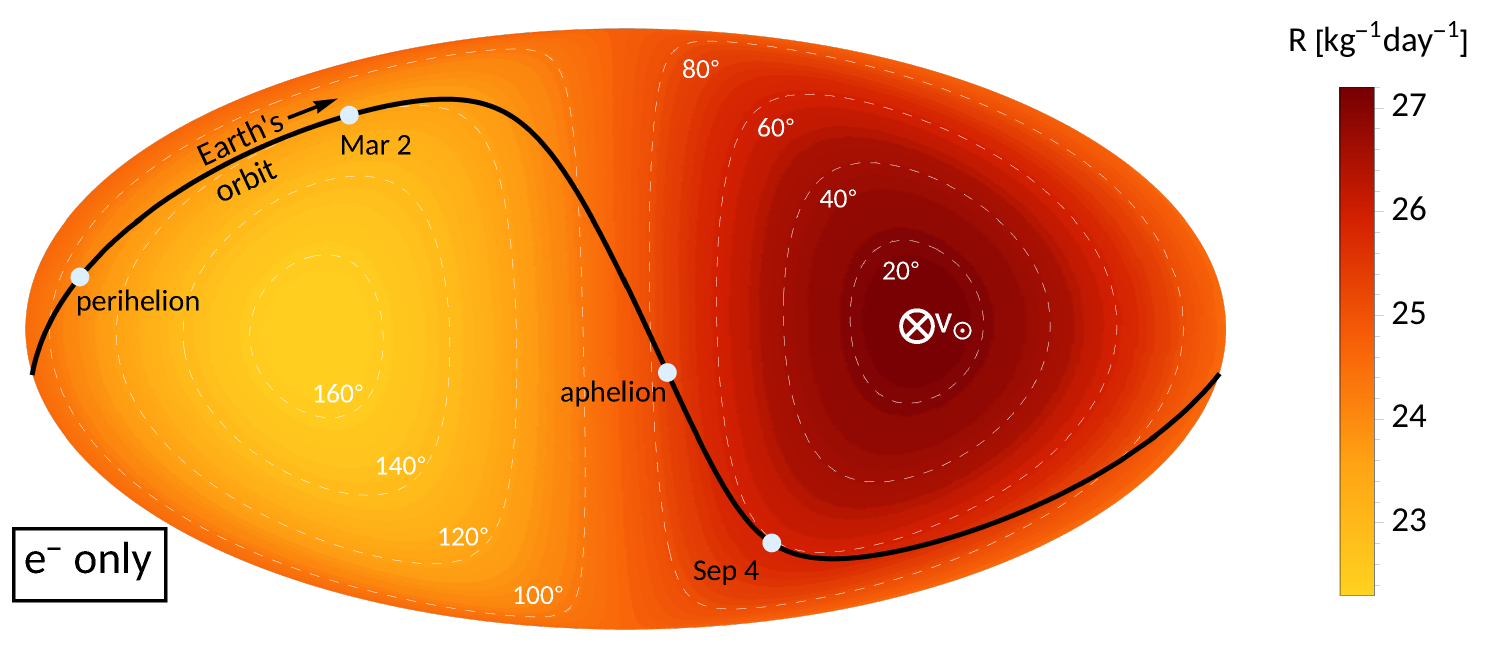}
    \includegraphics[width=0.45\textwidth]{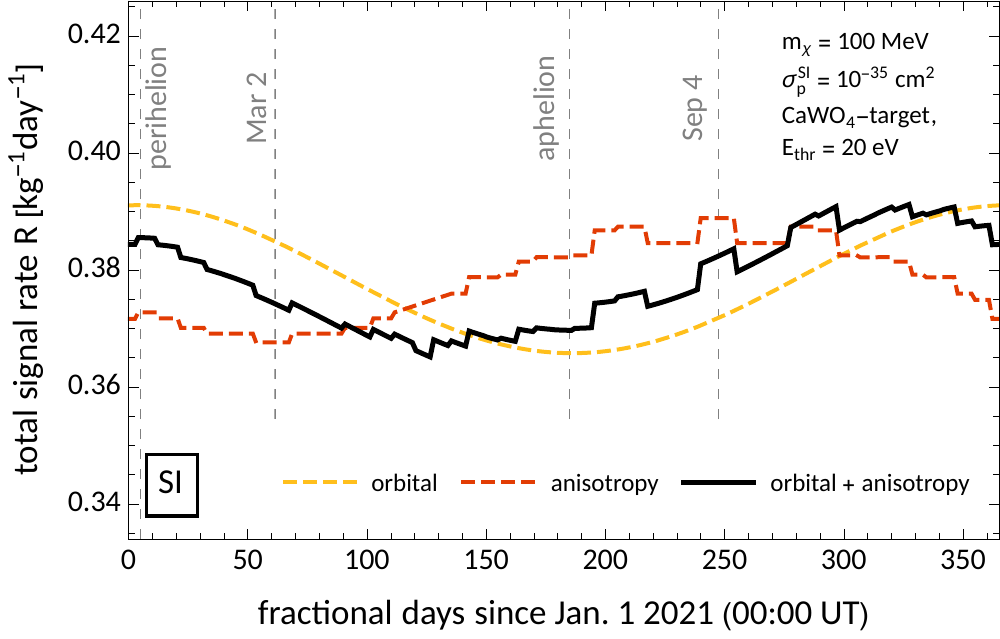}
    \includegraphics[width=0.44\textwidth]{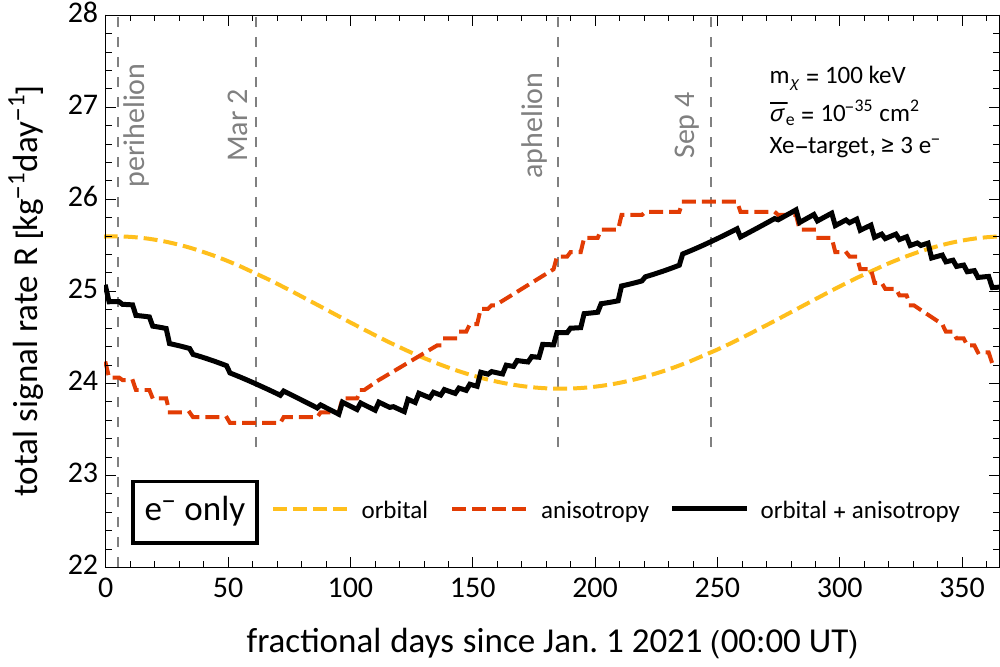}
    \caption{\textbf{First row:} The predicted SRDM~signal rate~$R$ at direct detection experiments for nuclear (left) and electron (right) interactions as a function of~$\theta$ obtained by means of MC~simulations with 35~(50) isoreflection rings. \textbf{Second row:} Mollweide projection (in the Sun's rest frame) of the event rate~$R$ with the black line showing the Earth's Kepler orbit. \textbf{Third row:} The annual modulation of the signal rate as a function of fractional days since beginning of 2021 is shown as a black line. It is obtained as a combination of the orbital and anisotropy modulation, depicted individually as yellow and red dashed lines.}
    \label{fig: annual modulation}
\end{figure*}

The situation is different when considering solar reflection via electron interactions, as seen in the bottom row of Fig.~\ref{fig: direct detection constraints}.
We show exclusion limits based on the experiments XENON10~\cite{Angle:2011th,Essig:2012yx,Essig:2017kqs} and XENON1T~\cite{Aprile:2019xxb} using liquid noble targets, as well as SENSEI@MINOS~\cite{Barak:2020fql} and CDMS-HVeV~\cite{Agnese:2018col} using silicon crystal targets.
\footnote{For a summary of the experimental setup, parameters, and the procedure to compute the exclusion limits, we refer to App.~\ref{app: experiments}.}
The inclusion of SRDM into the direct detection analysis extends the sensitivity of these experiments to the whole range of sub-MeV DM~particles.
The most stringent bounds are set by the XENON1T experiment and exclude DM~electron cross sections of~$\bar{\sigma}_e \gtrsim 3\times10^{-38}\text{ cm}^2$ for a DM~mass of~$m_\chi\approx 100\text{ keV}$.

Comparing the results for the dark photon model with the `electrons only' scenario, we find virtually no difference for DM~masses below 1~MeV, i.e. in the most relevant mass range.
For larger masses, where the SRDM~limits are weaker than halo constraints, we do find a significant difference.
We can understand the similarities and differences by remembering our discussion about the scattering rates in the dark photon model.
There we concluded that for keV (MeV)~masses, the scattering rate in the Sun is dominated by electron (nucleus) scatterings.
This leaves the dark photon model as indistinguishable from the leptophilic model for keV~masses, while for MeV~masses the scatterings on nuclei can both reduce or enlarge the excluded regions.
On the one hand, nuclear scatterings in the outer, cooler layers of the Sun might shield off the hot core and thereby weaken the SRDM~limits.
On the other hand, the addition of more target particles can also increase the SRDM~flux improving limits.
Both cases can be observed by comparing the left and right lower panels of Fig.~\ref{fig: direct detection constraints}.

In the absence of a DM~signal, future electron scattering experiments will improve upon these exclusion limits.
The gray lines in the lower panels of Fig.~\ref{fig: direct detection constraints} show projected limits for four experimental setups.
In anticipation of SENSEI@SNOLAB~\cite{Barak:2020fql} and DAMIC-M~\cite{Castello-Mor:2020jhd}, we assume two DM~detectors with a silicon semiconductor target, an exposure of~$100\text{ gr yr}$ and $1\text{ kg yr}$ respectively, and an observational threshold of 2 electron-hole pair for both setups.
Furthermore, we consider two direct detection experiments with liquid xenon as target.
The first is modelled after the proposed LBECA experiment, which is characterized by a low electron threshold (2~$e^-$)~\cite{Bernstein:2020cpc}.
For the exposure, we assume 100 kg~yr.
The XENONnT experiment is planned to have a larger exposure~\cite{Aprile:2020vtw}.
We assume a projected exposure of 2~t~yr for the S2-only search. 
Compared to LBECA, the XENONnT experiment is not mainly aiming at low-mass~DM and we fix the threshold to a higher value of 5 electrons.
Just as in the case of nuclear recoil experiments, these projections need to be understood as a lower boundary for future exclusion limits of a realistic xenon-target experiment run as we again assume an idealized scenario with zero-background.
Including a non-zero background is expected to weaken the projected limits by about 1-2 orders of magnitude depending on the solar neutrino background, see for example~\cite{Essig:2018tss}.

By including nuclear scatterings in the dark photon model, we find improved projection limits for DM~masses between 10 and 100 MeV for the xenon target experiments compared to the leptophilic scenario.
This is an interesting case, since the reflection and detection processes are not identical.
The DM~particles are reflected by nuclei, whereas the reflected particle flux is probed through its interaction with target electrons.

\vspace{10pt}\noindent\textbf{Signal modulations of SRDM: }
In Sec.~\ref{sss: expected signal modulation}, we anticipated that a potential SRDM~signal would feature an annual modulation.
This modulation would result as a combination of two effects.
For one, the variation of the Earth's distance causes a modulation of the particle flux through the detector.
Anisotropies of the SRDM~flux are a second source of annual modulations, as the Earth's orbit covers regions that are exposed to a varying DM~flux.
We already discussed and established the anisotropy of solar reflection in~\ref{ss: results spectrum} and in particular in Fig.~\ref{fig: anisotropy 1}.
The next step is to quantify the corresponding total signal modulation.

Again, we focus on SI~interactions with nuclei as well as electron interactions.
For nuclear interactions, we assume DM~particles with a mass of~$m_\chi=100\text{ MeV}$ and a cross section of~$\sigma_p^\mathrm{SI}=10^{-35}\text{ cm}^2$.
After getting reflected, they pass through a CRESST-type detector with a $\text{CaWO}_4$-target, an energy threshold of 20~eV, and an energy resolution of 4~eV.
For the leptophilic case, we determine the SRDM~flux of a 100~keV DM~particle with an interaction cross section of~$\bar{\sigma}_e=10^{-35}\text{cm}^2$ and its detection in a xenon-target experiment with an observational threshold of 3~electrons.
To study the anisotropy modulation of solar reflection via SI nuclear (electron) interactions, we perform MC simulations with 35~(50) isoreflection rings.

The top row of Fig.~\ref{fig: annual modulation} shows the obtained signal event rate as a function of the isoreflection angle~$\theta$ for SI and electron interactions on the left and right side respectively.
Again, we highlight the Earth orbit's $\theta$~coverage in gray.
The second row shows an alternative representation of the same information as a Mollweide projection in the Sun's rest frame which better visualize the directional dependence of the signal rate in the two experiments.
Here, a black line shows the Earth's orbit with certain special points highlighted.
Namely, the perihelion and aphelion where the orbital modulation assumes its maximum and minimum respectively, as well as the days of minimum (Sep 4) and maximum (Mar 2) isoreflection angle.
Note that the orbital modulation is not taken into account at this point yet.
We quantify a signal modulation in terms of the fractional modulation defined as
\begin{align}
    f_\mathrm{mod} &= {R_\mathrm{max}-R_\mathrm{min} \over R_\mathrm{max}+R_\mathrm{min}} \approx {R_\mathrm{max}-R_\mathrm{min} \over 2 \langle R \rangle}\, , \label{eq: fractional modulation}
\end{align}
where~$R_\mathrm{max}$ and~$R_\mathrm{min}$ are the maximum and minimum event rates over the year.
For example, the orbital modulation due to elliptical Kepler orbit can easily be estimated to be~$f_\mathrm{mod}^\mathrm{orbital} = {\ell_a^2 -\ell_p^2 \over \ell_a^2+\ell_p^2} \approx 3\%$, where $\ell_a(\ell_p)$ is the Earth's distance to the Sun at aphelion (perihelion).
This can be understood from the~$\ell^{-2}$ scaling of the SRDM~flux through the Earth alone.

For the nuclear recoil experiment, we find a small anisotropy modulation with~$f_\mathrm{mod}\approx 3\%$.
As such the anisotropy modulation is comparable to the orbital modulation.
This is illustrated in the left panel of the third row of Fig.~\ref{fig: annual modulation}, which depicts as a black line the signal rate as a function of time over the year 2021.
It also shows the individual modulations, anisotropy and orbital, as red and yellow dashed lines.

For solar reflection via electron scatterings only, we find a stronger dependence of the expected event rate on the isoreflection angle~$\theta$.
For a fixed distance from the Sun, the expected event rate varies by up to 10\% around its mean value depending on the direction.
For the region covered by the Earth's orbit, we obtained a fractional modulation of~$f_\mathrm{mod}\approx 5\%$.
The total modulation is therefore more determined by the effects of the SRDM~anisotropies, as seen in the right panel of the third row of Fig.~\ref{fig: annual modulation}.

As we have seen, the effect of the anisotropies can be as least as important as the orbital modulation due to its comparable amplitude which also results in a shift of the modulation's phase.
We conclude that an accurate prediction of the total annual modulation of SRDM~signals requires the knowledge of the anisotropies of the solar reflection particle flux, which can be computed using our MC simulations.

\section{Summary and conclusions}
\label{sec:discussion}

Solar reflection is the process of DM~particles from the galactic halo passing through the Sun and getting boosted by scatterings on hot solar targets.
Their increased kinetic energy facilitates their detection and can extend the sensitivity of terrestrial DM~searches towards lower masses.
In this paper, we studied the properties of the SRDM~flux ejected from the Sun and its detectability and phenomenology in laboratories on Earth.
All results presented in this paper are based on MC simulations of individual DM~particles passing through the Sun.
The MC~tool \texttt{DaMaSCUS-SUN} developed for this purpose and used to generate all results presented in this work is publicly available~\cite{Emken2021}.

We studied DM~models with dominant nuclear interactions (SI/SD) and electron interactions, as well as the dark photon model, which allows DM~particles to scatter on both nuclei and electrons.
For solar reflection via nuclear interactions, we were able to set exclusion limits based on the low-threshold experiments CRESST-III and CRESST-surface.
However, the experiments' low exposures do not yet suffice to extend their sensitivity to lower masses than accessible with halo~DM.
Unlike for standard direct detection limits, a higher exposure might probe not just weaker interactions but also lower masses in the context of solar reflection through nuclear scatterings.
Indeed, our projections with higher exposures show that gram-scale cryogenic calorimeters such as the~$\nu$-cleus experiment~\cite{Strauss:2017cuu} could extend their sensitivity to lower DM~masses by taking solar reflection into account provided that the exposure is above a few hundred gr~days.
Here, we assumed an energy threshold of~$20~eV$.
A reduction of this value would significantly improve the situation even further.
However, this conclusion is only valid for SI~nuclear interactions.
We studied DM with SD~nuclear interactions separately and found that even exposures as high as 100~kg~days are not sufficient to extend the probed DM~range.
As we discussed, this is due to the fact that the SD~cross sections probed by such an experiment are so high that the DM~particles get reflected by the cool outer layers.
For lower SD~cross sections, the SRDM flux can be comparable to the SI~case and very energetic, but the corresponding detection rates are suppressed.

For the scenario where DM~couples exclusively to electrons, we updated previous results~\cite{An:2017ojc} and derived new exclusion limits for sub-MeV~DM based on the leading experiments, most importantly XENON1T~\cite{Aprile:2019xxb}.
We also presented projected constraints of anticipated next-generation experiments with xenon and silicon targets.
With the dark photon model, we considered a case where solar reflection can be caused by scatterings on both electrons and nuclei.
However, for DM~masses below (above) an MeV, scatterings on electrons (nuclei) dominate the scattering rate.
In the case of our projected limits, we also found examples where the reflection and detection process were not the same, where DM~particles get reflected by solar nuclei and detected electron recoils.

We also investigated the annual modulation of a potential SRDM~signal, where we found two competing effects.
One source of modulation originated in the Earth's elliptical orbit around the Sun, while the other is caused by the anisotropy of the reflected DM~flux in the solar system.
Our simulation allowed us to quantify and compare both modulations and obtain the total modulation signature as a superposition of both effects.
The precise understanding of this modulation is important to distinguish a SRDM signal from both background or a signal caused by heavier halo~DM.

In this work, we focused on contact interactions, i.e. we assumed that the interaction mediator is much heavier than the momentum transfers of the scatterings.
It will be interesting to investigate a more general class of DM~interactions, in particular the scenario of ultralight mediators~\cite{SolarReflectionLightMediator}.
On the detection side, considering Migdal scatterings could be a way to probe nuclear interactions without the need for larger exposures.

Attempts to directly detect DM~particles of sub-GeV mass make up a growing field of active research with various strategies and proposals, both theoretical and experimental, to extend our experiments' discovery reach and cover more and more of the possible parameter space.
In this study, we showed that even without introducing additional assumptions or new detection technologies, solar reflection and its unique phenomenology can be a powerful aid in the search for light dark matter.

\acknowledgements
The author thanks Jelle Aalbers, Radovan Bast, Daniel Baxter, Riccardo Catena, Majken Christensen, Rouven Essig, Bradley J. Kavanagh, and Niklas Gr{\o}nlund Nielsen for valuable discussions on a variety of related topics and/or useful comments on the manuscript.
The author acknowledges that this work was produced as part of the Knut and Alice Wallenberg project Light Dark Matter (Dnr KAW 2019.0080).
The author was also supported by the Knut and Alice Wallenberg Foundation (PI, Jan Conrad).
The research presented in this paper made use of the following software packages, libraries, and tools: libphysica~\cite{Emken2021-3}, obscura~\cite{Emken2021-2}, WebPlotDigitizer~\cite{webplotdigitizer}, and Wolfram Mathematica~\cite{Mathematica}.
Most of the computations and simulations were enabled by resources provided by the Swedish National Infrastructure for Computing (SNIC) at the National Supercomputer Centre~(NSC) and the Chalmers Centre for Computational Science and Engineering~(C3SE).

\appendix

\section{Simulation details}
\label{app: simulation details}

\subsection{Equations of motion}
\label{app: equations of motion}
In between two scatterings inside the Sun, a DM~particle of mass~$m_\chi$ moves around the Sun's gravitational potential in a two-dimensional plane as the angular momentum~$\mathbf{J}$ is a constant of motion.
Using polar coordinates~$(r,\phi)$, the particle's motion in that plane is described by the Lagrangian
\begin{align}
	L &= \frac{1}{2}m_\chi\left( \dot{r}^2+r^2\dot{\phi}^2 \right) + \int\limits_{r}^{\infty}\dd r^\prime\;\frac{G_N m_\chi M_\odot(r^\prime)}{r^{\prime 2}}\, .\label{eq: sun lagrangian}
\end{align}
In App.~\ref{app:solar model}, we obtain the mass-radius relation~$M_\odot(r)$ as part of the Standard Solar Model.

The corresponding Euler-Lagrange equations,
\begin{subequations}
	\label{eq: equations of motion}
\begin{align}
	0&=\ddot{r}-r\dot{\phi}^2+\frac{G_NM_\odot(r)}{r^2}\, ,\\
	J&\equiv r^2\dot{\phi}=\text{const}\, ,
\end{align}
\end{subequations}
are the equations of motion describing orbits of freely falling particles.

In the MC~simulations of this paper, these equations are solved numerically, and the solutions are transformed back into three dimensions if necessary, for example when the particle scatters on a proton.
In order to do so, we need to keep track of the orientation of the coordinate axes in the Sun's rest frame.

Given a DM~particle's full configuration~$(t,\mathbf{r},\mathbf{v})$ in three dimensions, its angular momentum is~$\vec{J}=\vec{r}\times\vec{v}$, and the initial polar coordinates in the orbital plane and angular momentum are set to
\begin{align}
    r &= \left| \mathbf{r}\right|\, , \quad \phi = 0\, ,\quad \text{and } J = \norm{\vec{J}}\, ,
\end{align}
such that the corresponding coordinate axes are
\begin{align}
    	\unitvector{x} &= \frac{\vec{r}}{\left|\vec{r}\right|}\, ,\quad  \unitvector{z} =\vec{J}/\norm{\vec{J}} \, ,\quad\hat{\mathbf{y}} = \hat{\mathbf{z}}\times \hat{\mathbf{x}} \, .\label{eq: coordinate axes}
\end{align}
The hat~$\hat{\cdot}$ denotes unit vectors.

The equations of motion in Eqs.~\eqref{eq: equations of motion} are solved in two dimensions.
As a result, we obtain the new location of the DM~particle at a later time~$t^\prime>t$, for example the instance it scatters, as~$(r^\prime,\phi^\prime)$, which we want to translate back into three-dimensional vectors. 
Since we kept track of the axes given in Eq.~\eqref{eq: coordinate axes}, the new position and velocity vectors can be obtained as
\begin{subequations}
\begin{align}
	\mathbf{r}^\prime & = r^\prime\left(\cos \phi^\prime \; \hat{\mathbf{x}} + \sin\phi^\prime\;\hat{\mathbf{y}}\right)\, ,\\
	\mathbf{v}^\prime & = \left(\dot{r}^\prime\cos \phi^\prime-\frac{J}{r^{\prime}}\sin \phi^\prime\right)\hat{\mathbf{x}} \nonumber\\
	&+\left(\dot{r}^\prime \sin\phi^\prime+ \frac{J}{r^{\prime}}\cos\phi^\prime\right)\hat{\mathbf{y}}\, ,
\end{align}
\end{subequations}
Outside the Sun, the DM~particle follows a Kepler orbit, which we will review in the next section. In the solar interior, this is not the case, and the equations of motion need to be solved numerically.
We use the Runge-Kutta-Fehlberg method, which we summarize in Sec.~\ref{app:runge kutta fehlberg}.
This method requires the equations of motion to be expressed in terms of $1^{\rm st}$ order ordinary differential equations,
\begin{align}
	\dot{r} = v\, ,\quad \dot{v} = r\dot{\phi}^2-\frac{G_N M_\odot(r)}{r^2}\, ,\quad \dot{\phi} = \frac{J}{r^2}\, . \label{eq: 1st order eom}
\end{align}
Before we discuss the numerical solution, it will be useful to first consider the analytic solution of Eqs.~\eqref{eq: equations of motion} for unbound particles outside the Sun - hyperbolic Kepler orbits.

\subsection{Hyperbolic Kepler orbits}
\label{app: kepler}

\begin{figure}[h!]
    \centering
    \includegraphics[width=0.25\textwidth]{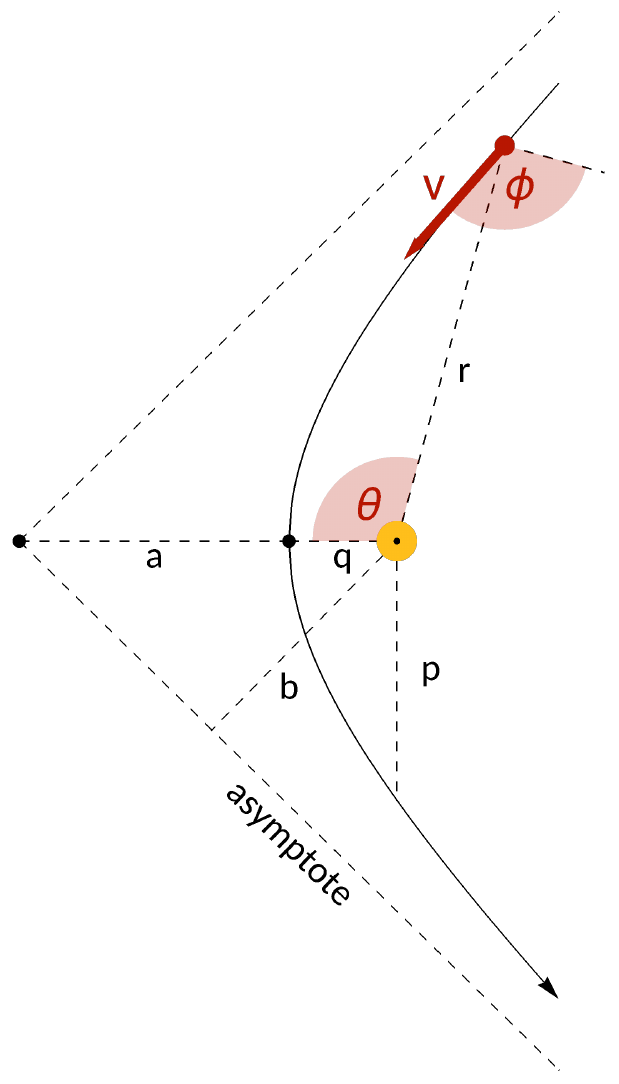}
    \caption{Hyperbolic Kepler orbit and its parameters}
    \label{fig: hyperbolic kepler orbit}
\end{figure}

Assuming a particle of mass~$m$ passing by a location~$\vec{r}$ at a time~$t$ with velocity~$\mathbf{v}$ inside the gravitational well of a central mass~$M_\odot$, its total energy is given by
\begin{align}
    E_\mathrm{tot}& = {m\over 2} \mathbf{v}^2 - {G_N M_\odot m \over r}\, .
\end{align}
If~$E_\mathrm{tot}>0$, the particle is unbound and moves along a hyperbolic Kepler orbit, illustrated in Fig.~\ref{fig: hyperbolic kepler orbit}.
As such, the orbit outside the Sun can be describe analytically and does not require numerical methods.

We review hyperbolic Kepler orbits at this point due to the usefulness of the analytic solution to Eq.~\eqref{eq: equations of motion}. It allows to shift a particle from its initial conditions asymptotically far away from the Sun close to its surface, where numerical methods have to take over.
In addition, a reflected particle can be propagated to the Earth's distance from the Sun.
In both cases, we save computational resources otherwise wasted to numerically reproduce a well-known solution.

We therefore want to compute the particle's new location~$\mathbf{r}^\prime$ and velocity~$\mathbf{v}^\prime$ at a later time~$t^\prime>t$, such that its distance to the central mass is~$r^\prime> R_\odot$.
In addition, the particle should not have passed by the orbit's periapsis~$q$, because it might lie in the Sun's interior ($q<R_\odot$), where the analytic solution does not apply.

The particle's orbit and position are characterized by a number of parameters:
\begin{subequations}
\label{eq:orbit parameter}
\begin{itemize}
\item the particle's distance to the Sun~$r$,
\item the particle's angular momentum with respect to the central mass,
    \begin{align}
        \mathbf{J} &= \mathbf{r} \times \mathbf{v}\, 
    \end{align}
    \item the particle's asymptotic speed,
    \begin{align}
        u &= \sqrt{\mathbf{v}^2-v_{\rm esc}^2(r)}\nonumber\\
        &= \sqrt{\mathbf{v}^2 - {2 G_N M_\odot \over r}}\, ,
    \end{align}
    \item its semimajor axis defined such that $a>0$,
    \begin{align}
       a &= \frac{G_N M_\odot}{u^2}\, , 
    \end{align}
    \item its semilatus rectum,
    \begin{align}
        p &= \frac{\mathbf{J}^2}{G_NM_\odot }\, ,
    \end{align}
     \item its eccentricity,
    \begin{align}
       e &= \sqrt{1+\frac{p}{a}}>1 \, ,
    \end{align}
     \item its periapsis,
    \begin{align}
        q &= (e-1)\,a\, ,\label{eq: periapsis}
    \end{align}
    \item the impact parameter,
    \begin{align}
        b &= \sqrt{e^2-1}\,a\, ,\label{eq: impact parameter}
    \end{align}
     \item the angle from the periapsis
    \begin{align}
       \cos \theta &=\frac{1}{e}\left(\frac{p}{r}-1\right) \, ,\label{eq: angle from periapsis}
    \end{align}
     \item its changing speed,
    \begin{align}
       v^2&=\frac{G_NM_\odot}{p}\left(1+e^2+2e\cos\theta\right) \, ,
    \end{align}
     \item the flight path angle between the velocity and the perpendicular of the radial direction,
    \begin{align}
        \tan \phi &=\frac{e\sin \theta}{1+e\cos\theta}\, ,\label{eq: flight path angle}
    \end{align}
     \item the eccentricity anomaly,
    \begin{align}
        \cosh F &=\frac{e+\cos \theta}{1+e\cos\theta}\, ,
    \end{align}
     \item the mean anomaly,
    \begin{align}
       M &=e\sinh F-F \, ,
    \end{align}
     \item and finally, the time from the periapsis,
    \begin{align}
        t-t_p &=\sqrt{\frac{a^3}{G_N M_\odot}}M\, .
    \end{align}
\end{itemize}
\end{subequations}

Finally, to obtain the final position and velocity vectors,~$\mathbf{r}^\prime$ and $\mathbf{v}^\prime$, we need to know the orientation of the Cartesian coordinate axes such that~$\theta=0$ actually corresponds to the periapsis,
\begin{subequations}
\begin{align}
	\unitvector{z} &= \vec{J}/\norm{\vec{J}}\, ,\\
	\unitvector{x} &=\cos\theta\;\hat{\mathbf{r}}+\sin\theta \ \hat{\mathbf{r}}\times\unitvector{z}  \, ,\\
	\unitvector{y} &= \unitvector{z}\times\unitvector{x}\, ,
\end{align}
\end{subequations}
where~$\hat{\mathbf{r}}\equiv \vec{r}/r$ and the angle from the periapsis~$\theta$ is given by Eq.~\eqref{eq: angle from periapsis}. 
Using the Eqs.~\eqref{eq:orbit parameter}, the particle's location~$\mathbf{r}^\prime$ and velocity~$\mathbf{v}^\prime$ at time~$t^\prime$ are given by
\begin{subequations}
\label{eq:Kepler shift}
\begin{align}
	t^\prime &= t +\text{sign}(r^\prime-r)\sqrt{\frac{a^3}{G_N M_\odot}} (M^\prime-M)\, ,\\
	\mathbf{r}^\prime &= r^\prime \left( \cos\theta^\prime\; \unitvector{x} +\sin\theta^\prime\;  \unitvector{y} \right)\, ,\\
	\mathbf{v}^\prime &=v^\prime\;\frac{e\sin\theta^\prime\;\hat{\vec{r}}^\prime+\left(1+e\cos\theta^\prime\right)\;\unitvector{z}\times\hat{\vec{r}}^\prime}{\sqrt{1+e^2+2e \cos\theta^\prime}}\, ,\nonumber\\
	&=\sqrt{\frac{G_NM_\odot}{p}}\left(e\sin\theta^\prime\;\hat{\mathbf{r}}^\prime+\left(1+e\cos\theta^\prime\right)\;\unitvector{z}\times\hat{\mathbf{r}}^\prime\right)\, , 
\end{align}
where again~$\hat{\mathbf{r}}^\prime=\vec{r}^\prime/r^\prime$ and the new angle from the periapsis is
\begin{align}
	\theta^\prime &= \text{sign}(r^\prime-r)\arccos\left[\frac{1}{e}\left(\frac{p}{r^\prime}-1\right)\right]\, .
\end{align}
\end{subequations}
These equations allow us to propagate a particle from large distances to the Sun's surface and vice versa without the need for numerical methods.

\subsection{Runge-Kutta-Fehlberg method}
\label{app:runge kutta fehlberg}

This chapter contains a brief review of the Runge-Kutta-Fehlberg method (RKF or RK45), an adaptive, iterative algorithm for the numerical solution of ordinary differential equations~(ODEs)~\cite{Fehlberg1969}.
In this work, we use this method to solve the equations of motion of the DM~particles as they move through the bulk mass of the Sun.

Given a $1^{\rm st}$ order ODE for an unknown function~$y(t)$,
\begin{align}
    {\dd y(t) \over \dd t} &= f(t, y(t) )\, ,
\end{align}
with some initial conditions~$y(t_0) = y_0$, we can find a numerical solution iteratively by computing the function's new value after a finite time step~$\Delta t$,
\begin{align}
    y_{k+1} &= y_k + \frac{25}{216}k_1+ \frac{1408}{2565}k_3+ \frac{2197}{4101}k_4- \frac{1}{5}k_5\, , \label{eq: RK4}\\
    t_{k+1} &= t_k + \Delta t\, ,
\end{align}
where the coefficients~$k_i$ are given by
\begin{subequations}
\label{eq: RK45 coefficients}
\begin{widetext}
\begin{align}
    k_1 &= \Delta t\; f\left(t_k, y_k\right) \, ,\\
	k_2 &= \Delta t\; f\left(t_k + \frac{1}{4}\Delta t, y_k+\frac{1}{4}k_1\right) \, ,\\
	k_3 &= \Delta t\; f\left(t_k + \frac{3}{8}\Delta t, y_k+\frac{3}{32}k_1+\frac{9}{32}k_2\right) \, ,\\
	k_4 &= \Delta t\; f\left(t_k + \frac{12}{13}\Delta t, y_k+\frac{1932}{2197}k_1-\frac{7200}{2197}k_2+\frac{7296}{2197}k_3\right) \, ,\\
	k_5 &= \Delta t\; f\left(t_k + \Delta t, y_k+\frac{439}{216}k_1-8k_2+\frac{3680}{513}k_3-\frac{845}{4104}k_4\right) \, ,\\
	k_6 &=\Delta t\; f\bigg(t_k + \frac{\Delta t}{2}, y_k-\frac{8}{27}k_1+2k_2-\frac{3544}{2565}k_3+\frac{1859}{4104}k_4-\frac{11}{40}k_5\bigg) \, .
\end{align}
\end{widetext}
\end{subequations}
The iteration step in Eq.~\eqref{eq: RK4} corresponds to a numerical solution by an ordinary ~$4^{\rm th}$ order Runge-Kutta method (RK4).
What makes the RK45 method so powerful is the fact that the same coefficients~$k_i$ of Eq.~\eqref{eq: RK45 coefficients} can simultaneously be combined to yield a~$5^{\rm th}$ order approximation of the solution,
\begin{align}
    \tilde{y}_{k+1} &=y_k + \frac{16}{135}k_1+ \frac{6656}{12825}k_3+ \frac{28561}{56430}k_4\nonumber\\
    &- \frac{9}{50}k_5+\frac{2}{55}k_6\, . \label{eq: RK5}
\end{align}
Therefore, the RK45 method is a combination of two ordinary Runge-Kutta methods, namely RK4 and RK5, yet only requiring the same number of function evaluations of~$f(t,y(t))$ as a RK6 method.
The~$5^{\rm th}$ order solution of Eq.~\eqref{eq: RK5} gives a direct estimate of the error of the RK4 solution in Eq.~\eqref{eq: RK4}.
This estimate is given by $\left| y_{k+1}-\tilde{y}_{k+1}\right|$ and allows to \emph{adapt} the time step~$\Delta t$ based on a specified error tolerance~$tol$ of $y$.
\begin{align}
	\Delta t_{k+1} = 0.84 \left(\frac{\text{tol}}{\left| y_{k+1}-\tilde{y}_{k+1}\right|}\right)^{1/4}\Delta t_k\, . \label{eq: adaptive delta t}
\end{align}
The new time step is used for the next iterative step, unless the error exceeded the tolerance, i.e.~$\left| y_{k+1}-\tilde{y}_{k+1}\right| > tol$.
In the latter case, the previous step is repeated using the new, smaller step size~$\Delta t_{k+1}$.
This way, the RK45 method guarantees that the solution's error is bounded by the tolerance while the step size may change adaptively, increasing the efficiency of our trajectory simulations.

We impose an additional upper bound on~$\Delta t$ ensuring that the time step is short compared to the time scale of scatterings,
\begin{align}
    \Delta t < 0.1 \times\tau(r,w)\, , \label{eq: adaptive delta t 2}
\end{align}
where the mean free time~$\tau(r,w)$ is given by Eq.~\eqref{eq: mean free time}.

In our simulations, the RKF method is used to solve the three equations of motion given in \eqref{eq: 1st order eom}. The used values for the error tolerances are
\begin{align}
    \text{tol}_r &= 1\,\text{km}\, , \\
    \text{tol}_v &= 10^{-3}\,\text{km s}^{-1}\, , \\
    \text{tol}_\phi &= 10^{-7}\, .
\end{align}
These values were found through experimental simulations of trajectories outside the Sun, where the analytic solutions can be used to verify that the RKF results are accurate.
In addition, it was confirmed that a further decrease of tolerance does not alter the solar reflection results.
Finally, we abort the simulation of a gravitationally bound particle after~$10^7$ time steps without scattering, and consider the particle as ``captured'' as discussed at the end of chapter~\ref{ss: trajectory simulation}.

\subsection{Initial conditions}
\label{app:initial conditions}

The initial conditions for our simulations should accurately describe DM~particle of the galactic halo whose trajectory is about to cross the Sun's surface.
Up to this crossing point, the particles follow unbound, hyperbolic Keplerian orbits.
The initial distribution of DM~particles is encapsulated in the halo model, which includes the local DM~energy density and velocity distribution far away from the Sun.
However, as the particles approach the Sun, they get focused and accelerated by the Sun's gravitational pull and their distribution gets distorted.
Therefore, we need to generate the initial particle location and velocity at a far distance from the Sun where the halo model applies and propagate the particle along its Kepler orbit toward the Sun.

The procedure of generating initial conditions is split into two parts.
At first, we need to generate a particle's location and velocity asymptotically far from the Sun\footnote{In practice, we use an initial distance of 1000 AU.} such that 
\begin{itemize}
    \item[(a)] the initial positions are effectively distributed homogeneously in space, and 
    \item[(b)] the resulting trajectory is guaranteed to cross the solar surface.
\end{itemize}
In a second step, we have to describe the particle's fall toward the Sun accounting for acceleration and focussing.
Using the results of Sec.~\ref{app: kepler}, this process can be described analytically up to surface crossing, where the orbit ceases to be a Kepler orbit and where the numerical RKF procedure takes over.

\paragraph*{Initial velocity~$\vec{u}$:}
The initial velocities of DM~particles far away from the Sun that are about to enter the star do not simply follow the velocity distribution of the SHM given in Eq.~\eqref{eq: SHM boost}.
Compared to the SHM, the velocity distribution of infalling DM~particles is distorted due to two independent facts:
\begin{enumerate}
    \item Faster particles enter the Sun with increased rate.
    \item Slower particles enter the Sun with greater numbers, since gravitational focusing can pull in slower particles from a greater volume than faster particles.
\end{enumerate}
The resulting velocity distribution is proportional to the differential entering rate, see Eq.~\eqref{eq: total entering rate}, and given in Eq.~\eqref{eq: IC velocity distribution}.
To sample a velocity~$\vec{u}$ from~$f_\mathrm{IC}(\vec{u})$, we start by sampling the asymptotic speed~$u$ from the speed distribution,
\begin{align}
    f_\mathrm{IC}(u) &=  \int \dd \Omega\, u^2 f_\mathrm{IC}(\mathbf{u})\nonumber\\
    &=\mathcal{N}_\mathrm{IC}\, \left(u + {\vesc{R_\odot}^2 \over u} \right) f_\odot(u)\, , \label{eq: IC speed distribution}
\end{align}
where $f_\odot(u)$ is the speed distribution of the SHM in the solar rest frame.
We sample~$u$ using rejection sampling.
Given a value of the speed, we need to sample the direction of~$\vec{u}$ next.
For this purpose, we express~$\vec{u}$ in spherical coordinates, i.e.~$\vec{u}(u,\theta,\phi)$ with the Sun's velocity~$\mathbf{v}_\odot$ along the $z$-axis.
With this choice of coordinate system, the velocity distribution~$f_\odot(\vec{u})$, and thereby~$f_\mathrm{IC}(\vec{u})$, only depend on the speed~$u$ and the polar angle~$\theta$ (or rather~$\cos\theta$), and not on the azimuthal angle~$\phi$ which therefore follows the uniform distribution~$\mathcal{U}_{[0,2\pi]}$.
We sample the value of~$\cos\theta$ via rejection sampling from the conditional distribution function for a fixed value of~$u$,
\begin{align}
    f_\mathrm{IC}(\cos\theta) &=\left.\frac{\int_0^{2\pi}\dd\phi\, f_\mathrm{IC}(\vec{u})}{\int\dd\Omega\, f_\mathrm{IC}(\vec{u})}\right|_{u \text{ fixed}}\nonumber\\
    &=\left.\frac{\int_0^{2\pi}\dd \phi\, f_\odot(\vec{u})}{f_\odot(u) / u^2}\right|_{u \text{ fixed}}\nonumber\\
    &= \left.\underbrace{\frac{2\pi}{f_\odot(u) / u^2}}_{\equiv \mathcal{N}_\mathrm{IC}^\prime}\; f_\odot(\vec{u})\right|_{u \text{ fixed}}\, . \label{eq: IC distribution cos theta}
\end{align}
The prefactor~$\mathcal{N}_\mathrm{IC}^\prime$ is another normalization constant.
The domain of~$\cos\theta$ depends on the fixed value of~$u$ and is given by
\begin{subequations}
\begin{align}
    \cos\theta &\in [-1, \cos\theta_\mathrm{max}]\, 
    \intertext{with}
    \cos\theta_\mathrm{max} &= \text{min}\left(1, \frac{v_\mathrm{gal}^2-v_\odot^2 - u^2}{2u v_\odot}\right)\, .
\end{align}
\end{subequations}
This is due to our choice of our coordinate system and the fact that the fastest particles only approach the Sun from a particular direction.

Finally, by sampling the azimuthal angle~$\phi\in(0,2\pi)$ from a uniform distribution, we can construct the full velocity vector~$\vec{u}$.

\paragraph*{Initial position~$\vec{r}$:}
After knowing the initial velocity, choosing the initial position is crucial, since we want to ensure that the particle does not fail to hit the solar surface but is instead on an collision course with the Sun.
In other words, we want to ensure that the orbit's perihelion~$q$, given by Eq.~\eqref{eq: periapsis}, is smaller than the solar radius~$R_\odot$.
This will put a constraint on the viable impact parameter~$b$ of Eq.~\eqref{eq: impact parameter}, which can be used to sample the initial location.

\begin{figure}[t!]
    \centering
    \includegraphics[width=0.4\textwidth]{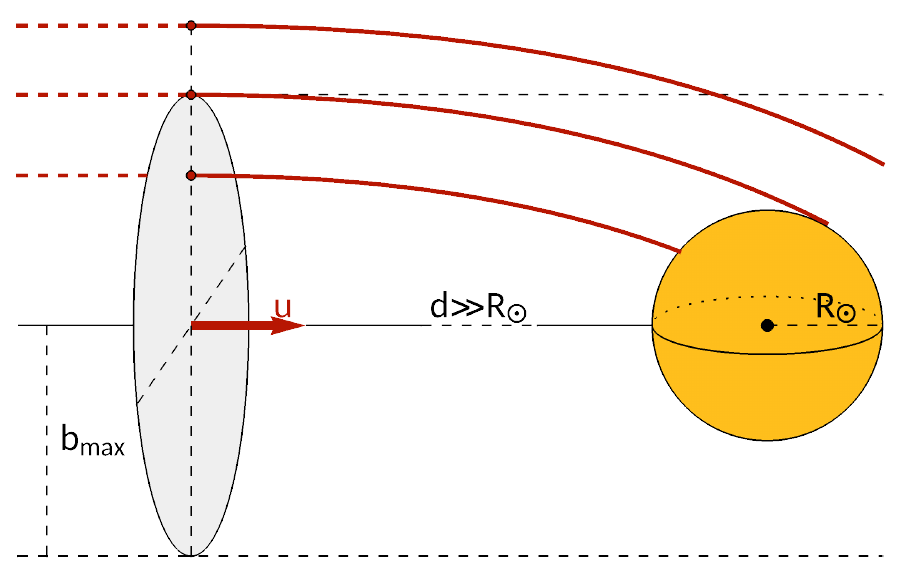}
    \caption{Initial conditions and constraint on impact parameter~$b$}
    \label{fig: initial conditions}
\end{figure}

For a given initial location~$\mathbf{r}$ at a far distance, the angular momentum of the DM~particle is given by
\begin{align}
    J = \left| \mathbf{r} \times \mathbf{u}\right| = r u \sin \alpha \, .
\end{align}
At the perihelion~$q$, the angle~$\alpha$ is equal to $90^\circ$, and the conserved angular momentum can be written as
\begin{align}
    J = q w(u, q)\, ,
\end{align}
where $w(u, q)=\sqrt{u^2 + {2 G_N M_\odot \over q}}$.
If the particle's angular momentum is too high, its orbit's perihelion lies outside the Sun which we want to avoid.
Therefore, we require the initial angular momentum to be bounded by
\begin{align}
    J < R_\odot w(u,R_\odot) \, ,
\end{align}
which is equivalent to~$q < R_\odot$.
Looking at Eq.~\eqref{eq: impact parameter}, this imposes an upper bound for the impact parameter~$b=J/u$,
\begin{align}
    b &< b_\mathrm{max}\equiv {R_\odot w(u,R_\odot) \over u} \nonumber\\
    &= \sqrt{1+{\vesc{R_\odot}^2\over u^2}}R_\odot\, . \label{eq: hit condition}
\end{align}
The fact that~$b>R_\odot$ reflects the effect of gravitational focussing.
In practice, since we sample initial positions at a large yet \textit{finite} distance~$r\gg R_\odot$, we modify the maximum impact parameter slightly to
\begin{align}
   b_\mathrm{max}&=\sqrt{\frac{u^2+\vesc{R_\odot}^2}{u^2+\vesc{r}^2}}R_\odot\, .
\end{align}
By setting~$\vesc{r}=0$, we recover Eq.~\eqref{eq: hit condition}.

We can now sample the initial position from a homogeneous distribution with the constraint that the particle's impact parameter satisfies Eq.~\eqref{eq: hit condition}, as illustrated in Fig.~\ref{fig: initial conditions}.
It is given by
\begin{align}
	\vec{r} = d \;\hat{\mathbf{z}} +\sqrt{\xi}\;b_\mathrm{max}\left(\cos \varphi \;\hat{\mathbf{x}}+\sin \varphi \;\hat{\mathbf{y}} \right)\, ,\label{eq:initial position}
\end{align}
where we used~$\unitvector{z} \equiv -\frac{\mathbf{u}}{u}$, such that~$\unitvector{x}$ and~$\unitvector{y}$ span the plane perpendicular to~$\vec{u}$.
Furthermore,~$\xi$ and~$\varphi$ are sampled values of the uniform distributions~$\mathcal{U}_{[0,1]}$ and $\mathcal{U}_{[0,2\pi]}$ respectively.
The resulting locations are distributed uniformly on the two-dimensional plane with~$b < b_\mathrm{max}$.\\[1ex]

At this point, the sampled initial conditions~$(t,\vec{r},\vec{u})$ describe a DM~particle far outside the solar system, which is going to cross the solar surface.
After it is propagated to a radius~$r^\prime \gtrsim R_\odot$ analytically (see Sec.~\ref{app: kepler}), the equations of motion will be solved numerically as described in Sec.~\ref{app:runge kutta fehlberg}.

\subsection{Target velocity sampling}
\label{app: target velocity sampling}
In this appendix, we review the algorithm to sample the velocity~$\vec{v}_T$ of a solar target following Ref.~\cite{Romano2018}.

We assume a DM~particle moving with velocity~$\vec{v}_\chi$ through a plasma of temperature~$T$ when it scatters on a target~$i$ of mass~$m_i$.
While the velocity~$\vec{v}_i$ of the solar nuclei and electrons follows a Maxwell-Boltzmann distribution~$f_i(\mathbf{v}_T,T)$ given by Eq.~\eqref{eq: maxwell boltzmann}, we have to weigh each target velocity by its contribution to the scattering probability which depends on the relative speed
\begin{align}
    v_\mathrm{rel}\equiv|\vec{v}_\chi - \vec{v}_i| = \sqrt{v_\chi^2 + v_i^2 -2v_\chi v_i\mu}\, ,
\end{align}
as we can see from Eq.~\eqref{eq: total scattering rate}.
Here, we defined the cosine of the angle between the two velocities as~$\mu \equiv \cos\theta=  \frac{\vec{v}_\chi\cdot\vec{v}_i}{v_\chi v_i}$.
This means that the PDF of the scattering target velocity~$\vec{v}_T$ is given by
\begin{align}
    f(\vec{v}_T) &= {v_\mathrm{rel} \sigma_i(v_\mathrm{rel}) f_i(\vec{v}_T,T) \over \int\dd^3 \mathbf{v}^\prime_T\; v_\mathrm{rel}^\prime \sigma_i(v_\mathrm{rel}^\prime) f_i(\vec{v}_T^\prime,T)}\, . \label{eq: target velocity pdf}
\end{align}
In the DM~models assumed in this work, the total cross section never depends on the target velocity, and we thus find
\begin{align}
    f(\vec{v}_T) &= {v_\mathrm{rel} \over \langle v_\mathrm{rel} \rangle}f_i(\vec{v}_T,T)\, .
\end{align}
Furthermore, we use the isotropy of the Maxwell-Boltzmann distribution and separate the target speed~$v_T$ from the direction determined by~$\mu$ and the azimuthal angle~$\varphi$,
\begin{align}
    f(v_T,\mu,\varphi) &= {v_\mathrm{rel} \over \langle v_\mathrm{rel} \rangle}f_i(v_T,T) f(\mu) f(\varphi)\, ,
\end{align}
At this point we also introduced the speed distribution corresponding to Eq.~\eqref{eq: maxwell boltzmann},
\begin{align}
    f_i(v_T,T)&= 4\pi v_T^2 f_i(\vec{v}_T,T) =\frac{4\kappa^3}{\sqrt{\pi}}v_T^2e^{-\kappa^2 v_T^2}\, , \label{eq: maxwell boltzmann speed}
    \intertext{as well as the angles' PDFs}
    f(\mu) &= \frac{1}{2}\, , \text{ for }\mu\in[-1,1]\, ,\\
    f(\varphi)&= \frac{1}{2\pi}\, , \text{ for }\varphi\in[0,2\pi)\, .
\end{align}
The azimuthal angle~$\varphi$ can be sampled independently from the uniform distribution~$\mathcal{U}_{[0,2\pi]}$, but~$\mu$ and~$v_T$ cannot as the relative speed~$v_\mathrm{rel}$ depends on both.
This leaves us with
\begin{align}
    f(v_T,\mu) &= {\sqrt{v_T^2+v_\chi^2-2 v_T v_\chi\mu} \over \langle v_\mathrm{rel} \rangle}f_i(v_T,T) f(\mu)\label{eq: target velocity pdf 2}
\end{align}
In order to sample the random variables~$\mu$ and~$v_T$ from this bivariate distribution, we use the following theorem~\cite{devroye:1986}.
Random variables~$\mathbf{x}=(x_1,x_2,...)$ with a PDF of form
\begin{align}
f(\mathbf{x}) = C g(\mathbf{x})\psi(\mathbf{x})\, ,    \label{eq: pdf form}
\end{align}
where~$C>0$, $g(\mathbf{x})$ is itself a PDF, and~$0<\psi(\mathbf{x})<1$, can be sampled by first drawing~$\mathbf{x^\prime}$ of the distribution $g(\mathbf{x})$ and accepting it with a probability of~$\psi(\mathbf{x}^\prime)$.
If~$\mathbf{x^\prime}$ gets rejected, the procedure simply repeats until an accepted value for~$\mathbf{x^\prime}$ is found.

We can re-shape Eq.~\eqref{eq: target velocity pdf 2} into the form of Eq.~\eqref{eq: pdf form} via
\begin{widetext}
\begin{align}
    f(v_T,\mu) &= C \underbrace{\frac{\kappa v_\mathrm{rel}}{x+y}}_{\equiv \psi(\mathbf{x})} \underbrace{f(\mu) \left[ \left(\frac{\sqrt{\pi}y}{\sqrt{\pi}y+2}\right)\frac{4}{\sqrt{\pi}}x^2e^{-x^2}+\left(\frac{2}{\sqrt{\pi}y+2}\right)2x^3e^{-x^2} \right]}_{\equiv g(\mathbf{x})}\, ,
\end{align}
\end{widetext}
where we define the dimensionless parameters~$x\equiv \kappa v_T$ and~$y\equiv \kappa v_\chi$.

\begin{figure*}
    \centering
    \includegraphics[width=0.49\textwidth]{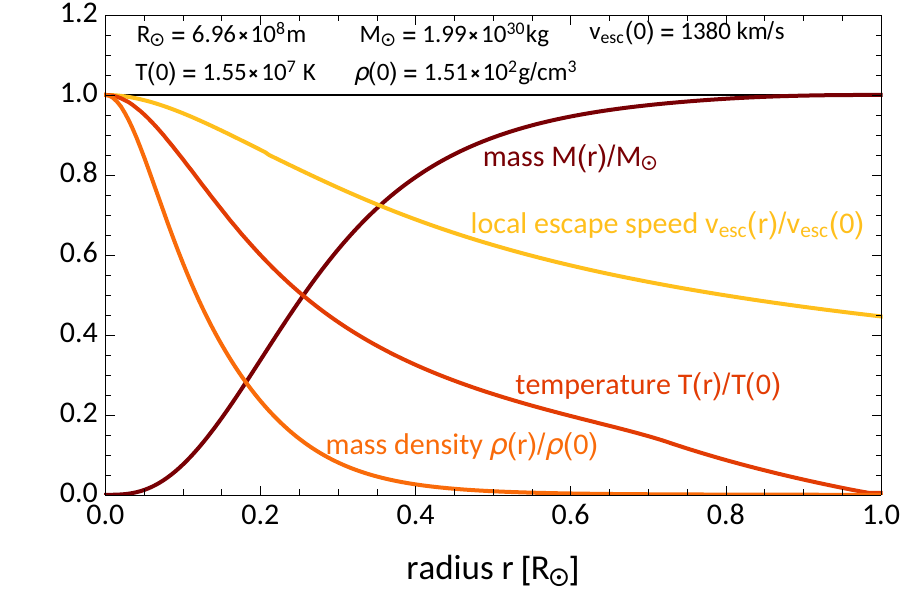}
    \includegraphics[width=0.49\textwidth]{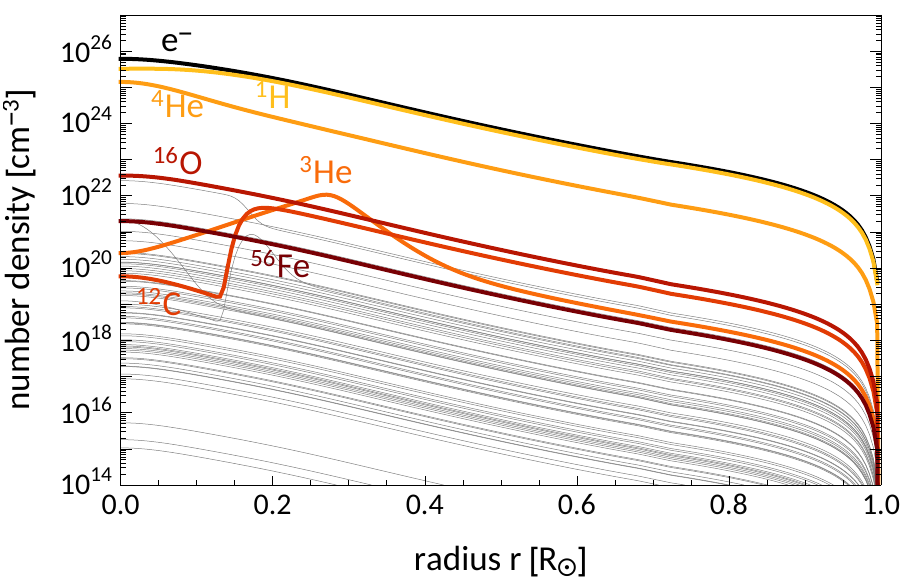}
    \caption{The Standard Solar Model: The left panel shows the Sun's local escape velocity, mass density, temperature and mass profiles. The right panel depicts the number density of the electrons and nuclei of the solar plasma. The number density of electrons is determined by demanding charge neutrality, see Eq.~\eqref{eq: solar model number density electrons}.}
    \label{fig: solar model}
\end{figure*}

In practice, we can sample~$v_T$ and~$\mu$ using this theorem through the following algorithm.
First, we draw~$\mu$ from a uniform distribution~$\mathcal{U}_{[-1,1]}$.
Next, we sample~$x$.
With a probability of $\frac{2}{\sqrt{\pi}y+2}$ we sample from the distribution $2x^3e^{-x^2}$ via
\begin{align}
    x = \sqrt{-\ln \xi_1 \xi_2}\, ,
\end{align}
where~$\xi_i$ are independently drawn from~$\mathcal{U}_{[0,1]}$.
Otherwise $x$ follows the distribution $\frac{4}{\sqrt{\pi}}x^2e^{-x^2}$, which we can sample by
\begin{align}
    x = \sqrt{-\ln \xi_1 - \cos^2\left(\frac{\pi}{2}\xi_2\right)\ln \xi_3}\, .
\end{align}
We accept the values~$(\mu=\cos\theta, v_T=x/\kappa)$ with a probability of~$\frac{\kappa v_\mathrm{rel}}{x+y}$.
In the case of a rejection, we repeat the procedure until a pair of~$(\mu,v_T)$ is accepted.

Finally, given the target speed~$v_T$ and the angles~$\varphi$ and $\theta=\arccos\mu$, the target velocity can be constructed via
\begin{align}
    \vec{v}_T = v_T \begin{pmatrix}
    u \cos\theta + \frac{\sin\theta}{s}\left(u w\cos\varphi-v\sin\varphi\right)\\
    v \cos\theta + \frac{\sin\theta}{s}\left(v w\cos\varphi+u\sin\varphi\right)\\
    w \cos\theta - s \sin\theta\cos\varphi
    \end{pmatrix}\, ,
\end{align}
where $(u,v,w)$ are the components of the unit vector pointing into the direction of the DM~velocity, i.e.~$\vec{v}_\chi/v_\chi$, and~$s\equiv \sqrt{1-w^2}$.

\section{Standard Solar Model}
\label{app:solar model}

For the simulation of DM~trajectories through the Sun, we rely on a solar model for the properties of the star's interior.
While there are many solar models in the literature, the solar reflection results depend only very weakly on this choice~\cite{An:2017ojc}.
Just like in the previous work~\cite{Emken:2017hnp}, we choose the Standard Solar Model~(SSM) AGSS09~\cite{Asplund2009,Serenelli:2009yc}.
It provides the mass-radius relation~$M(r)$, the temperature profile~$T(r)$, the density profile~$\rho(r)$, and the mass fractions~$f_i(r)$ for the 29 most common solar isotopes.
The mass, temperature, and density profile are shown in the left panel of Fig.~\ref{fig: solar model}.

For the calculation of the DM~scattering rate in Eq.~\eqref{eq: total scattering rate}, we can find the number density of the isotope~$i$ of mass~$m_i$ via
\begin{align}
    n_i(r) = f_i(r) \frac{\rho(r)}{m_i}\, . \label{eq: solar model number density nuclei}
\end{align}
In order to obtain the number density of electrons in the solar plasma, we impose charge neutrality,
\begin{align}
    n_e(r) = \sum_{i=1}^{29} n_i(r) Z_i\, , \label{eq: solar model number density electrons}
\end{align}
where~$Z_i$ is the atomic number of the respective isotope.
The number densities are depicted in the right panel of Fig.~\ref{fig: solar model}, where we highlighted the most relevant targets.

Another quantity we can derive from the solar model's parameters it the local escape velocity~$v_\mathrm{esc}(r)$, which is also shown in the left panel of Fig.~\ref{fig: solar model}.
\begin{align}
	v_{\rm esc}^2(r) =\frac{2G_NM_\odot}{R_{\odot}}\left[1+\frac{R_\odot}{M_\odot}\int\limits_{r}^{R_\odot}\dd r^\prime\;\frac{M(r^\prime)}{r^{\prime 2}}\right] \, , \label{eq: sun escape velocity}
\end{align}
where we use the Sun's radius~$R_\odot = 6.957\cdot 10^8 \mathrm{m}$ and mass~$M_\odot = 1.98848 \cdot 10^{30}\mathrm{kg}$.
Outside the Sun, the escape velocity reduces to $v_{\rm esc}^2(r) =\frac{2G_NM_\odot}{r}$.

One more important property of the Sun, which is not part of the SSM, is its velocity~$\mathbf{v}_\odot$ in the galactic rest frame.
It consists of two components,
\begin{subequations}
\label{eq: Sun velocity}
\begin{align}
    \mathbf{v}_\odot = \mathbf{v}_r + \mathbf{v}_s\, .
\end{align}
The two components are the local galactic rotation velocity~\cite{Kerr:1986hz},
\begin{align}
     \mathbf{v}_r = \begin{pmatrix} 0\\v_0 \\0\end{pmatrix}=\begin{pmatrix} 0\\220 \\0\end{pmatrix} \text{ km/s}\, ,
\end{align}
and the Sun's peculiar motion relative to the local standard of rest~\cite{Schonrich2010},
\begin{align}
     \mathbf{v}_s = \begin{pmatrix} 11.1\\12.2 \\7.3\end{pmatrix} \text{ km/s}\, .
\end{align}
\end{subequations}
As described in Sec.~\ref{app:initial conditions}, knowledge of~$\mathbf{v}_\odot$ is required for the generation of initial conditions in our simulations and for the definition of the isoreflection angle in Eq.~\eqref{eq: isoreflection angle}.

\section{Details on direct detection experiments and exclusion limits}
\label{app: experiments}

In this section, we summarize the experimental details of the different direct detection experiments, for which we derive exclusion limits in Sec.~\ref{ss: results direct detection}.

\subsection{Nuclear recoil experiments}
For SI and SD nuclear interactions, we derive exclusion limits based on the experiments CRESST-III~\cite{Abdelhameed:2019hmk,Abdelhameed:2019mac} and CRESST-surface~\cite{Angloher:2017sxg}.
The procedure to compute the exclusion limits for these experiments has been summarized in App. B of~\cite{Emken:2018run}, which also contain all experimental parameters for CRESST-surface explicitly, such as exposure, threshold, and energy resolution.
In the case of CRESST-III, the experiment consists of a $\text{CaO}_4\text{W}$ target and had an exposure of 5.6 kg days, an energy threshold of 30.1~eV, an energy resolution of 4.6~eV, and an overall flat efficiency 50\%, as summarized in Table 1 of~\cite{Abdelhameed:2019mac}.
In addition, the energy data and target-specific efficiencies were released alongside~\cite{Abdelhameed:2019mac}.

In short,  we derive the CRESST limits in this paper using Yellin's maximum gap method~\cite{Yellin:2002xd} in combination with the energy data released by the CRESST collaboration, which reproduces the official limits to good accuracy.

\subsection{Electron recoil experiments}
For DM-electron scatterings, the exclusion limits presented in Sec.~\ref{ss: results direct detection} are based on the experiments XENON10~\cite{Angle:2011th,Essig:2012yx,Essig:2017kqs}, XENON1T~\cite{Aprile:2019xxb}, SENSEI@MINOS~\cite{Barak:2020fql}, and SuperCDMS~\cite{Amaral:2020ryn}.

The exact procedure for the derivation of the XENON10/XENON1T based constraints has previously been summarized in App.~E of~\cite{Catena:2019gfa}, while a summary of the corresponding information for SENSEI@MINOS can be found in Sec. IV D of~\cite{Catena:2021qsr}.
For the latter, we use Poisson statistics to derive an exclusion limit based on each electron-hole pair ($Q)$ bin and identify the strongest of these bounds. For SuperCDMS, we follow the same steps as for SENSEI@MINOS. The reported exposure was reported to be 1.2 gram day. We assume a threshold of a single electron-hole pair and include the first six $Q$~bins in our analysis.
The number of events in each bin are (178800,1320,248,64,19,6) as extracted from table I of~\cite{Barak:2020fql}.

 \bibliography{ref}

\end{document}